\documentclass[footinbib,twocolumn,amsmath,amstex,amssymb,mathfonts,superscriptaddress,prx,longbibliography]{revtex4-1}
\usepackage{graphicx}
\usepackage[usenames, dvipsnames]{color}
\usepackage{bm}
\usepackage{verbatim}
\usepackage{amsmath}
\usepackage{amssymb}
\usepackage{amsthm}
\usepackage{amsfonts}
\usepackage{float}
\usepackage{soul}
\usepackage[caption=false]{subfig}

\def\blue{\textcolor{black}}
\def\GJ{\textcolor{black}}

\usepackage{bbm}

\begin{document}

\title{Unraveling non-Hermitian pumping: emergent spectral singularities and anomalous responses}

\author{Ching Hua Lee}
\email{phylch@nus.edu.sg}
\affiliation{Department of Physics, National University of Singapore, Singapore, 117542.}
\author{Linhu Li}
\affiliation{Department of Physics, National University of Singapore, Singapore, 117542.}
\author{Ronny Thomale}
\affiliation{ Institute for Theoretical Physics and Astrophysics, University of W\"urzburg, Am Hubland, D-97074 W\"urzburg, Germany}
\author{Jiangbin Gong}
\affiliation{Department of Physics, National University of Singapore, Singapore, 117542.}

\date{\today}
\begin{abstract}
\noindent Within the expanding field of non-Hermitian physics, non-Hermitian pumping has emerged as a key phenomenon, epitomized through the skin effect via extensive boundary mode accumulation modifying the conventional Bloch picture. Beyond redefining bulk-boundary correspondences, we show that non-Hermitian pumping induces an unprecedented type of spectral topology: it admits a classification in terms of graph topology, which is distinct from conventional topological 
classifications of the eigenstate or energy Riemann surface. Each topological class is characterized by a conformally invariant configuration of spectral branching singularities, with gap-preserving transitions giving rise to emergent band geometry and Berry curvature discontinuities physically manifested as anomalous response kinks. 
By placing all Hermitian and non-Hermitian lattice Hamiltonians on equal footing, our comprehensive framework also enables the first analytic construction of topological phase diagrams in the presence of multiple non-reciprocal coupling scales, as prototypically demonstrated for the extended non-Hermitian Chern and Kitaev models. 
Based on general algebraic geometry properties of the energy dispersion, our framework can be directly generalized to multiple bands, dimensions, and even interacting systems. Overall, it reveals the conspiracy of band representations, spectral topology, and complex geometry
as it unfolds in directly measurable quantities.
\end{abstract}

\maketitle

\section{Introduction} 
The realm of non-Hermiticity harbors a variety of spectacular yet non-intuitive phenomena. Novel Fermi surface properties such as enhanced amplification emerge when energy bands intersect along highly degenerate exceptional manifolds~\cite{berry2004physics,rotter2009non,1751-8121-45-44-444016,ashida2017parity,hassan2017dynamically,xu2017weyl,zhou2019exceptional,shen2018topological,carlstrom2018exceptional,moors2019disorder,lee2018tidal,Wang2019non,Yang2019non,carlstrom2019knotted,yoshida2019symmetry,yashida2019exceptional,okugawa2019exceptional}, and likewise new topological classes appear when the constraints of Hermiticity are relaxed~\cite{gong2018topological,kawabata2019symmetry,Liu2019nonHclass,zhou2019periodic,Li2019geometric,liu2019topological,kawabata2019classification,okuma2019topological,borgnia2019non,hockendorf2019non}. Capturing the attention of much recent theoretical and experimental effort is the phenomenon of non-Hermitian pumping (skin effect)~\cite{yao2018edge,xiong2018does,yao2018non,alvarez2018non,kunst2018biorthogonal,jin2018bulk,lee2019anatomy,yokomizo2019nonbloch,nobuyuki2019gbz,kai2019gbz,zhesen2019agbz,lee2020many}, where eigenmodes are relentlessly ``pumped" towards the boundaries due to effectively asymmetric gain/loss~\footnote{Non-Hermitian pumping also exists in fully reciprocal systems, where the asymmetry only appears when the effective description is restricted to certain momentum subspaces.}. As intuitively expected, this pumping results in extreme sensitivity to finite-size and boundary effects, as well as the intensely studied modifications of topological bulk-boundary correspondences (BBCs). 



The physical and formal implications of non-Hermitian pumping, however, extend far beyond modified band structures or topological descriptions. As we shall reveal, non-Hermitian pumping also leads to a new type of classification in terms of graph topology, marked by physical signatures such as response kinks. It generically deforms the complex band structure into a graph-like structure with characteristic branching singularities, with topological transitions corresponding to unconventional anomalous linear responses. Characterized by a graph-theoretic spectral classification, this newly-defined type of topology is distinct from conventional topological characterizations~\cite{shen2018topological,gong2018topological} i.e. winding properties of either the eigenstates or the energy Riemann surface, which in the simplest instances give rise to $\mathbb{Z}$ windings and vorticities. Transitions between our topological singularities manifest as discontinuities in the Fubini Study metric of the eigenbands, whose imaginary part corresponds to the Berry curvature. Adding to their enigmatic allure is that such complex singularity transitions, which can result in anomalous response behavior, do not necessitate band gap closures, unlike conventional topological transitions which rely on gap closures for any discontinuous evolution of the bands.



With the rise of experimental platforms such as topological lasers, photonic crystals, mechanical frameworks and circuits for non-Hermitian phenomena~\cite{longhi2015robust,midya2018non,PhysRevB.100.075423,hofmann2019chiral,zhao2019non,longhi2019nonlaser,schomerus2019nonreciprocal,brandenbourger2019non,fruchart2019dualities,tobias2019observation,tobias2019reciprocal,ananya2019observation,lei2019observation}, and non-Hermitian pumping in particular~\cite{tobias2019observation,tobias2019reciprocal,ananya2019observation,lei2019observation}, a comprehensive understanding of these exotic singularities and responses is of practical and theoretical exigence. 
As such, we devise a universal framework that puts generic non-Hermitian lattice Hamiltonians on equal footing as their Hermitian counterparts, which are immune to the skin effect. Specifically, we formulate a unitary restoration procedure that maps any non-Hermitian model to a \emph{quasi-reciprocal} surrogate Hamiltonian at the \emph{operator} level, such that the effectively restored bulk Hilbert space allows topological invariants and metrics to faithfully predict topological phase boundaries just like in genuinely reciprocal or Hermitian systems. Literally, this procedure ``unravels'' non-Hermitian pumping through a redefined non-local basis where the accumulated eigenmodes appear equilibrated (Fig.~1a). Figuratively, it illuminates the deeper implications of non-Hermitian pumping beyond what can be predicted from simply defining a generalized Brillouin zone (BZ)~\cite{yao2018edge,lee2019anatomy,yokomizo2019nonbloch,nobuyuki2019gbz,kai2019gbz,zhesen2019agbz}. Most salient are the non-perturbative effects implied by additional couplings: While we ordinarily expect weak couplings across distant sites to at most trivially modify the band structure, they, no matter how weak, can generically produce more complex topological singularities when non-Hermitian pumping is present. Such enigmatic behavior is a consequence of the emergent non-locality that also underscores Berry curvature discontinuities in the absence of band touchings.

For concreteness, we shall illustrate our findings with two quintessential non-Hermitian models that have so far eluded rigorous characterization: the extended non-Hermitian Kitaev chain and the extended non-Hermitian Chern insulator. To explore the interesting singularities, we introduce in both models asymmetric couplings beyond nearest neighbors (NNs), which are also physically relevant in realistic photonics and plasmonics setups governed by non-compact orbitals or long-ranged Coulomb forces. In the minimal description of the extended non-Hermitian Kitaev chain, which is of D$^\dagger$-class topology \footnote{It is distinct from the non-Hermitian SSH model which also possesses $\mathbb{Z}_2$ topology.} ($\mathbb{Z}_2$ with conjugated particle-hole symmetry (cPHS) and a real line-gap)~\cite{kawabata2019symmetry,Li2019geometric}, both NN and next-nearest-neighbor (NNN) couplings are in fact necessary and hence essential aspects of this non-Hermitian topological class. Our extended non-Hermitian Chern insulator, which is pedagogically designed to reduce to a minimal 1D description with two effective non-reciprocal couplings, describes the only other known singularity class (besides the well-known NN non-Hermitian Chern insulator~\cite{yao2018non}) where the surrogate Berry curvature, which reliably predicts Chern edge modes, can be analytically computed.

This paper is organized as follows. In Section II, we review known properties of non-Hermitian pumping, and subsequently introduce the concept of a quasi-reciprocal surrogate Hamiltonian that implements the generalized BZ at the operator level. The consequent emergent non-locality of the surrogate basis will be a recurring theme of this work. In Section III, we continue with a detailed treatment of a few common spectral singularities, followed by a discussion of their classification and topological transitions. In particular, we illustrate our formalism with two detailed examples: the 1D non-Hermitian extended Kitaev model and the 2D extended non-Hermitian Chern model, where the construction of the surrogate basis with nontrivial generalized BZ proves crucial for topological characterization and extraction of Berry curvature discontinuities. In Section IV, we elaborate on such discontinuities and their role as phase transitions that occur without any gap closure. Finally, we conclude in Section V that we have developed the appropriate classification scheme of non-Hermitian physics from the viewpoint of non-Hermitian pumping, which elucidates and encodes interesting principal phenomena expected to emanate from non-Hermitian systems.

\section{Unraveling non-Hermitian pumping} 

\subsection{Preliminaries}

We first briefly review the rudiments of non-Hermitian pumping in non-Hermitian lattices. 
Consider a 1D effective Hamiltonian described by 
\begin{equation}
H=\sum\limits_{ij;\alpha\beta}h^{\alpha\beta}_{ij}c^\dagger_{i\alpha}c_{j\beta}=\sum\limits_{k;\alpha\beta}H^{\alpha\beta}(k)c^\dagger_{k\alpha}c_{k\beta},
\end{equation}
where $h^{\alpha\beta}_{ij}$ and $H^{\alpha\beta}(k)$ are its real space and momentum space representations, and $ij$, $k$, $\alpha\beta$ indexes unit cells, momentum and intra-cell orbitals, respectively. Non-Hermitian pumping, also known as the non-Hermitian skin effect, is an extensive accumulation of eigenmodes that occur when \emph{all} eigenmodes are ``pumped'' towards the boundaries under open boundary conditions (OBCs). Intuitively, it occurs when the 1D effective description contains gain/loss terms that couple asymmetrically in real space.  Indeed, it can be shown that~\cite{lee2019anatomy} the necessary condition for non-Hermitian pumping is that the effective 1D description $H$ is simultaneously non-Hermitian and non-reciprocal~\footnote{These conditions are generically sufficient unless there is additional symmetry obstruction, see Ref.~\cite{nobuyuki2019gbz}.}, which can be respectively expressed as the first and second of the following inequalities: $h_{ji}^{\beta\alpha}\neq h_{ij}^{\alpha\beta}\neq [h_{ji}^{\beta\alpha}]^*$. In momentum space, these conditions take the form $H^T(-k)\neq H(k)\neq H^\dagger(k)$. In other words, there must either exist coupling terms whose magnitudes are direction-dependent, or there must be the simultaneous presence of magnetic flux and on-site gain/loss~\cite{li2019topology}. 
In 2D or higher, $H(k)$ also implicitly contain momentum parameters in other directions perpendicular to the boundary, and it is possible that a fully reciprocal (but still non-Hermitian) lattice can still exhibit non-reciprocity in the effective 1D description~\cite{tobias2019reciprocal} ($H^T(-k)\neq H(k)$). Thereafter, we shall refer to this mode accumulation only as non-Hermitian pumping, with the implicit understanding that it occurs only when the 1D description is also non-reciprocal.

The conditions of non-reciprocity and non-Hermiticity, which leads to non-Hermitian pumping, also has intuitive interpretations in terms of the energy spectrum. In Hermitian cases, the periodic boundary condition (PBC) spectrum $\epsilon(k)$, $k\in [0,2\pi)$ is confined to the real line, and in reciprocal cases, $\epsilon(k_0-k)=\epsilon(k_0+k)$ mandates a degenerate loop, $k_0$ a fixed system-dependent parameter. But the simultaneous presence of non-reciprocity and non-Hermiticity relaxes both of these constraints, allowing $\epsilon(k)$ to generically trace closed loops with non-vanishing areas in the complex energy plane (Fig.~\ref{fig:skin}b,c). Since $k$ is a periodic parameter label, these loops are necessarily closed even if the eigenenergy bands switch partners after every period~\GJ{\cite{lee2019anatomy,kai2019gbz,longhi2019probing,Li2019geometric}}.

Non-Hermitian pumping under OBCs causes extensive boundary accumulation which is not in line with the conventional Bloch picture. As such, we expect the OBC spectrum to deviate considerably from the PBC spectrum, which corresponds to Bloch-type eigensolutions at real momenta (Fig.~\ref{fig:skin}). This ostensible breakdown of BBC is the hallmark of non-Hermitian pumping~\cite{lee2019anatomy}. Mathematically, it can be expressed as the extensive spectra flow $\epsilon(k)\rightarrow \bar\epsilon(k)$ into the interior of PBC loop/s $\{\epsilon(k)\}$ as we interpolate between PBCs and OBCs (or, more generally, by adding spatial non-uniformity). In the thermodynamic limit, $\bar\epsilon(k)$, $k\in [0,2\pi)$ converges to the OBC spectrum, excluding its topological modes which are isolated protected eigensolutions~\cite{lee2019anatomy}. Although we have explicitly referred to $H(k)$ and $\epsilon(k),\bar\epsilon(k)$ as the Hamiltonian and eigenenergies, the above conceptual review and most of the rest of this paper applies equally well to generic operators, i.e., also the circuit Laplacian and their eigenspectra.



\subsection{Quasi-reciprocal surrogate Hamiltonian} 

To study the effects of non-Hermitian pumping, we introduce the \emph{surrogate} Hamiltonian 
\begin{equation}
\bar H(k)=H(k+i\kappa(k)),
\label{kappa}
\end{equation}
where $\kappa(k)$ is defined such that the $\bar H(k)$ eigenstate of interest experiences no non-Hermitian pumping when it is put under OBCs - our so-called property of \emph{quasi-reciprocity}~\footnote{A quasi-reciprocal system is not necessarily reciprocal, since it can still be non-reciprocal while exhibiting unbroken bulk-boundary correspondence i.e. with Hermitian flux and no gain/loss. } 
(Fig.~1). To be specific, $\kappa(k)$ is given by the smallest (magnitude-wise) complex deformation of the momentum $k$ such that the eigenvalues $\bar\epsilon(k)$ of $\bar H(k)$ lie at the endpoint of the PBC-OBC spectral flow. This formalism fully encodes the effects of non-Hermitian pumping at the operator level, beyond existing works~\cite{yao2018edge,lee2019anatomy,yokomizo2019nonbloch,nobuyuki2019gbz,kai2019gbz,zhesen2019agbz} that introduce a generalized BZ (complex analytic continuation of the momentum) for finding the skin eigenmodes. In other words, given any physical Hamiltonian $H(k)$, we define a surrogate Hamiltonian $\bar H(k)$ possessing almost identical OBC spectra~\footnote{except for subextensive topological modes} but avoids the complications of non-Hermitian pumping. This is further elaborated on in Sect.~\ref{sec:surrogatebasis}. Physically, $\bar H(k)$ provides an effective description of the OBC system after the non-reciprocally pumped modes have ``equilibrated'' at the boundaries. Most importantly, it experiences no further pumping, and is hence characterizable by \emph{all} approaches valid for reciprocal or Hermitian systems which obey the BBC.
By representing the effects of non-Hermitian pumping as a generically non-analytic momentum deformation $\kappa(k)$, we shall soon uncover manifold exotic non-analytic behavior not present in the simplest case of constant $\kappa$ as in commonly studied models~\cite{yao2018edge,yao2018non}. We emphasize that the OBC and PBC systems possess their own distinct eigenspaces, and it has to be the OBC $\bar H(k)$, not the PBC $H(k)$, that determines all physical responses of a bounded system (which is under OBCs by definition), even those concerning ``bulk" properties such as the Berry curvature. 

\begin{figure}
 \centering
\includegraphics[width=.99\linewidth]{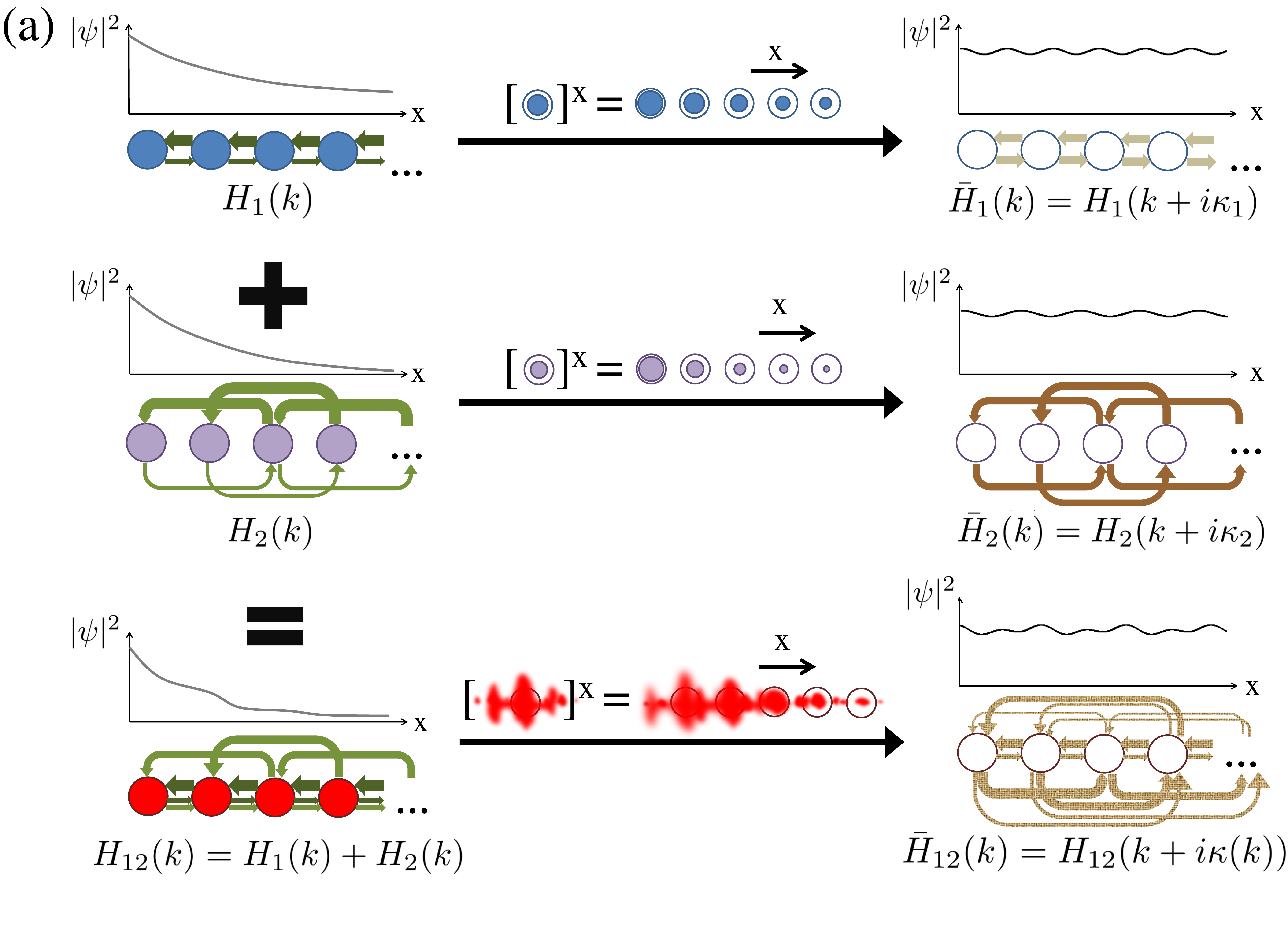}\\
\includegraphics[width=.99\linewidth]{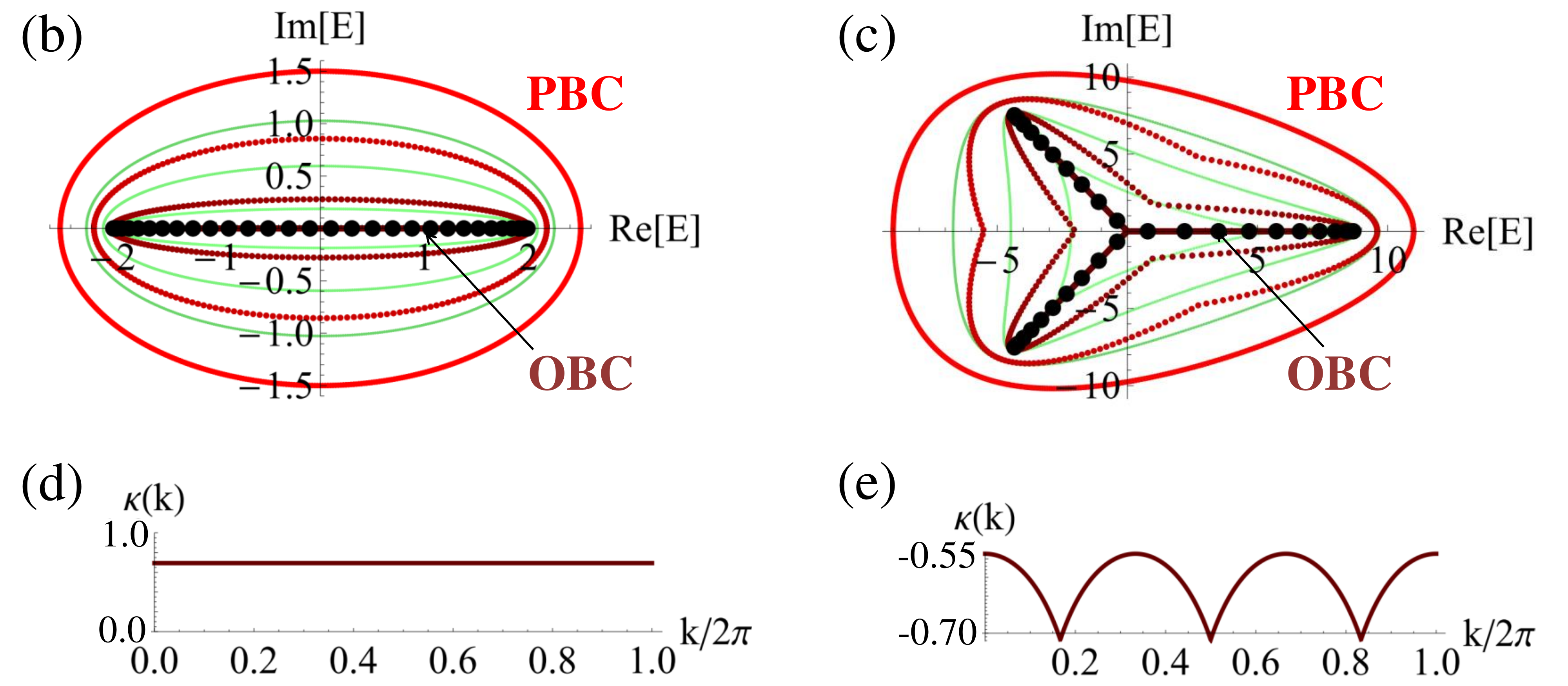}
\caption{(a) Construction of quasi-reciprocal Hamiltonians $\bar H_1(k),\bar H_2(k),\bar H_{12}(k)$ (Right Column) from various physical Hamiltonians $H_1(k),H_2(k),H_{12}(k)$ (Left Column) which are afflicted by non-Hermitian pumping. With non-reciprocal couplings across only one distance scale (Top or Middle Row, corresponding to NNs in $H_1$ or NNNs in $H_2$), spatial eigenmode accumulation can be nullified with a simple spatial basis rescaling corresponding to a constant $\kappa=\kappa_1$ or $\kappa_2$. But the equilibration of the combination of two or more pumping length scales in $H_{12}=H_1+H_2$ (Bottom Row) requires a non-local basis redefinition (Bottom Center) dictated by nonconstant $\kappa(k)$ which is not equivalent to $\kappa_1$ or $\kappa_2$, resulting in a non-local $\bar H_{12}(k)$. (b-c) PBC $\epsilon(k)$ (red), OBC $\bar \epsilon(k)=\epsilon(k+i\kappa(k))$ (black) spectra and their interpolations $\epsilon(k+ir\kappa(k))$ (various shades of brown) where $r=0.4,0.8,1$. (b) and (c) depict $E=2z+\frac1{2z}$ and $E=z^2+\frac{10}{z}$ where $z=e^{ik}$, representing cases with one and two pumping scales respectively. Background light green curves are contours of constant $\kappa$. (d-e) $\kappa(k)$ for cases (b) and (c), contrasting the constant $\kappa=\log\,2\approx 0.69$ cases (d) with a non-constant $\kappa(k)$ with cusps that indicate emergent non-locality (e). }
\label{fig:skin}
\end{figure}

We now describe how the quintessential complex deformation $\kappa(k)$ can be computed. Intuitively, it is the $k$-dependent deformation of the PBC spectral loops~\footnote{Up to exponentially small corrections in system size, see Refs.~\cite{alvarez2018non,lee2019anatomy}.} $\epsilon(k)\rightarrow \epsilon(k+i\kappa(k))=\bar\epsilon(k)$ such that $\bar\epsilon(k)$, $k\in[0,2\pi)$ collapses into one or more arcs or lines (Fig.~\ref{fig:skin}b,c) which cannot be contracted even further. More precisely, $\kappa(k)$ can be determined from the
characteristic Laurent polynomial of the eigenenergy equation $\text{Det}\,[H(z)-E\,\mathbb{I}]=0$, which generically can be written as a bivariate polynomial $P(E,z)=0$. We shall use $E$ to refer to the energy as an algebraic variable, and use $\epsilon(k),\bar\epsilon(k)$ when it is also an eigenenergy of the original/surrogate Hamiltonian. For particle-hole symmetric 2-component Hamiltonians with only off-diagonal entries for instance, the eigenenergy equation
assumes the form~\cite{suppmat}
\begin{equation}
E^2=\sum_{n=-l_L}^{l_R} t_n z^n
\label{Poly}
\end{equation}
with $z=e^{i(k+i\kappa(k))}$ and $l_L,l_R$ the sum of the maximal ranges of the left/right couplings over both components/sublattices. While the $t_n$'s coincide with the physical couplings in single-component systems, they are sums of product of the couplings. In more general cases, $t_n$ also depends on $E$, and thus do not directly correspond to any particular group of couplings. 

For each $k$, $\kappa(k)$ is thus the \emph{smallest} complex deformation for which there exists another momentum $k'$ such that \emph{both} $z=e^{ik}e^{-\kappa(k)}$ and $z'=e^{ik'}e^{-\kappa(k)}$ are roots satisfying the same eigenenergy $E$. In other words, $\kappa(k)=\kappa(k')$ are the symmetric deformations necessary to make the eigenenergies of $\bar H(k)$ and $\bar H(k')$ coincide, as geometrically evident from Fig.~\ref{fig:skin}b,c. The loci of $E$ where this occurs precisely constitute the OBC pumped spectra $\bar\epsilon(k)$.

Note that while $\bar H(k)$ is quasi-reciprocal, i.e., immune to non-Hermitian pumping, it is not necessarily reciprocal. Reciprocity requires symmetric physical couplings ($h_{ij}^{\alpha\alpha}=h_{ji}^{\beta\alpha}$) for all pairs $ij$ and $\alpha\beta$, and is a stronger condition than quasi-reciprocity, which requires $\kappa(k)=0$ $\forall k$, a constraint\footnote{Non-Hermitian pumping occurs whenever $t^*_n\neq t_n\neq t_{-n}$ for at least one $n$.} on the relatively small number of $t_n$ coefficients from Eq.~\ref{Poly} formed by products and sums of the physical couplings.

An important point that will be shortly elaborated on is that $\kappa(k)$ has nontrivial $k$-dependence whenever couplings exist across a range of distances such that $t_n\neq t_{-n}$ for more than one $n$. This is actually the case in most realistic systems, since non-Hermiticity from just one type of coupling is enough to cause $t_n\neq t_{-n}$ for many different $n$. Such $\kappa(k)$ dependences are nonanalytic in general, and lead to various forms of spectral singularities depending on the algebraic form of the characteristic polynomial.

\subsection{Surrogate non-local basis}
\label{sec:surrogatebasis}

We next discuss the physical interpretation of the $k\rightarrow k+i\kappa(k)$ deformation in terms of the Hilbert space basis, going beyond existing generalized BZ descriptions. A central motivation of our framework is that this deformation can be regarded not just as an esoteric BZ redefinition, but as a \emph{physical} change of basis orbitals. This is because OBCs allow for much greater freedom in basis transforms than PBCs, which require the Bloch nature (translation invariance) to be preserved. More precisely, the surrogate Hamiltonian $\bar H$ (its PBC version was defined in Eq.~\ref{kappa}) admits a similarity transform $S$ such that
\begin{equation}
H^\text{OBC}= S^{-1}\bar H^\text{OBC}S\simeq S^{-1}\bar H^\text{PBC}(k)S,
\end{equation}
where $S$ undoes, or ``unravels'', the complex deformation by implementing the complex gauge transform associated with $k\rightarrow k-i\kappa(k)$. In terms of eigenenergies, we have
\begin{equation}
\{\varepsilon^\text{OBC}\}=\{\bar\varepsilon^\text{OBC}\}\simeq\{\bar\varepsilon^\text{PBC}(k)\},
\label{energies}
\end{equation}
where the $\simeq$ sign denotes an approximate equivalence that projects out, i.e., excludes isolated topological eigenmodes. In other words, spectrum of $H^\text{OBC}$, which is an OBC spectrum of a system that has been subject to non-Hermitian pumping, is exactly equivalent to the spectrum of $\bar H^\text{OBC}$ without the pumping, which is further equivalent up to a set of measure zero (in the thermodynamic limit) to that of $\bar H^\text{PBC}$. The upshot of this discussion is that, due of the existence of $S$, the surrogate Hamiltonian $\bar H(k)$ describes a bona-fide physical lattice system whose bulk properties are not susceptible to non-Hermitian pumping, and can be used to predict the topology and responses of the original Hamiltonian $H(k)$. Note that this procedure is not applicable to disordered systems which also experience non-Hermitian pumping due to spatial non-homogeneity, since the pumping can no longer be ``gauged'' away by a unique $\kappa(k)$.

While $S$ can be numerically computed by taking the quotient of the matrices that diagonalize $H^\text{OBC}$ and $\bar H^\text{OBC}$, insight into its physical ramifications can be gleaned from Fourier expanding the rescaling factor $e^{-\kappa(k)}$. A nonconstant $\kappa(k)$ renders the eigenequation nonanalytic in $e^{ik}$, leading to emergent non-locality in real-space that is difficult to realize in models with few hoppings~\footnote{Some exceptions arise in Floquet systems, with rapid quenching behavior giving rise to high temporal and spatial harmonics or dynamical instabilities~\cite{mikami2016brillouin,zhou2018recipe,li2018realistic,lee2020ultrafast}.}. To see that, note that at each $x$ site, $k\rightarrow k+i\kappa(k)$ replaces the Bloch prefactor $e^{ikx}$ by 
\medmuskip=-2mu
\thickmuskip=-2mu
\begin{equation}
\left(e^{ik}e^{-\kappa(k)}\right)^x\approx \left(e^{ik}\sum_{l=-l_m}^{l_m} \Gamma_l e^{ilk}\right)^x=\sum_{l'=-l_mx}^{l_mx}\Gamma'_{l'}(x)e^{i(l'+x)k},
\label{realbasis}
\end{equation}
\medmuskip=2mu
\thickmuskip=2mu
with $\Gamma_l$ denoting the Fourier coefficients of $e^{-\kappa(k)}$ and $\Gamma'_l(x)$ their multinomial sum. $\Gamma_l$ are generically power-law decaying~\footnote{
See for instance Ref.~\cite{he2001exponential,lee2015free} for a treatment of how Fourier coefficient decay rates depend on complex analytic structure.} due to the non-analyticity, which we can truncate at large orders $\pm l_m$ for convenient numerical treatment. 
Hence we can alternatively interpret the complex deformation as a non-local basis redefinition, where $\bar H(k)$ is re-interpreted as the non-deformed $H(k)$ acting in a non-local basis with each site replaced by a linear combination of sites according to Eq.~\ref{realbasis}, each rescaled by $\Gamma'_{l'}(x)$ (Fig.~\ref{fig:skin}a).


\blue{\section{quasi-reciprocal solutions for multiple non-reciprocal length scales} }
Having discussed the formal aspects of our framework, we next provide a few canonical illustrations on how to
\blue{solve the surrogate Hamiltonian for systems with multiple non-reciprocal length scales,
where specific singularities emerge in the OBC $\bar \epsilon(k)$}.  \GJ{We shall connect their branching patterns with the number of coexisting non-reciprocal length scales.
On the other hand, the quasi-reciprocal surrogate Hamiltonians we obtain are also essential for locating topological phase transitions featuring topological boundary modes,
as we illustrate with two concrete examples below.}



\blue{\subsection{One non-reciprocal length scale}}
As a warm-up, we consider the simplest case where the characteristic polynomial $P(E,z)$ can be separated into parts containing $E$ and $z$ separately:
\begin{equation}
F(E)=t_+z+\frac{t_-}{z}\,+t_0,
\label{Poly2}
\end{equation}
with $t_+\neq t_-$, and $t_0$ denoting an unimportant energy offset. We refer to this as the case with one non-reciprocal length scale because asymmetry only occurs in the $z,z^{-1}$ terms, and no other higher powers. $F(E)$ is an arbitrary function of $E$ which will turn out to have no nontrivial bearing on the singularity. For single-component models, $F(E)=E$, and $t_+,t_-$ are the asymmetric (non-reciprocal) right and left couplings. Note, however, that for multi-band models, $t_+,t_-$ may not directly correspond to the bare couplings. More complicated models that possess one non-reciprocal length scale can be described by Eq.~\ref{Poly2} with modified $F(E)$ i.e. the non-Hermitian SSH model~\cite{suppmat} $H_\text{SSH}(z)=(t_-+z)\sigma_++(t_++z^{-1})\sigma_-$, where $\sigma_\pm=(\sigma_x\pm i\sigma_y)/2$ are linear combinations of the Pauli matrices. For $H_\text{SSH}(z)$, the characteristic polynomial reads $E^2=t_+z+\frac{t_-}{z}+t_+t_-+1$, such that $F(E)=E^2-t_+t_--1$.  
 
To determine $\kappa(k)$, we transform Eq.~\ref{Poly2} to a more convenient form by substituting $z=e^{-\kappa(k)}w$, $|w|=1$, such that 
\begin{equation}
F(E)= w\,e^{-\kappa(k)}\sqrt{\frac{t_+}{t_-}}+w^*e^{\kappa(k)}\sqrt{\frac{t_-}{t_+}},
\label{Poly3}
\end{equation}
where $F(E)=\frac{E^2-t_0}{\sqrt{t_+t_-}}$. Evidently, $w=e^{ik}$ and $w^*=e^{-ik}$ will take symmetrical roles if $\kappa(k)$ is equal to a constant $\kappa_0$ defined by $e^{\kappa_0}=\sqrt{\frac{t_+}{t_-}}$. In this case, both $z=w\,e^{-k_0}$ and $z'=w^*e^{-k_0}$ are simultaneously roots of Eq.~\ref{Poly2} for the same $E$. As such, we obtain a constant complex deformation $k\rightarrow k+i\kappa_0$ with
\begin{equation}
\kappa(k)=\kappa_0 =\log\sqrt{\frac{t_+}{t_-}}.
\end{equation}
From Eq.~\ref{realbasis}, we find that there exists only one nonzero basis redefinition coefficient $\Gamma_0=e^{-\kappa_0}$ resulting from one non-reciprocal length scale. This simple result can be visualized as a spatial exponential rescaling~\cite{yao2018edge} $\sim e^{-\kappa_0 x}$ that counteracts the non-Hermitian pumping (Fig.~\ref{fig:skin}a), which geometrically takes the form of a nonconformal contraction of the spectral loop in the complex energy plane (Fig.~\ref{fig:skin}b).
\begin{figure*}
 \centering
\subfloat[]{\includegraphics[width=.61\linewidth]{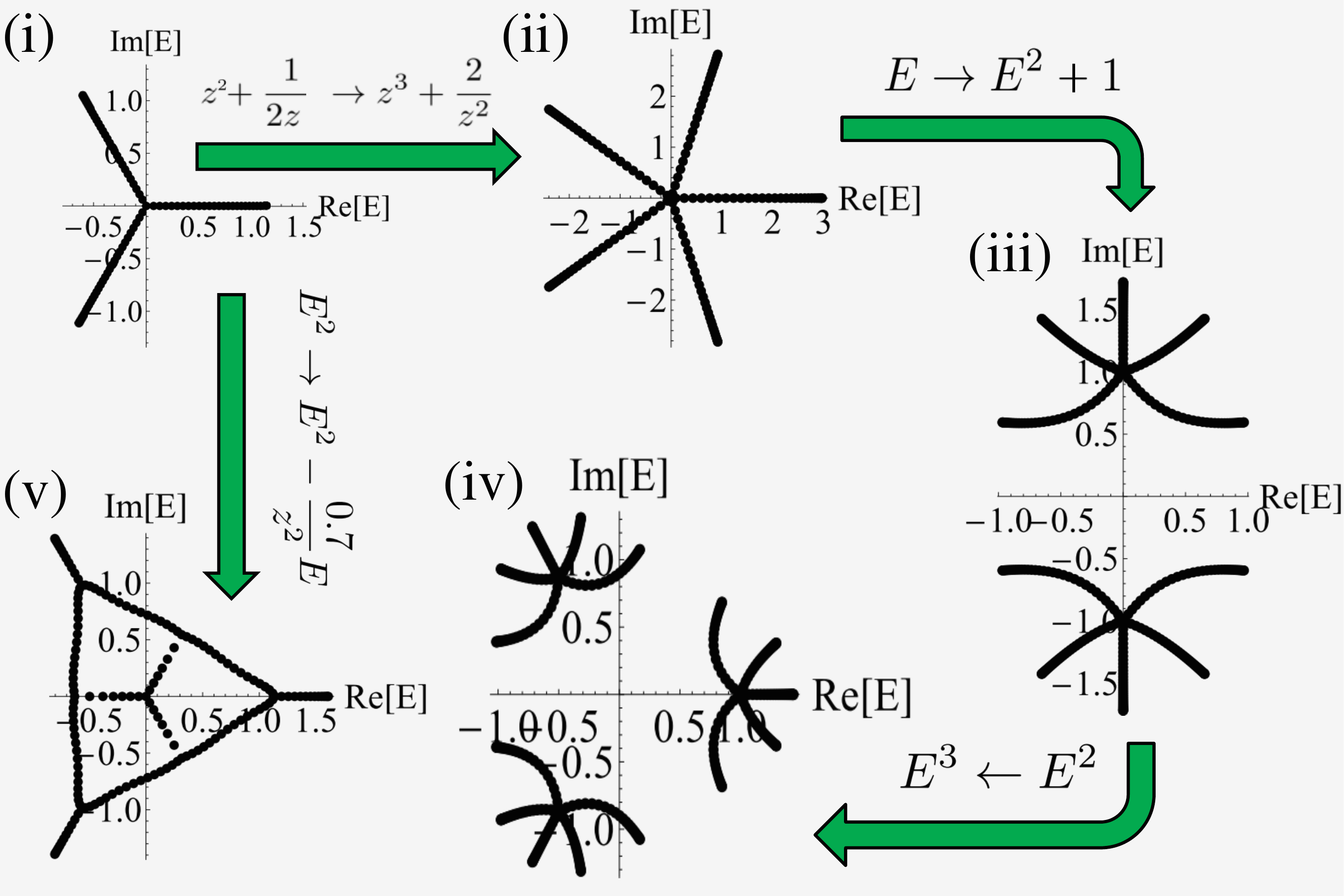}}
\subfloat[]{\includegraphics[width=.39\linewidth]{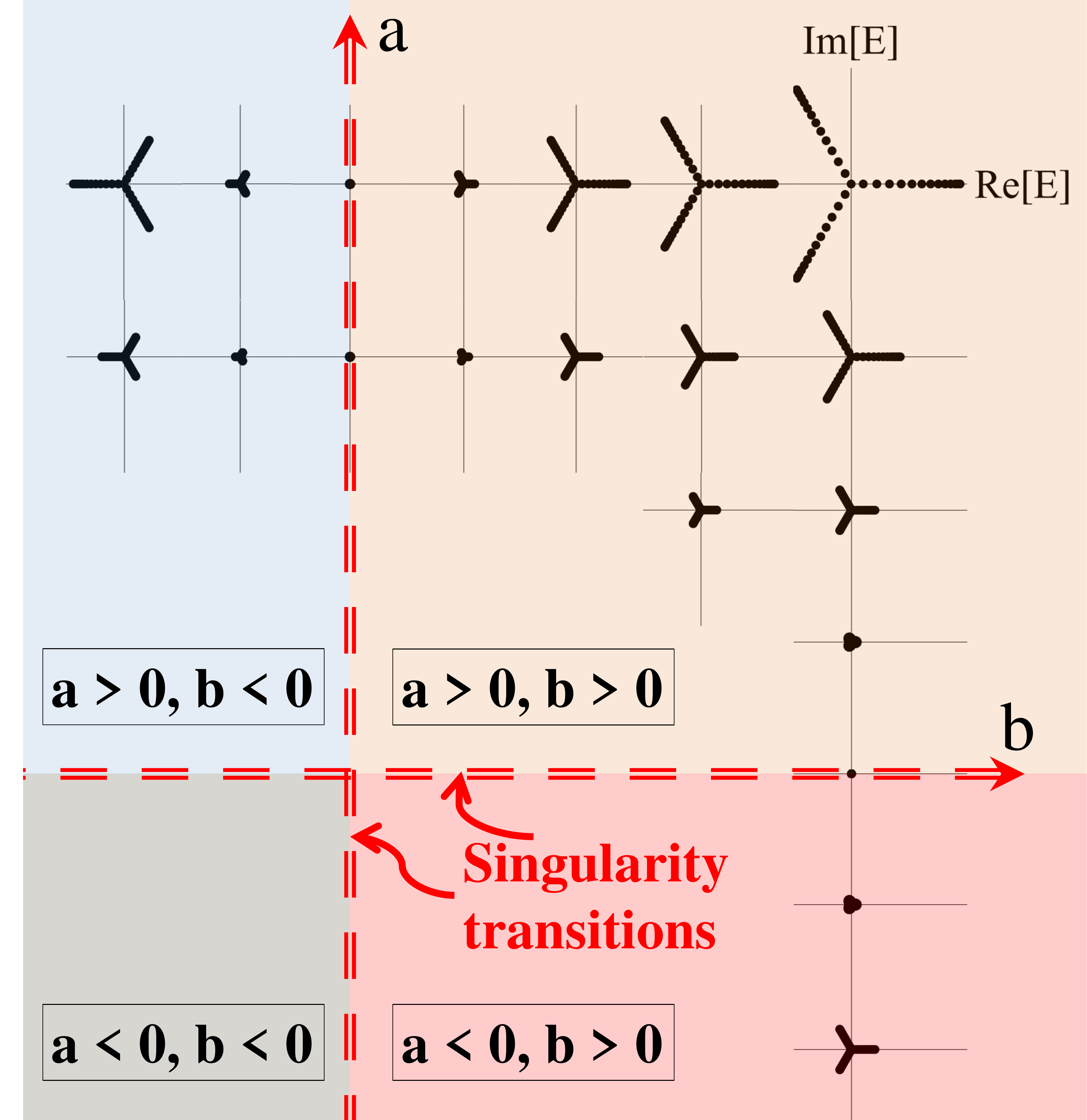}}
\caption{(a) OBC spectra of various Hamiltonians with characteristic polynomials related by complex mappings. Starting from $E=z^2+\frac1{2z}$ in (i), the branching number is increased to $3+2=5$ in (ii), whose spectrum is then split into $N=2$ and then $N=3$ with $E^N=-1+z^3+\frac{2}{z^2}$ (iii and iv). However, mappings of $E$ that also contain $z$ lead to much more complicated OBC spectral graphs, as in (v) with a graph cycle and $7$ three-fold singularities. (b) Illustration of singularity transitions with $E=az^2+b/z$, where the OBC spectra $\bar\epsilon$ shrinks to a point and morphs into possibly different shapes or orientations along lines $a=0$ and $b=0$.  }
 \label{fig:quasi}
\end{figure*}
\\
\\

\blue{\subsection{Multiple non-reciprocal length scales and spectral singularities}}
\noindent\underline{1. Two non-reciprocal length scales:}
Going beyond analytic characterizations in the existing literature~\cite{yao2018edge,lee2019anatomy,zhesen2019agbz}, we consider the next simplest characteristic polynomial with broken reciprocity at two scales (two different powers of $z,z^{-1}$), where $P(E,z)$ takes the form:
\begin{equation}
F(E)=z^2+\frac{b}{z},
\label{Poly3}
\end{equation}
which is also schematically illustrated in the bottom row of Fig.~\ref{fig:skin}a. Eq.~\ref{Poly3} represents the simplest classes of non-Hermitian Hamiltonians with couplings beyond NNs, as in our two illustrative models presented later. As before, we have collected all dependence on $E$ in the function $F(E)$, whose detailed form will be irrelevant for the branching topology of $\bar\epsilon(k)$. To find the $\bar\epsilon(k)$ spectral loci and hence $\kappa(k)$, we search for $E=\bar\epsilon(k)$ where there exist roots $z,z'$ of Eq.~\ref{Poly3} with the same $\kappa(k)$, i.e. satisfying $|z|=|z'|$. By analytically solving the cubic polynomial as detailed in Appendix A3, one obtains the $\bar\epsilon(k)$ loci as values of $E$ satisfying
\begin{equation}
F(E)\propto \left(\frac{b}{2}\right)^{2/3}\omega_j,
\label{Poly4}
\end{equation}
i.e., 3 straight lines radiating from the origin of the complex $F(E)$ plane aligned along the cube roots of unity $\omega_j$, $j=0,1,2$ (Fig.~\ref{fig:skin}c and ~\ref{fig:quasi}a). To find the deformation $\kappa(k)$ needed to bring $\epsilon(k)$ to $\bar\epsilon(k)$, we particularize the eigenenergy equation Eq.~\ref{Poly3} to the $F(\bar\epsilon(k))$ loci we have just derived (Eq.~\ref{Poly4}). The resultant expression can be elegantly expressed as $\text{Im}\left[\left(e^{2\kappa(k)}e^{2ik}+b\,e^{-\kappa(k)}e^{-ik}\right)\omega^{-j}\right]=0$, which yields
\begin{equation}
\kappa(k)=-\frac1{3}\log\left|{\frac{b}{2\cos\left(k-2\pi j /3\right)}}\right|,
\label{Newk}
\end{equation}
where $j$ is chosen to give the branch the smallest complex deformation $|\kappa(k)|$ (Fig.~\ref{fig:skin}c). Notably, the form of $F(E)$ explicitly appears neither in the OBC $\bar\epsilon(k)$ loci nor even $\kappa(k)$; our procedure of restoring reciprocity is only aware of the structure of the couplings as reflected in the eigenenergy Laurent polynomial, rendering other information from $F(E)$, e.g., the number of bands irrelevant. What $F(E)$ controls is the explicit energetics, which can be recovered by conformally mapping the actual $\bar\epsilon(k)$ loci onto the equally spaced $Y$-shaped junction (Eq.~\ref{Poly4}) that forms the signature OBC spectral singularity of a Hamiltonian with two non-reciprocal length scales (Eq.~\ref{Poly3}).
\\ \\
\noindent\underline{2. Generic non-reciprocal couplings:} By generalizing the above arguments, it can be shown that for characteristic polynomials of the form
\begin{equation}
E^N=az^p+\frac{b}{z^q}
\label{Poly6}
\end{equation}
where $p,q>0$, the OBC $\bar\epsilon(k)$ spectrum takes the shape of a $N(p+q)$-pointed star, generalizing the above-mentioned $p=2,q=1$ case which gives a 3-pointed OBC star (Fig.\ref{fig:quasi}a(ii)). This result can also be intuitively obtained by regarding $p+q$ as the number of times the BZ is folded. 

Most generally, the characteristic polynomial is a bivariate polynomial\footnote{Negative powers of $z$ can be made positive through an appropriate multiplicative factor.}
\begin{equation}
P(E,z)=\sum_{m,n}p_{m,n}E^mz^n=0.
\label{Poly5}
\end{equation}
that contains multiple coefficients of $m$ and $n$, and may not be separable into parts that depend separately on $E$ and $z$, as in Eqs.~\ref{Poly2} and \ref{Poly3}. The exact correspondence between the graph topology of the OBC singularity and the algebro-geometric properties of its associated $P(E,z)$ is an open problem. However, from a single well-understood case, one can already understand that all other cases relate to each other via a conformal transformation of $E$. As illustrated in Fig.\ref{fig:quasi}a(ii) to (iv), $E\rightarrow E^2+1=(E+i)(E-i)$ splits the 5-pointed OBC star into two stars centered at $E=\pm i$, while $E^2\rightarrow E^3$ produces three images of the star from two. These mappings can be easily implemented by increasing the number of components. For instance, to map an arbitrary single-component (band) $E=E_0(z)$ into $E^2+1=E_0(z)$, one turns to the Hamiltonian $H(z) = \left(\begin{matrix} 0 & E_0(z)-1 \\ 1 & 0\end{matrix}\right)$. Likewise, to map it to $E^3+1=E_0(z)$, one can enlarge the Hamiltonian to $H(z) = \left(\begin{matrix} 0 & 0 & E_0(z)-1 \\ 1 & 0& 0 \\ 0 & 1 & 0\end{matrix}\right)$. More generally, given a Hamiltonian with complicated $P(E,z)$, the trick will be to attempt to bring it into a simpler known form through a conformal transformation of $E$, with branch cuts introducing multiple Riemann sheets corresponding to multiple images of the original (both OBC and PBC) spectrum. 

Still, there of course exist many exotic possibilities not transformable to simple star patterns. Consider going from the model in Fig.\ref{fig:quasi}a(i) to (v) via a mapping $E^2\rightarrow E^2-\frac{0.7}{z^2}E$ which involves $z=e^{ik}$ as well. Since that modifies $\kappa(k)$, the OBC spectrum of (v) cannot be understood in terms of that in (i), and in fact forms a different pattern, containing even a closed graph cycle~\cite{suppmat}. The resultant characteristic polynomial $P(E,z)=E^2-\frac{0.7}{z^2}E-z^2-\frac1{2z}=0$ can be obtained, for instance, from a Hamiltonian of the form $H(z) = \left(\begin{matrix} 0.7/z^2 & z^2+1/(2z) \\ 1 & 0\end{matrix}\right)$, which also includes a term on the diagonal. Note, however, that graph cycles in the OBC spectrum $\bar\epsilon$ do not necessary require complete $E$ dependence, and can in fact arise in single-component models with multiple powers of $z$, for instance $E=(z^3+2z^2+z+z^{-1}+4z^{-2})/2$ from Ref.~\cite{kai2019gbz}. We conclude this discussion by reiterating that the graph topology of the OBC spectrum is, in its essence, a property of the characteristic polynomial $P(E,z)$, not the Hamiltonian per se, with the exact nature of this graph topology being an open topic for future studies.


\subsection{Singularity transitions}

We have just seen how the OBC spectral graph can be drastically modified as the characteristic polynomial $P(E,z)$ varies. When the spectral graph topology changes discontinuously, at least part of the OBC spectrum shrinks to a point i.e. assumes a complex ``flat-band''. The tuning of physical parameters that affect such transitions will generate a phase diagram containing regions of different spectral graph topologies, as in Fig.~\ref{fig:quasi}b where $E=az^2+b/z$. Singularity transitions occur when $a=0$ or $b=0$, since the OBC spectrum shrinks to a point and, optionally, flip across these transitions. More sophisticated transitions are possible in other models, like in those appearing in Fig.~\ref{fig:quasi}a. 

Notably, these topological transitions of the OBC spectral graph generically do \emph{not} coincide with OBC bandgap closures, which occur when two or more components of the graphs (i.e. stars in Fig.~\ref{fig:quasi}a) intersect. Yet, because of the emergent non-locality, the eigenstates get to converge non-analytically and mix at the transition degeneracy, eigenstate properties like the Berry curvature can still change discontinuously, \blue{as elaborated latter in Sec. \ref{sec:berry_curvature}.}


\GJ{\subsection{Topological phase boundaries}}

We next elaborate on two models where our surrogate Hamiltonian formalism is essential for \GJ{locating the topological phase boundaries associated with  in-gap boundary modes}. While the presence of topological boundary modes is conceptually unrelated to the OBC spectral graph topology, topological phase boundaries are determined by gap closures (intersections) of these OBC spectral graphs. As such, we emphasize that it is the OBC quasi-reciprocal surrogate Hamiltonian $\bar H(k)$, not the PBC $H(k)$, that should be used to compute topological invariants which then correctly predict the presence of boundary modes. 
\bigskip

\subsubsection{1D: Non-Hermitian extended Kitaev chain} 


\GJ{Our first example is a non-Hermitian version of the extended Kitaev model~\cite{kitaev,li2016Z2}, with
a minimal model Hamiltonian given by $H_D={\bm h}(k)\cdot{\bm \sigma}$, where
\begin{eqnarray}
h_x&=&\Delta_2\sin\phi\sin 2k+ig_x\notag\\
h_y&=&\Delta_2\cos\phi\sin 2k+\Delta_1 \sin k+ig_y\notag\\
h_z&=&m-t_1\cos k -t_2 \cos 2k.
\label{kitaev}
\end{eqnarray}
Note that this model Hamiltonian minimally contains both NN and NNN coupling terms. As explained below, the necessary presence of both NN and NNN couplings results in a complicated characteristic polynomial with more than one non-reciprocal length scale, whose identification of topological phase diagram requires our surrogate Hamiltonian formalism.
 These couplings contain unequal phase factors to break its chiral symmetry into particle-hole symmetry (PHS) described by $\mathcal{C}H^T(k)\mathcal{C}^{-1}=-H(-k)$ with $\mathcal{C}$ a unitary matrix, such that $Z$ topology is broken into $Z_2$ (D-class) topology~\cite{li2016Z2}. 
As complex conjugation does not coincide with transposition for non-Hermitian systems, one can alternatively define a conjugated PHS (cPHS) as $\mathcal{C}H^*(k)\mathcal{C}^{-1}=-H(-k)$~\cite{kawabata2019symmetry}. Here we shall consider an example with cPHS belonging to the D$^\dagger$-class, as PHS does not allow non-Hermitian pumping with simple constant non-Hermitian terms~\cite{Li2019geometric}.
In particular, the
cPHS define here enforces $\sigma_xH^*(k)\sigma_x=-H(-k)$, allowing only two types of constant~\footnote{More possibilities exist for $k$-dependent non-Hermitian terms, i.e. $\sin k\, \sigma_z$.} non-Hermitian terms: $ig_x\sigma_x$ and $ig_y\sigma_y$, so that cPHS would be broken if any of the parameters $m,t_1,t_2,g_x,g_y$ become complex. Unlike in the SSH model, both the $ig_x\sigma_x$ and $ig_y\sigma_y$ terms can separately lead to the skin effect, since $\sin 2k$ from the NNN couplings appear in both $h_x$ and $h_y$.}

In principle, there is no further restriction on the parameters of $H_D$ (Eq.~\ref{kitaev}), whose characteristic polynomial generically takes the form $E^2=P_8(z)/z^4$, with $P_8(z)$ an 8th-order polynomial in $z$. However, for the purpose of analytically obtaining $\kappa(k)$ and hence the surrogate Hamiltonian $\bar H_D(k)$, we shall normalize $m=1$ and impose the conditions~\cite{suppmat} $t_1=\Delta_1\cos\phi$, $t_2=\Delta_2$ and
\begin{equation}
\Delta_1^2=-\frac{2\Delta_2\left[2Ag_y+(g_y^2+A^2)(g_y\cos \phi+g_x\sin\phi)\right]}{g_yA\sin^2\phi}
\label{constraint}
\end{equation}
with $A=(\Delta_2-1)\cos\phi$.
Collectively, these constraints allow the characteristic polynomial to take a simple quadratic form as detailed in Appendix B, with only $\Delta_2$, $\phi$, $g_x$ and $g_y$ as independent parameters. With them, the surrogate Hamiltonian $\bar{H}_D(k)={H}_D(k+i\kappa(k))$ can be defined via the constant complex deformation $k\rightarrow k+i\kappa(k)$ with 
\begin{eqnarray}
\kappa(k)=\kappa=-\log\left|-\frac{(\Delta_2-1)\cos \phi- g_y}{(\Delta_2-1)\cos \phi+ g_y}\right|.
\end{eqnarray}

\begin{figure}
\includegraphics[width=1\linewidth]{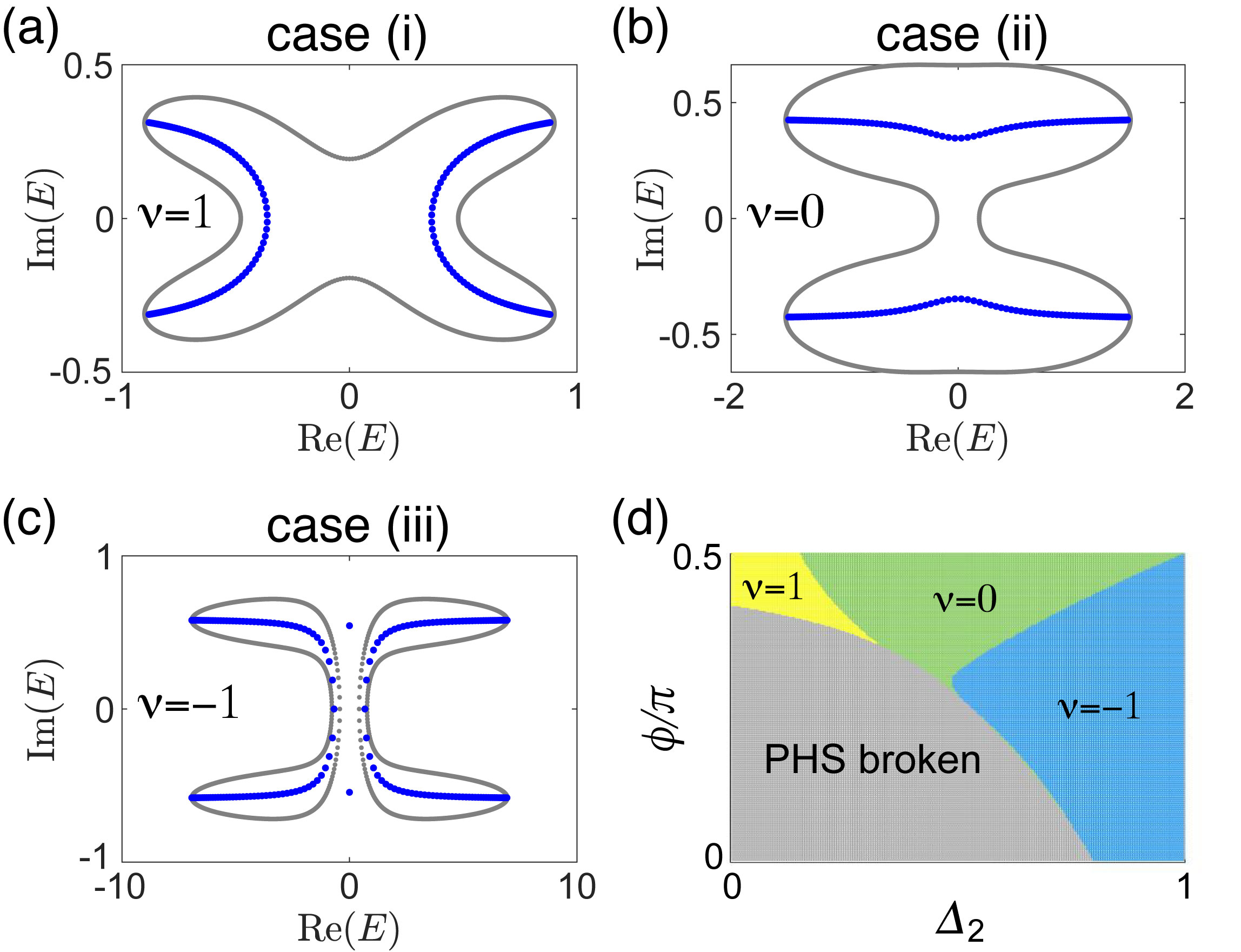}
\caption{(a-c)  OBC spectra (blue) enclosed by the PBC spectra (gray) of $H_D$ for cases (i) to (iii) corresponding to parameters $\phi=0.45\pi$, $g_x=g_y=0.6$ and
$\Delta_2=0.1,0.3$ and $0.9$ respectively. These three cases respectively possess a real line gap, imaginary line gap and real line gap with topological modes, as classified by their
distinct topological invariants $\nu=1,0,-1$ defined by Eq.~\ref{eqnu}. (d) Phase diagram of $H_D$ with $g_x=g_y=0.6$ kept constant, with phase boundaries analytically solvable via the surrogate pseudospin we introduced. The PHS broken regime occurs when some of the other parameters in $H_D$ take complex values, as mandated by the constraints given by Eq.~\ref{constraint} and its preceding discussion.
}\label{kitaev_phases}
\end{figure}

\begin{figure*}
 \centering
\subfloat[]{\includegraphics[width=.33\linewidth]{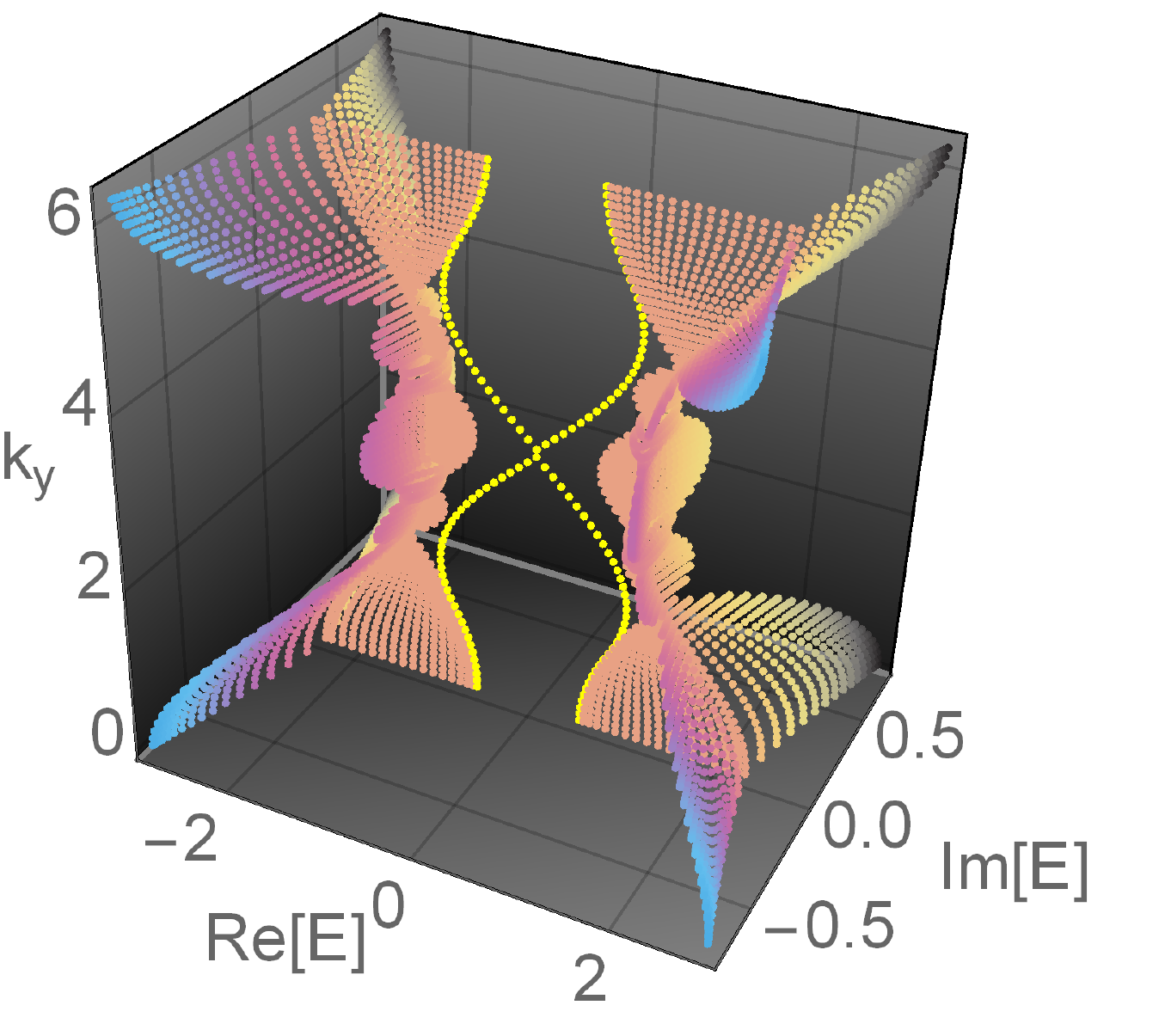}}
\subfloat[]{\includegraphics[width=.33\linewidth]{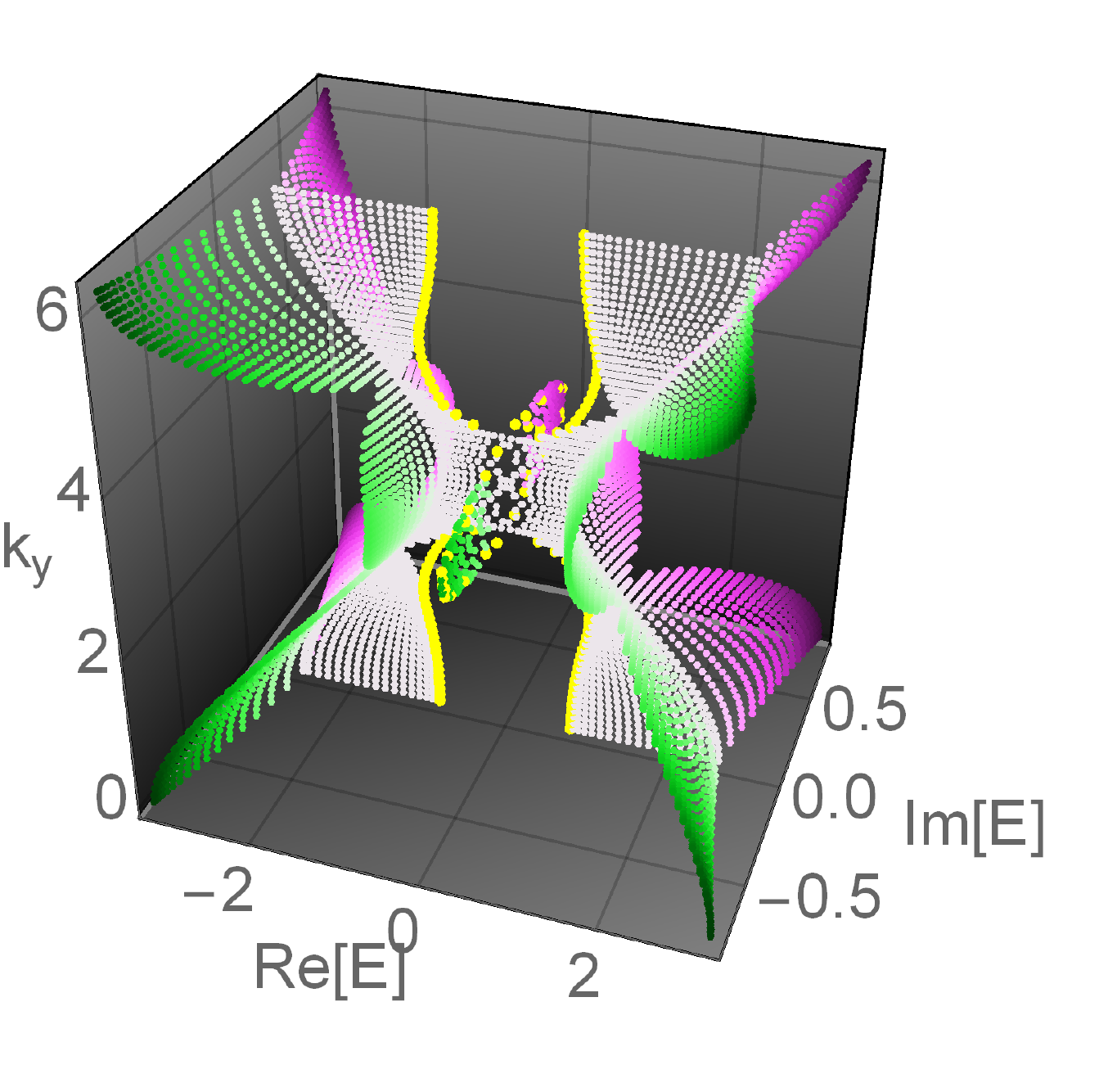}}
\subfloat[]{\includegraphics[width=.33\linewidth]{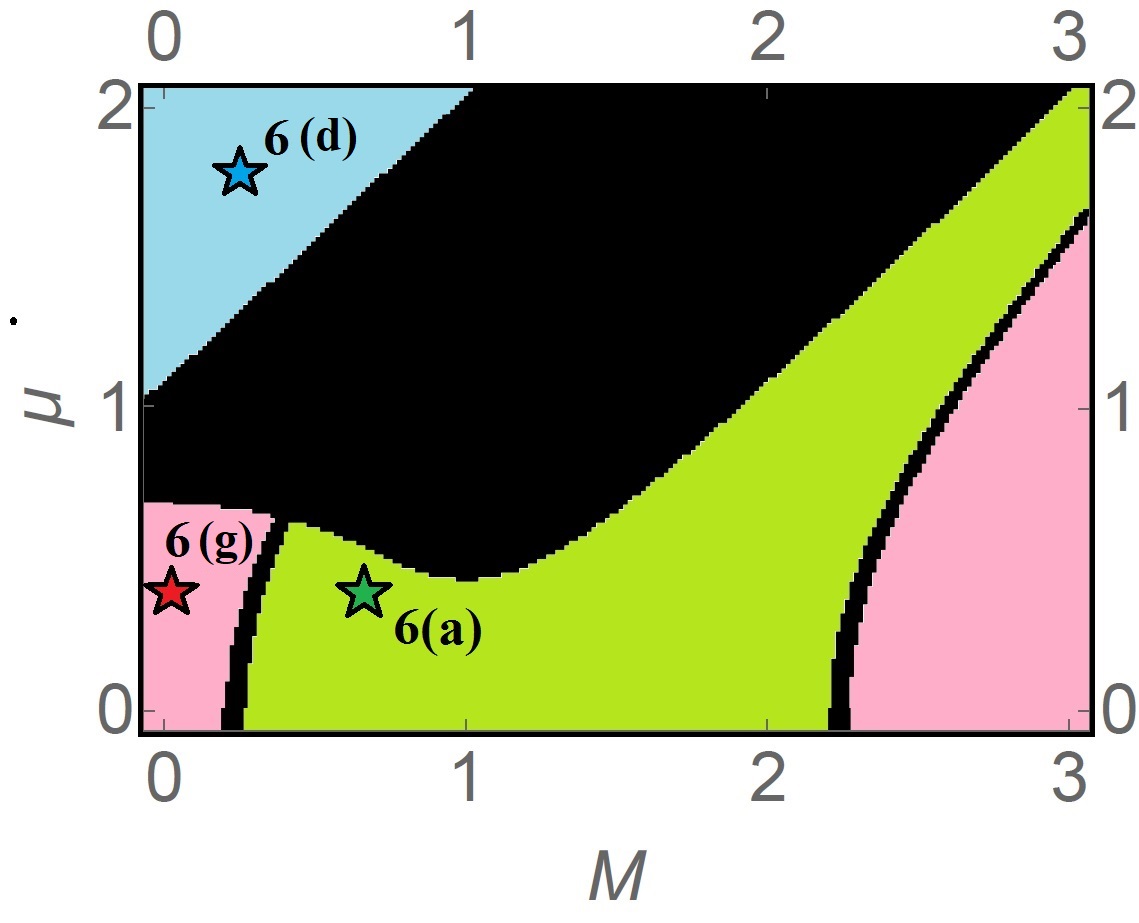}}
\caption{$x$-OBC spectra of $H_\text{Ch}$ for (a) a topologically nontrivial case $M=0.5,\mu=0.3,v_0=2$ with $\bar c=1$ edge mode and (b) a gapless case $M=1,\mu=0.5,v_0=1.5$ without well-defined separate bands. (c) Phase diagram for $v_0=1.3$, with black gapless regions separating phases with topologically nontrivial edge modes (green), trivial edge modes (blue) and no edge modes (pink). The stars denote representative cases presented in the following figure.  }
 \label{fig:chern}
\end{figure*}

\noindent Since PHS still holds after the complex deformation, the topology of this non-Hermitian extended Kitaev model can be characterized by its surrogate pseudospin expectation vectors at the high symmetric points $k=0,\pi$. As a 2-level non-Hermitian system, it possesses two qualitatively distinct pseudospin vectors for each eigenenergy band $\bar E(k)$: the physical pseudospin expectation~\cite{Jiang2018nochiral} $\bar {S}^{RR}_\mu=\langle\bar\psi^R|\sigma_\mu|\bar\psi^R\rangle$ and the biorthogonal pseudospin $\bar {S}^{LR}_\mu=\langle\bar\psi^L|\sigma_\mu|\bar\psi^R\rangle$, where $\mu=x,y,z$ and $\bar H_D|\bar \psi^R\rangle=\bar E|\bar\psi^R\rangle$, $\bar H^\dagger_D|\bar \psi^L\rangle=\bar E^*|\bar\psi^L\rangle$. Taking both of the bands into consideration, PHS ensures that the trajectories of $\bar {\bold S}^{RR}(k)$ are mirror-symmetric about the equator of the Bloch sphere. Furthermore, each band of $\bar H_D$ can either be mirror-symmetric with itself 
or both bands can form mirror-symmetric partners of each other. 
In Appendix \ref{sec:kitaev}, these two possible types of configurations are shown to correspond to (spectrally) topologically distinct cases with imaginary [Fig.~\ref{kitaev_phases}(b)] and real line gaps [Fig.~\ref{kitaev_phases}(a,c)] respectively. 



To furthermore predict the presence of topological boundary modes, we turn to the biorthogonal pseudospin $\bar {\bold S}^{LR}(k)$. In the cases with imaginary line gap, $\bar{E}(0)$ and $\bar{E}(\pi)$ are imaginary and so are $\bar S^{LR}_z(0)$ and $\bar S^{LR}_z(\pi)$. 
The case when $\bar{E}(0)$ and $\bar{E}(\pi)$ are both real i.e. with real line gap, is more interesting, containing the possibility of hosting topological modes. It can be shown that ${\rm Sign}[\bar{S}^{LR}_z(0)]=\epsilon\,{\rm Sign}[\bar{S}^{LR}_z(\pi)]$, with $\epsilon=1(-1)$ corresponding to the scenario without (with) topological edge states [Figs.~\ref{kitaev_phases}(b) and (c)]. All in all, there are three distinct phases $\nu=1,0,-1$ characterized by the topological invariant~\cite{Li2019geometric}
\begin{equation}
\nu={\rm Sign}\{[\text{Re}[\bar{S}^{LR}_z(0)] \text{Re}[\bar{S}^{LR}_z(\pi)]\},
\label{eqnu}
\end{equation}
as mapped out by the phase diagram of Fig.~\ref{kitaev_phases}(d):
\begin{itemize}
\item Case (i), the $\nu=1$ phase with real line gap and no topological boundary mode. 
\item Case (ii), the $\nu=0$ phase with imaginary line gap and no topological boundary mode. 
\item Case (iii), the $\nu=-1$ phase with real line gap and isolated topological boundary modes. 
\end{itemize}

These three phases $\nu=-1,0,1$ are all the possible gapped phases of this PH-symmetric system, since $\bar{E}(0)$, $\bar{E}(\pi)$ must be both real or both imaginary.  We would like to highlight  that in computing $\nu$, we have made crucial use of the $\kappa(k)$ deformation introduced by our formalism, without which it is difficult to obtain the surrogate $\bar {\bold S}^{RR}(k)$ and $\bar {\bold S}^{LR}(k)$ that correctly predict the OBC behavior.
\GJ{As a final note, NNN couplings here and hence two or more non-reciprocal length scales are necessary for realizing the variety of topological configurations afforded by the symmetry class considered here}. 





\begin{figure*}
 \centering
\includegraphics[width=\linewidth]{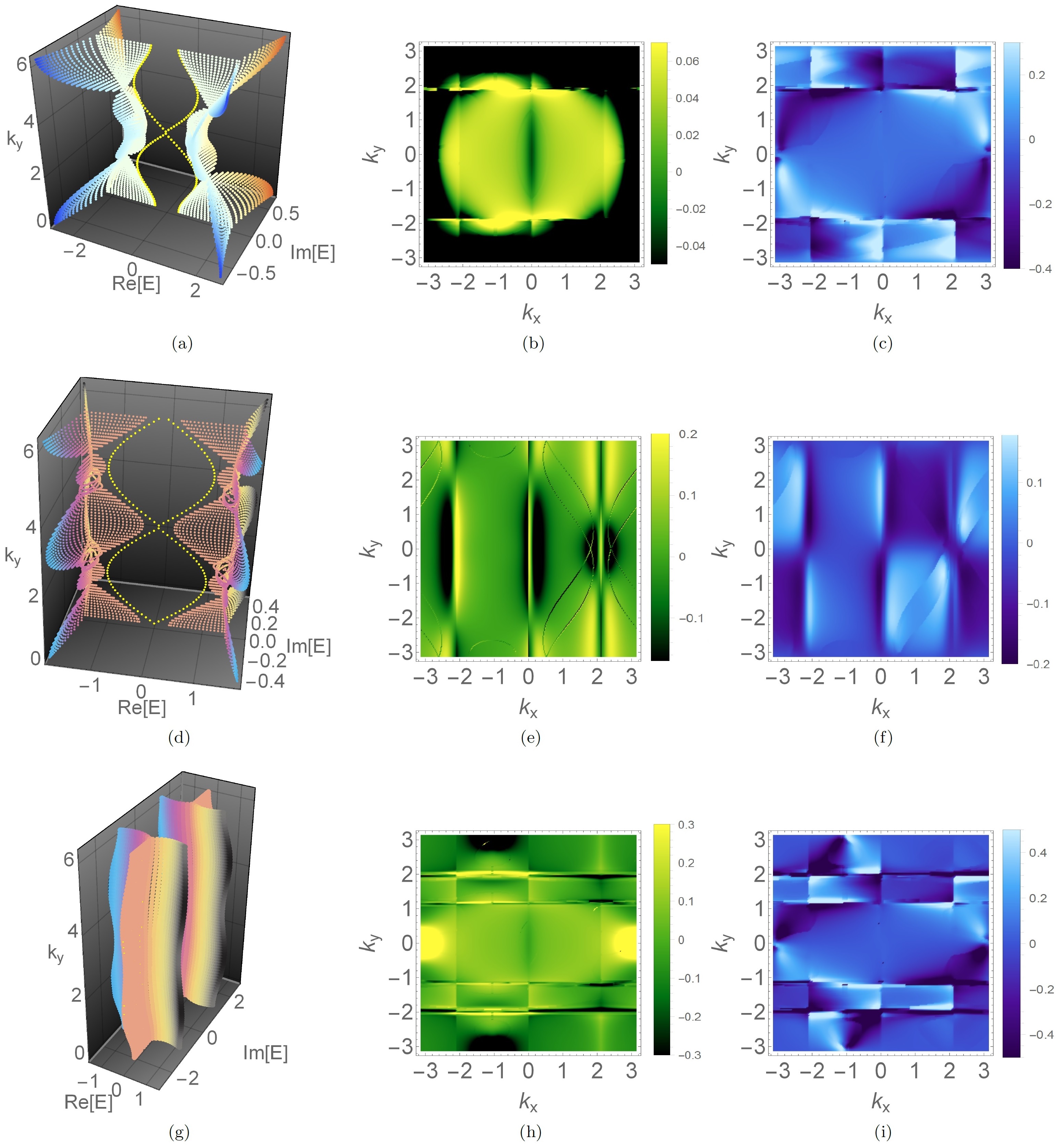}
\caption{$x$-OBC spectra (Left), surrogate Berry curvature (Center) and FS metric trace \GJ{{\rm Tr}}$\,g$ (Right) of $H_\text{Ch}$ for parameters in three different regimes: $M=0.7,\mu=0.4,v_0=1.3$ (Top), $M=0.2,\mu=1.8,v_0=1.3$ (Middle) and $M=0,\mu=0.4,v_0=1.3$ (Right). While all three scenarios contain cubic singularities, only the Top scenario is topologically nontrivial. The Middle scenario also contains edge modes (yellow), but is deformable to the trivial case. While all three scenarios exhibit Berry curvature of FS metric discontinuities at $k_x=0,\pm 2\pi/3$, only the Top and Bottom cases possess discontinuities at $k_y=\pm \cos^{-1}(\mu-M)$ (The Middle case admits no such $k_y$ solutions). Despite their discontinuities, the Berry curvatures integrate to quantized Chern numbers $1,0$ and $0$ respectively. 
}
 \label{fig:chern2}
\end{figure*}
\subsubsection{2D: Extended non-Hermitian Chern insulator}

We now illustrate the use of quasi-reciprocal surrogate quantities in a 2D setting, where quantities like band geometry and Berry curvature can be modified by non-Hermitian pumping through the non-analytic complex deformation. In the case of cylindrical boundary conditions, we have without loss of generality OBCs in the $x$-direction and PBCs in the $y$-direction, such that $k_y$ is still a well-defined parameter. For each $k_y$-slice, the surrogate Hamiltonian is defined by $\bar H(k_x;k_y)=H(k_x+i\kappa_x(\bold k),k_y)$. In the case of OBCs in both directions, which we shall not consider in-depth here, quasi-reciprocity also has to be restored in the $y$-direction, giving rise to $\bar{\bar H}(\bold k)=\bar H(k_x+i\kappa_x(\bold k),k_y)=H(k_x+i\kappa_x(\bold k), k_y + i\kappa_y(\bold k))$ where $\kappa_x$ is taken as a spectator parameter in the second iteration, and $\kappa_y$ determined by how $H$ depends on $k_y$ both explicitly and through $\kappa_x(\bold k)$.  

Focusing on cylindrical boundary conditions from now on, we see that different $k_y$ slices of the \emph{same} system can possess different OBC spectral graph topologies, with discontinuous gapped transitions between them as further discussed in the next section. Below, we introduce a minimal and analytically tractable example of such a non-Hermitian model with both Chern topology and signature Y-shaped spectral topology:
\begin{equation}
H_\text{Ch}(\bold k)= (v+z^{-1})\sigma_++(u+z-v\,z^2)\sigma_-+\sin k_y\,\sigma_z ,
\label{Chern}
\end{equation}
where $z=e^{ik_x}$ and $u=M+\cos k_y-\mu$ and $v=v_0(M+\cos k_y +\mu)$. 
As contrasted with the NN non-Hermitian Chern model commonly studied in the literature~\cite{yao2018non}, our model Eq.~\ref{Chern} contains fundamentally asymmetric couplings (detailed in Appendix C), and is \emph{not} adiabatically connected to any Hermitian Chern model. In other words, the role of its NNN couplings in the $x$-direction (coefficient of $z^2$) is not just to perturb away from the known phase of the NN Chern model, but to define a new Chern phase existing on a Y-shaped spectral graph.

By design, the characteristic polynomial of $H_\text{Ch}$ with $x$-OBCs assumes the classic form of Eq.~\ref{Poly3}: 
\begin{equation}F_\text{Ch}(E)=z^2+b/z,
\label{ChernF}
\end{equation}
with $F_\text{Ch}(E)=(1+\sin^2k_y+uv-E^2)/v^2$ and $b=-u/v^2$. Hence, as in Fig.~\ref{fig:skin}c, each $k_y$-slice of the OBC skin spectrum $\bar\epsilon(k_x;k_y)$ consists of two cubic singularities, with the size and origin of each Y-shaped star controlled by the $k_y$-dependent $b$ and $F_\text{Ch}(E)$ respectively (Figs.~\ref{fig:chern} and \ref{fig:chern2}). Additionally, isolated in-gap topological modes can also exist. Their existence in each 1D $k_y$-slice can be predicted by the generalized topological criterion of Ref.~\cite{lee2019anatomy} or chiral-symmetric winding number of $\bar H_\text{Ch}(k_x;k_y)$ over $k_x$. However, most important for our context is whether they traverse the gap over a full cycle of $k_y$, and that is determined by the Chern number $\bar c=\frac1{2\pi}\int \bar\Omega_{xy}d^2\bold k$, where 
$\bar \Omega_{xy}=\text{Im}\,\bar\Gamma_{xy}$ is the biorthogonal Berry curvature corresponding to the imaginary part of the gauge-invariant quantity
\begin{equation}
\bar \Gamma_{\mu\nu}=\langle \partial_\mu\bar\psi_L|\bar Q|\partial_\nu \bar\psi_R\rangle.
\label{Gamma}
\end{equation}
Here $\bar Q=\mathbb{I}-\bar P$ where $\bar P=|\bar\psi_R\rangle\langle \bar\psi_L|$ the biorthogonal projector onto the band of lower $\text{Re}\,E$ that is biorthogonally spanned~\cite{brody2013biorthogonal,herviou2019entanglement} by left/right eigenvectors $\bar\psi_L,\bar\psi_R$. 

Since our $\bar H_\text{Ch}$ is quasi-reciprocal, the Chern number of its occupied eigenstate must always be a quantized integer corresponding to the number of gap-traversing topological edge modes, at least when the gap is well-defined. Fig.~\ref{fig:chern}a shows a gapped case with $\bar c=1$ topological edge mode (yellow curve), while Fig.~\ref{fig:chern}b shows a gapless case with no well-defined edge mode. Note that gapless cases can occur as typically as gapped cases in non-Hermitian systems, as illustrated by their extended black regions in the phase diagram of Fig.~\ref{fig:chern}c. 

\section{Discontinuous Berry curvature and band metric}\label{sec:berry_curvature}

Very interestingly, we observe discontinuities in both the Berry curvature $\bar \Omega_{xy}$ and the trace of the Fubini-Study (FS) metric \GJ{{\rm Tr}}$\,\bar g=\text{Re}\,\bar\Gamma$, as shown for three contrasting cases in the Center and Right columns of Fig.~\ref{fig:chern2} respectively. Both of these quantities are derived from $\bar\Gamma$, which contain momentum-space derivatives that pick up qualitative transitions in the behavior of the eigenstate. Since the branching behavior of the OBC spectrum $\bar \epsilon(\bold k)$ is controlled by $\kappa(\bold k)$ (i.e. in Eq.~\ref{Newk}) which in turn enters the eigenstate, a singularity transition will qualitatively modify the form of $\kappa(\bold k)$ and lead to non-analytic discontinuities. Physically, these non-analyticities arise from the emergent non-locality induced by the non-Hermitian pumping. 


Note that because $\kappa(k)$ is continuous at its kinks, $\bar \Gamma_{\mu\nu}$ do not diverge at its discontinuities. How discontinuities of $\bar\Gamma_{\mu\nu}$ arise from $\bm\kappa(\bold k)$ discontinuities can be understood from the functional dependence of the momenta through $\bm \kappa(\bold k)$:
\begin{equation}
|\bar\psi_R\rangle=|\bar\psi_R(\bold k+i\bm \kappa(\bold k))\rangle
\end{equation} 
and likewise for the left eigenvector. As such, writing $\bm p=\bm k+i\bm\kappa(\bm k)$,
\begin{eqnarray}
|\partial_\nu\bar\psi_R\rangle&=&\frac{d}{dk_\nu}|\bar\psi_R(\bm p)\rangle\notag\\
&=& \frac{d|\bar\psi_R(\bm p)\rangle}{d\bm p}\cdot\frac{d}{dk_\nu}\left(\bm k+i\bm\kappa(\bm k)\right) \notag\\
&=& \frac{d|\bar\psi_R(\bm p)\rangle}{d p_\nu}+i\frac{d|\bar\psi_R(\bm p)\rangle}{d\bm p}\cdot\frac{d\kappa(\bm k)}{dk_\nu}.
\label{gradphiR}
\end{eqnarray}
The notation above is somewhat subtle: In line one, the partial derivative $\partial_\nu$ on the LHS refers to a derivative with respect to $k_\nu$ that treats other momenta as independent. However, on the RHS, we have rewritten it as a total derivative in $k_\nu$ to emphasize its total derivative nature with respect to $\bm p=\bm k+i\bm \kappa(\bm k)$.

The key takeaway from Eq.~\ref{gradphiR} is that $|\partial_\nu\bar\psi_R\rangle$ contains discontinuities from $\frac{d\kappa(\bm k)}{dk_\nu}$, which is discontinuous whenever there is a kink $\bm \kappa(\bm k)$ i.e. as depicted in Fig.~1. Yet, the gradient $\frac{d\kappa(\bm k)}{dk_\nu}$ never diverges as long as there are no essential singularities in the complex band structure. This finiteness is inherited in the OBC response quantities from $\bar\Gamma_{\mu\nu}=\langle \partial_\mu\bar\psi_L|\bar Q|\partial_\nu\bar\psi_R\rangle$.

For our model Eq.~\ref{Chern}, singularity transitions occur at $b=-u/v^2=0$ (Eq.~\ref{ChernF}), i.e. at $k_y=\cos^{-1}(\mu-M)$. Furthermore, $\kappa(\bold k)$ also exhibits kinks at $k_x=0,\pm 2\pi/3$, where the OBC spectrum jumps from one branch to the next~\cite{suppmat}. Indeed, these two sets of lines \GJ{exactly} correspond to the discontinuities in $\bar \Omega_{xy}$ and $\GJ{{\rm Tr}}\,\bar g$ in Fig.~\ref{fig:chern2}. In the $\bar c=1$ case of the Top row ($M=0.7,m=0.4,v_0=1.3$), the discontinuities along $k_y$ is also seen to correspond to flips (reflections) in the Y-shaped spectra, corroborating with results depicted in Fig.~\ref{fig:quasi}b. No true flips and hence discontuinities along $k_y$ exist for the Center row case ($M=0.2,m=1.8,v_0=1.3$), for which $k_y=\pm\cos^{-1}(\mu-M)$ admits no real solution. Despite the discontinuities, it is remarkable that the Berry curvature in all gapped cases integrate to integer multiples of $2\pi$.

Discontinuous transitions between different spectral singularity classes do not close the gap, and represent a new type of transition that can still be physically detect through observables that depend on the momentum-space \emph{gradients} of eigenstates. The simplest examples are the gauge-invariant quantity $\bar\Gamma_{\mu\nu}=\langle \partial_\mu\bar\psi_L|\bar Q|\partial_\nu \bar\psi_R\rangle$, $Q=\mathbb{I}-|\bar \psi_R\rangle\langle \bar\psi_L|$ introduced earlier, whose real and imaginary parts respectively correspond to the FS metric trace $Tr\,\bar g=\text{Re}\,Tr\,\bar Q$ which controls the locality of effective interactions and non-linear response~\cite{lee2014lattice, lee2017band,li2020detection}, and the Berry curvature which appears in the Kubo formula for linear response. In PT-symmetric quantum mechanics, the symmetrized form $\frac1{2}(\bar\Gamma_{\mu\nu}+\bar\Gamma_{\nu\mu})$ has also been proposed as a quantum geometric tensor~\cite{zhang2019quantum}. Quantities containing higher order gradients correspond to higher-order cumulants in the noise spectrum, and are expected to exhibit discontinuities too. 



\section{Discussion}
In our quest for a quasi-reciprocal picture of non-Hermitian systems where non-Hermitian pumping is eliminated, non-locality and its concomitant non-analyticity \GJ{emerge} as unavoidable consequences. These effects lead to enigmatic properties such as discontinuous Berry curvature and band geometry, which can result in anomalous transport and noise responses in generic systems with non-Hermitian descriptions.  \GJ{It would be fascinating to actually observe the physical consequences of discontinuous Berry curvatures discovered in this work. }

At the level of formalism, we developed a restoration procedure to map any non-Hermitian model to its quasi-reciprocal surrogate model with reinstalled bulk-boundary correspondence, from where its topological nature unfolds unambiguously. By encoding the equilibration behavior of accumulated pumped states as non-holomorphic complex momentum deformations, the effective non-locality leads to not only gap-preserving topological transitions, but also ultimately a topological classification of OBC spectra related to the classification of algebraic varieties.

Our approach applies universally to any system whose characteristic polynomial (energy dispersion) admits a surrogate non-local basis construction. Tailored for realistic setups with multiple effective components and long-ranged couplings, it uncovers possibly non-perturbative topological contributions, unique to non-Hermitian systems, that would not be revealed by oversimplified short-ranged representations. \GJ{Through analogous considerations repeatedly applied to each dimension}, it can be extended to higher dimensional lattices which support exceptional nodal structures and generalized skin-topological modes~\cite{lee2018tidal,lee2019hybrid}. Being based on unitary transformations, our formalism remains valid in the realm of interacting systems, and can shed light on the interesting interplay between the non-Hermitian skin effect and many-body phenomena such as emergent Fermi surfaces~\cite{mu2019emergent,lee2019many,ashida2017parity}.



\section{Acknowledgements}
We thank Hannah Price, Mohammad Hafezi, Bo Yang, Cristiane M. Smith, Kohei Kawabata, Xizheng Zhang and Mu Sen for stimulating discussions. RT is funded by the Deutsche Forschungsgemeinschaft (DFG, German Research Foundation) through project-id 258499086 - SFB 1170 and through the W\"urzburg-Dresden Cluster of Excellence on Complexity and Topology in Quantum Matter –ct.qmat project-id 39085490 - EXC 2147. JG is funded by  the Singapore National Research Foundation (NRF) Grant No. NRF-NRFI201704 (WBS No. R-144-000-378-281). Part of this work resulted from discussions at the KITP program ``Topological Quantum Matter: From Concepts to Realizations''.
{\it Note added:} Towards the completion of our manuscript, we learnt of a new tangentially related manuscript Ref.~\cite{zhesen2019agbz}. While Ref.~\cite{zhesen2019agbz} focused on the mathematical construction of the generalized BZ for models with multiple hoppings and bands, our work focuses on the novel physical consequences of such systems, such as discontinuous Berry curvature, response kinks, and emergent spectral classification in terms of graph topology.

\bibliography{references}

\begin{thebibliography}{84}%
\makeatletter
\providecommand \@ifxundefined [1]{%
 \@ifx{#1\undefined}
}%
\providecommand \@ifnum [1]{%
 \ifnum #1\expandafter \@firstoftwo
 \else \expandafter \@secondoftwo
 \fi
}%
\providecommand \@ifx [1]{%
 \ifx #1\expandafter \@firstoftwo
 \else \expandafter \@secondoftwo
 \fi
}%
\providecommand \natexlab [1]{#1}%
\providecommand \enquote  [1]{``#1''}%
\providecommand \bibnamefont  [1]{#1}%
\providecommand \bibfnamefont [1]{#1}%
\providecommand \citenamefont [1]{#1}%
\providecommand \href@noop [0]{\@secondoftwo}%
\providecommand \href [0]{\begingroup \@sanitize@url \@href}%
\providecommand \@href[1]{\@@startlink{#1}\@@href}%
\providecommand \@@href[1]{\endgroup#1\@@endlink}%
\providecommand \@sanitize@url [0]{\catcode `\\12\catcode `\$12\catcode
  `\&12\catcode `\#12\catcode `\^12\catcode `\_12\catcode `\%12\relax}%
\providecommand \@@startlink[1]{}%
\providecommand \@@endlink[0]{}%
\providecommand \url  [0]{\begingroup\@sanitize@url \@url }%
\providecommand \@url [1]{\endgroup\@href {#1}{\urlprefix }}%
\providecommand \urlprefix  [0]{URL }%
\providecommand \Eprint [0]{\href }%
\providecommand \doibase [0]{http://dx.doi.org/}%
\providecommand \selectlanguage [0]{\@gobble}%
\providecommand \bibinfo  [0]{\@secondoftwo}%
\providecommand \bibfield  [0]{\@secondoftwo}%
\providecommand \translation [1]{[#1]}%
\providecommand \BibitemOpen [0]{}%
\providecommand \bibitemStop [0]{}%
\providecommand \bibitemNoStop [0]{.\EOS\space}%
\providecommand \EOS [0]{\spacefactor3000\relax}%
\providecommand \BibitemShut  [1]{\csname bibitem#1\endcsname}%
\let\auto@bib@innerbib\@empty
\bibitem [{\citenamefont {Berry}(2004)}]{berry2004physics}%
  \BibitemOpen
  \bibfield  {author} {\bibinfo {author} {\bibfnamefont {Michael~V}\
  \bibnamefont {Berry}},\ }\bibfield  {title} {\enquote {\bibinfo {title}
  {Physics of nonhermitian degeneracies},}\ }\href@noop {} {\bibfield
  {journal} {\bibinfo  {journal} {Czechoslovak journal of physics}\ }\textbf
  {\bibinfo {volume} {54}},\ \bibinfo {pages} {1039--1047} (\bibinfo {year}
  {2004})}\BibitemShut {NoStop}%
\bibitem [{\citenamefont {Rotter}(2009)}]{rotter2009non}%
  \BibitemOpen
  \bibfield  {author} {\bibinfo {author} {\bibfnamefont {Ingrid}\ \bibnamefont
  {Rotter}},\ }\bibfield  {title} {\enquote {\bibinfo {title} {A non-hermitian
  hamilton operator and the physics of open quantum systems},}\ }\href@noop {}
  {\bibfield  {journal} {\bibinfo  {journal} {Journal of Physics A:
  Mathematical and Theoretical}\ }\textbf {\bibinfo {volume} {42}},\ \bibinfo
  {pages} {153001} (\bibinfo {year} {2009})}\BibitemShut {NoStop}%
\bibitem [{\citenamefont {Heiss}(2012)}]{1751-8121-45-44-444016}%
  \BibitemOpen
  \bibfield  {author} {\bibinfo {author} {\bibfnamefont {W~D}\ \bibnamefont
  {Heiss}},\ }\bibfield  {title} {\enquote {\bibinfo {title} {The physics of
  exceptional points},}\ }\href
  {http://stacks.iop.org/1751-8121/45/i=44/a=444016} {\bibfield  {journal}
  {\bibinfo  {journal} {Journal of Physics A: Mathematical and Theoretical}\
  }\textbf {\bibinfo {volume} {45}},\ \bibinfo {pages} {444016} (\bibinfo
  {year} {2012})}\BibitemShut {NoStop}%
\bibitem [{\citenamefont {Ashida}\ \emph {et~al.}(2017)\citenamefont {Ashida},
  \citenamefont {Furukawa},\ and\ \citenamefont {Ueda}}]{ashida2017parity}%
  \BibitemOpen
  \bibfield  {author} {\bibinfo {author} {\bibfnamefont {Yuto}\ \bibnamefont
  {Ashida}}, \bibinfo {author} {\bibfnamefont {Shunsuke}\ \bibnamefont
  {Furukawa}}, \ and\ \bibinfo {author} {\bibfnamefont {Masahito}\ \bibnamefont
  {Ueda}},\ }\bibfield  {title} {\enquote {\bibinfo {title}
  {Parity-time-symmetric quantum critical phenomena},}\ }\href@noop {}
  {\bibfield  {journal} {\bibinfo  {journal} {Nature communications}\ }\textbf
  {\bibinfo {volume} {8}},\ \bibinfo {pages} {1--6} (\bibinfo {year}
  {2017})}\BibitemShut {NoStop}%
\bibitem [{\citenamefont {Hassan}\ \emph {et~al.}(2017)\citenamefont {Hassan},
  \citenamefont {Zhen}, \citenamefont {Solja\ifmmode \check{c}\else
  \v{c}\fi{}i\ifmmode~\acute{c}\else \'{c}\fi{}}, \citenamefont {Khajavikhan},\
  and\ \citenamefont {Christodoulides}}]{hassan2017dynamically}%
  \BibitemOpen
  \bibfield  {author} {\bibinfo {author} {\bibfnamefont {Absar~U.}\
  \bibnamefont {Hassan}}, \bibinfo {author} {\bibfnamefont {Bo}~\bibnamefont
  {Zhen}}, \bibinfo {author} {\bibfnamefont {Marin}\ \bibnamefont
  {Solja\ifmmode \check{c}\else \v{c}\fi{}i\ifmmode~\acute{c}\else
  \'{c}\fi{}}}, \bibinfo {author} {\bibfnamefont {Mercedeh}\ \bibnamefont
  {Khajavikhan}}, \ and\ \bibinfo {author} {\bibfnamefont {Demetrios~N.}\
  \bibnamefont {Christodoulides}},\ }\bibfield  {title} {\enquote {\bibinfo
  {title} {Dynamically encircling exceptional points: Exact evolution and
  polarization state conversion},}\ }\href {\doibase
  10.1103/PhysRevLett.118.093002} {\bibfield  {journal} {\bibinfo  {journal}
  {Phys. Rev. Lett.}\ }\textbf {\bibinfo {volume} {118}},\ \bibinfo {pages}
  {093002} (\bibinfo {year} {2017})}\BibitemShut {NoStop}%
\bibitem [{\citenamefont {Xu}\ \emph {et~al.}(2017)\citenamefont {Xu},
  \citenamefont {Wang},\ and\ \citenamefont {Duan}}]{xu2017weyl}%
  \BibitemOpen
  \bibfield  {author} {\bibinfo {author} {\bibfnamefont {Yong}\ \bibnamefont
  {Xu}}, \bibinfo {author} {\bibfnamefont {Sheng-Tao}\ \bibnamefont {Wang}}, \
  and\ \bibinfo {author} {\bibfnamefont {L.-M.}\ \bibnamefont {Duan}},\
  }\bibfield  {title} {\enquote {\bibinfo {title} {Weyl exceptional rings in a
  three-dimensional dissipative cold atomic gas},}\ }\href {\doibase
  10.1103/PhysRevLett.118.045701} {\bibfield  {journal} {\bibinfo  {journal}
  {Phys. Rev. Lett.}\ }\textbf {\bibinfo {volume} {118}},\ \bibinfo {pages}
  {045701} (\bibinfo {year} {2017})}\BibitemShut {NoStop}%
\bibitem [{\citenamefont {Zhou}\ \emph {et~al.}(2019)\citenamefont {Zhou},
  \citenamefont {Lee}, \citenamefont {Liu},\ and\ \citenamefont
  {Zhen}}]{zhou2019exceptional}%
  \BibitemOpen
  \bibfield  {author} {\bibinfo {author} {\bibfnamefont {Hengyun}\ \bibnamefont
  {Zhou}}, \bibinfo {author} {\bibfnamefont {Jong~Yeon}\ \bibnamefont {Lee}},
  \bibinfo {author} {\bibfnamefont {Shang}\ \bibnamefont {Liu}}, \ and\
  \bibinfo {author} {\bibfnamefont {Bo}~\bibnamefont {Zhen}},\ }\bibfield
  {title} {\enquote {\bibinfo {title} {Exceptional surfaces in pt-symmetric
  non-hermitian photonic systems},}\ }\href@noop {} {\bibfield  {journal}
  {\bibinfo  {journal} {Optica}\ }\textbf {\bibinfo {volume} {6}},\ \bibinfo
  {pages} {190--193} (\bibinfo {year} {2019})}\BibitemShut {NoStop}%
\bibitem [{\citenamefont {Shen}\ \emph {et~al.}(2018)\citenamefont {Shen},
  \citenamefont {Zhen},\ and\ \citenamefont {Fu}}]{shen2018topological}%
  \BibitemOpen
  \bibfield  {author} {\bibinfo {author} {\bibfnamefont {Huitao}\ \bibnamefont
  {Shen}}, \bibinfo {author} {\bibfnamefont {Bo}~\bibnamefont {Zhen}}, \ and\
  \bibinfo {author} {\bibfnamefont {Liang}\ \bibnamefont {Fu}},\ }\bibfield
  {title} {\enquote {\bibinfo {title} {Topological band theory for
  non-hermitian hamiltonians},}\ }\href@noop {} {\bibfield  {journal} {\bibinfo
   {journal} {Phys. Rev. Lett.}\ }\textbf {\bibinfo {volume} {120}},\ \bibinfo
  {pages} {146402} (\bibinfo {year} {2018})}\BibitemShut {NoStop}%
\bibitem [{\citenamefont {Carlstr{\"o}m}\ and\ \citenamefont
  {Bergholtz}(2018)}]{carlstrom2018exceptional}%
  \BibitemOpen
  \bibfield  {author} {\bibinfo {author} {\bibfnamefont {Johan}\ \bibnamefont
  {Carlstr{\"o}m}}\ and\ \bibinfo {author} {\bibfnamefont {Emil~J}\
  \bibnamefont {Bergholtz}},\ }\bibfield  {title} {\enquote {\bibinfo {title}
  {Exceptional links and twisted fermi ribbons in non-hermitian systems},}\
  }\href@noop {} {\bibfield  {journal} {\bibinfo  {journal} {Physical Review
  A}\ }\textbf {\bibinfo {volume} {98}},\ \bibinfo {pages} {042114} (\bibinfo
  {year} {2018})}\BibitemShut {NoStop}%
\bibitem [{\citenamefont {Moors}\ \emph {et~al.}(2019)\citenamefont {Moors},
  \citenamefont {Zyuzin}, \citenamefont {Zyuzin}, \citenamefont {Tiwari},\ and\
  \citenamefont {Schmidt}}]{moors2019disorder}%
  \BibitemOpen
  \bibfield  {author} {\bibinfo {author} {\bibfnamefont {Kristof}\ \bibnamefont
  {Moors}}, \bibinfo {author} {\bibfnamefont {Alexander~A.}\ \bibnamefont
  {Zyuzin}}, \bibinfo {author} {\bibfnamefont {Alexander~Yu.}\ \bibnamefont
  {Zyuzin}}, \bibinfo {author} {\bibfnamefont {Rakesh~P.}\ \bibnamefont
  {Tiwari}}, \ and\ \bibinfo {author} {\bibfnamefont {Thomas~L.}\ \bibnamefont
  {Schmidt}},\ }\bibfield  {title} {\enquote {\bibinfo {title} {Disorder-driven
  exceptional lines and fermi ribbons in tilted nodal-line semimetals},}\
  }\href {\doibase 10.1103/PhysRevB.99.041116} {\bibfield  {journal} {\bibinfo
  {journal} {Phys. Rev. B}\ }\textbf {\bibinfo {volume} {99}},\ \bibinfo
  {pages} {041116} (\bibinfo {year} {2019})}\BibitemShut {NoStop}%
\bibitem [{\citenamefont {Lee}\ \emph {et~al.}(2018)\citenamefont {Lee},
  \citenamefont {Li}, \citenamefont {Liu}, \citenamefont {Tai}, \citenamefont
  {Thomale},\ and\ \citenamefont {Zhang}}]{lee2018tidal}%
  \BibitemOpen
  \bibfield  {author} {\bibinfo {author} {\bibfnamefont {Ching~Hua}\
  \bibnamefont {Lee}}, \bibinfo {author} {\bibfnamefont {Guangjie}\
  \bibnamefont {Li}}, \bibinfo {author} {\bibfnamefont {Yuhan}\ \bibnamefont
  {Liu}}, \bibinfo {author} {\bibfnamefont {Tommy}\ \bibnamefont {Tai}},
  \bibinfo {author} {\bibfnamefont {Ronny}\ \bibnamefont {Thomale}}, \ and\
  \bibinfo {author} {\bibfnamefont {Xiao}\ \bibnamefont {Zhang}},\ }\bibfield
  {title} {\enquote {\bibinfo {title} {Tidal surface states as fingerprints of
  non-hermitian nodal knot metals},}\ }\href@noop {} {\bibfield  {journal}
  {\bibinfo  {journal} {arXiv:1812.02011}\ } (\bibinfo {year}
  {2018})}\BibitemShut {NoStop}%
\bibitem [{\citenamefont {Wang}\ \emph {et~al.}(2019)\citenamefont {Wang},
  \citenamefont {Ruan},\ and\ \citenamefont {Zhang}}]{Wang2019non}%
  \BibitemOpen
  \bibfield  {author} {\bibinfo {author} {\bibfnamefont {Huaiqiang}\
  \bibnamefont {Wang}}, \bibinfo {author} {\bibfnamefont {Jiawei}\ \bibnamefont
  {Ruan}}, \ and\ \bibinfo {author} {\bibfnamefont {Haijun}\ \bibnamefont
  {Zhang}},\ }\bibfield  {title} {\enquote {\bibinfo {title} {Non-hermitian
  nodal-line semimetals with an anomalous bulk-boundary correspondence},}\
  }\href {\doibase 10.1103/PhysRevB.99.075130} {\bibfield  {journal} {\bibinfo
  {journal} {Phys. Rev. B}\ }\textbf {\bibinfo {volume} {99}},\ \bibinfo
  {pages} {075130} (\bibinfo {year} {2019})}\BibitemShut {NoStop}%
\bibitem [{\citenamefont {Yang}\ and\ \citenamefont {Hu}(2019)}]{Yang2019non}%
  \BibitemOpen
  \bibfield  {author} {\bibinfo {author} {\bibfnamefont {Zhesen}\ \bibnamefont
  {Yang}}\ and\ \bibinfo {author} {\bibfnamefont {Jiangping}\ \bibnamefont
  {Hu}},\ }\bibfield  {title} {\enquote {\bibinfo {title} {Non-hermitian
  hopf-link exceptional line semimetals},}\ }\href {\doibase
  10.1103/PhysRevB.99.081102} {\bibfield  {journal} {\bibinfo  {journal} {Phys.
  Rev. B}\ }\textbf {\bibinfo {volume} {99}},\ \bibinfo {pages} {081102}
  (\bibinfo {year} {2019})}\BibitemShut {NoStop}%
\bibitem [{\citenamefont {Carlstr\"om}\ \emph {et~al.}(2019)\citenamefont
  {Carlstr\"om}, \citenamefont {St\aa{}lhammar}, \citenamefont {Budich},\ and\
  \citenamefont {Bergholtz}}]{carlstrom2019knotted}%
  \BibitemOpen
  \bibfield  {author} {\bibinfo {author} {\bibfnamefont {Johan}\ \bibnamefont
  {Carlstr\"om}}, \bibinfo {author} {\bibfnamefont {Marcus}\ \bibnamefont
  {St\aa{}lhammar}}, \bibinfo {author} {\bibfnamefont {Jan~Carl}\ \bibnamefont
  {Budich}}, \ and\ \bibinfo {author} {\bibfnamefont {Emil~J.}\ \bibnamefont
  {Bergholtz}},\ }\bibfield  {title} {\enquote {\bibinfo {title} {Knotted
  non-hermitian metals},}\ }\href {\doibase 10.1103/PhysRevB.99.161115}
  {\bibfield  {journal} {\bibinfo  {journal} {Phys. Rev. B}\ }\textbf {\bibinfo
  {volume} {99}},\ \bibinfo {pages} {161115} (\bibinfo {year}
  {2019})}\BibitemShut {NoStop}%
\bibitem [{\citenamefont {Yoshida}\ \emph {et~al.}(2019)\citenamefont
  {Yoshida}, \citenamefont {Peters}, \citenamefont {Kawakami},\ and\
  \citenamefont {Hatsugai}}]{yoshida2019symmetry}%
  \BibitemOpen
  \bibfield  {author} {\bibinfo {author} {\bibfnamefont {Tsuneya}\ \bibnamefont
  {Yoshida}}, \bibinfo {author} {\bibfnamefont {Robert}\ \bibnamefont
  {Peters}}, \bibinfo {author} {\bibfnamefont {Norio}\ \bibnamefont
  {Kawakami}}, \ and\ \bibinfo {author} {\bibfnamefont {Yasuhiro}\ \bibnamefont
  {Hatsugai}},\ }\bibfield  {title} {\enquote {\bibinfo {title}
  {Symmetry-protected exceptional rings in two-dimensional correlated systems
  with chiral symmetry},}\ }\href {\doibase 10.1103/PhysRevB.99.121101}
  {\bibfield  {journal} {\bibinfo  {journal} {Phys. Rev. B}\ }\textbf {\bibinfo
  {volume} {99}},\ \bibinfo {pages} {121101} (\bibinfo {year}
  {2019})}\BibitemShut {NoStop}%
\bibitem [{\citenamefont {Yoshida}\ and\ \citenamefont
  {Hatsugai}(2019)}]{yashida2019exceptional}%
  \BibitemOpen
  \bibfield  {author} {\bibinfo {author} {\bibfnamefont {Tsuneya}\ \bibnamefont
  {Yoshida}}\ and\ \bibinfo {author} {\bibfnamefont {Yasuhiro}\ \bibnamefont
  {Hatsugai}},\ }\bibfield  {title} {\enquote {\bibinfo {title} {Exceptional
  rings protected by emergent symmetry for mechanical systems},}\ }\href
  {\doibase 10.1103/PhysRevB.100.054109} {\bibfield  {journal} {\bibinfo
  {journal} {Phys. Rev. B}\ }\textbf {\bibinfo {volume} {100}},\ \bibinfo
  {pages} {054109} (\bibinfo {year} {2019})}\BibitemShut {NoStop}%
\bibitem [{\citenamefont {Okugawa}\ and\ \citenamefont
  {Yokoyama}(2019)}]{okugawa2019exceptional}%
  \BibitemOpen
  \bibfield  {author} {\bibinfo {author} {\bibfnamefont {Ryo}\ \bibnamefont
  {Okugawa}}\ and\ \bibinfo {author} {\bibfnamefont {Takehito}\ \bibnamefont
  {Yokoyama}},\ }\bibfield  {title} {\enquote {\bibinfo {title} {Topological
  exceptional surfaces in non-hermitian systems with parity-time and
  parity-particle-hole symmetries},}\ }\href {\doibase
  10.1103/PhysRevB.99.041202} {\bibfield  {journal} {\bibinfo  {journal} {Phys.
  Rev. B}\ }\textbf {\bibinfo {volume} {99}},\ \bibinfo {pages} {041202}
  (\bibinfo {year} {2019})}\BibitemShut {NoStop}%
\bibitem [{\citenamefont {Gong}\ \emph {et~al.}(2018)\citenamefont {Gong},
  \citenamefont {Ashida}, \citenamefont {Kawabata}, \citenamefont {Takasan},
  \citenamefont {Higashikawa},\ and\ \citenamefont
  {Ueda}}]{gong2018topological}%
  \BibitemOpen
  \bibfield  {author} {\bibinfo {author} {\bibfnamefont {Zongping}\
  \bibnamefont {Gong}}, \bibinfo {author} {\bibfnamefont {Yuto}\ \bibnamefont
  {Ashida}}, \bibinfo {author} {\bibfnamefont {Kohei}\ \bibnamefont
  {Kawabata}}, \bibinfo {author} {\bibfnamefont {Kazuaki}\ \bibnamefont
  {Takasan}}, \bibinfo {author} {\bibfnamefont {Sho}\ \bibnamefont
  {Higashikawa}}, \ and\ \bibinfo {author} {\bibfnamefont {Masahito}\
  \bibnamefont {Ueda}},\ }\bibfield  {title} {\enquote {\bibinfo {title}
  {Topological phases of non-hermitian systems},}\ }\href@noop {} {\bibfield
  {journal} {\bibinfo  {journal} {Physical Review X}\ }\textbf {\bibinfo
  {volume} {8}},\ \bibinfo {pages} {031079} (\bibinfo {year}
  {2018})}\BibitemShut {NoStop}%
\bibitem [{\citenamefont {Kawabata}\ \emph
  {et~al.}(2019{\natexlab{a}})\citenamefont {Kawabata}, \citenamefont
  {Shiozaki}, \citenamefont {Ueda},\ and\ \citenamefont
  {Sato}}]{kawabata2019symmetry}%
  \BibitemOpen
  \bibfield  {author} {\bibinfo {author} {\bibfnamefont {Kohei}\ \bibnamefont
  {Kawabata}}, \bibinfo {author} {\bibfnamefont {Ken}\ \bibnamefont
  {Shiozaki}}, \bibinfo {author} {\bibfnamefont {Masahito}\ \bibnamefont
  {Ueda}}, \ and\ \bibinfo {author} {\bibfnamefont {Masatoshi}\ \bibnamefont
  {Sato}},\ }\bibfield  {title} {\enquote {\bibinfo {title} {Symmetry and
  topology in non-hermitian physics},}\ }\href {\doibase
  10.1103/PhysRevX.9.041015} {\bibfield  {journal} {\bibinfo  {journal} {Phys.
  Rev. X}\ }\textbf {\bibinfo {volume} {9}},\ \bibinfo {pages} {041015}
  (\bibinfo {year} {2019}{\natexlab{a}})}\BibitemShut {NoStop}%
\bibitem [{\citenamefont {Liu}\ \emph {et~al.}(2019)\citenamefont {Liu},
  \citenamefont {Jiang},\ and\ \citenamefont {Chen}}]{Liu2019nonHclass}%
  \BibitemOpen
  \bibfield  {author} {\bibinfo {author} {\bibfnamefont {Chun-Hui}\
  \bibnamefont {Liu}}, \bibinfo {author} {\bibfnamefont {Hui}\ \bibnamefont
  {Jiang}}, \ and\ \bibinfo {author} {\bibfnamefont {Shu}\ \bibnamefont
  {Chen}},\ }\bibfield  {title} {\enquote {\bibinfo {title} {Topological
  classification of non-hermitian systems with reflection symmetry},}\ }\href
  {\doibase 10.1103/PhysRevB.99.125103} {\bibfield  {journal} {\bibinfo
  {journal} {Phys. Rev. B}\ }\textbf {\bibinfo {volume} {99}},\ \bibinfo
  {pages} {125103} (\bibinfo {year} {2019})}\BibitemShut {NoStop}%
\bibitem [{\citenamefont {Zhou}\ and\ \citenamefont
  {Lee}(2019)}]{zhou2019periodic}%
  \BibitemOpen
  \bibfield  {author} {\bibinfo {author} {\bibfnamefont {Hengyun}\ \bibnamefont
  {Zhou}}\ and\ \bibinfo {author} {\bibfnamefont {Jong~Yeon}\ \bibnamefont
  {Lee}},\ }\bibfield  {title} {\enquote {\bibinfo {title} {Periodic table for
  topological bands with non-hermitian symmetries},}\ }\href@noop {} {\bibfield
   {journal} {\bibinfo  {journal} {Physical Review B}\ }\textbf {\bibinfo
  {volume} {99}},\ \bibinfo {pages} {235112} (\bibinfo {year}
  {2019})}\BibitemShut {NoStop}%
\bibitem [{\citenamefont {Li}\ \emph {et~al.}(2019)\citenamefont {Li},
  \citenamefont {Lee},\ and\ \citenamefont {Gong}}]{Li2019geometric}%
  \BibitemOpen
  \bibfield  {author} {\bibinfo {author} {\bibfnamefont {Linhu}\ \bibnamefont
  {Li}}, \bibinfo {author} {\bibfnamefont {Ching~Hua}\ \bibnamefont {Lee}}, \
  and\ \bibinfo {author} {\bibfnamefont {Jiangbin}\ \bibnamefont {Gong}},\
  }\bibfield  {title} {\enquote {\bibinfo {title} {Geometric characterization
  of non-hermitian topological systems through the singularity ring in
  pseudospin vector space},}\ }\href {\doibase 10.1103/PhysRevB.100.075403}
  {\bibfield  {journal} {\bibinfo  {journal} {Phys. Rev. B}\ }\textbf {\bibinfo
  {volume} {100}},\ \bibinfo {pages} {075403} (\bibinfo {year}
  {2019})}\BibitemShut {NoStop}%
\bibitem [{\citenamefont {Liu}\ and\ \citenamefont
  {Chen}(2019)}]{liu2019topological}%
  \BibitemOpen
  \bibfield  {author} {\bibinfo {author} {\bibfnamefont {Chun-Hui}\
  \bibnamefont {Liu}}\ and\ \bibinfo {author} {\bibfnamefont {Shu}\
  \bibnamefont {Chen}},\ }\bibfield  {title} {\enquote {\bibinfo {title}
  {Topological classification of defects in non-hermitian systems},}\ }\href
  {\doibase 10.1103/PhysRevB.100.144106} {\bibfield  {journal} {\bibinfo
  {journal} {Phys. Rev. B}\ }\textbf {\bibinfo {volume} {100}},\ \bibinfo
  {pages} {144106} (\bibinfo {year} {2019})}\BibitemShut {NoStop}%
\bibitem [{\citenamefont {Kawabata}\ \emph
  {et~al.}(2019{\natexlab{b}})\citenamefont {Kawabata}, \citenamefont
  {Bessho},\ and\ \citenamefont {Sato}}]{kawabata2019classification}%
  \BibitemOpen
  \bibfield  {author} {\bibinfo {author} {\bibfnamefont {Kohei}\ \bibnamefont
  {Kawabata}}, \bibinfo {author} {\bibfnamefont {Takumi}\ \bibnamefont
  {Bessho}}, \ and\ \bibinfo {author} {\bibfnamefont {Masatoshi}\ \bibnamefont
  {Sato}},\ }\bibfield  {title} {\enquote {\bibinfo {title} {Classification of
  exceptional points and non-hermitian topological semimetals},}\ }\href@noop
  {} {\bibfield  {journal} {\bibinfo  {journal} {Phys. Rev. Lett.}\ }\textbf
  {\bibinfo {volume} {123}},\ \bibinfo {pages} {066405} (\bibinfo {year}
  {2019}{\natexlab{b}})}\BibitemShut {NoStop}%
\bibitem [{\citenamefont {Okuma}\ and\ \citenamefont
  {Sato}(2019)}]{okuma2019topological}%
  \BibitemOpen
  \bibfield  {author} {\bibinfo {author} {\bibfnamefont {Nobuyuki}\
  \bibnamefont {Okuma}}\ and\ \bibinfo {author} {\bibfnamefont {Masatoshi}\
  \bibnamefont {Sato}},\ }\bibfield  {title} {\enquote {\bibinfo {title}
  {Topological phase transition driven by infinitesimal instability: Majorana
  fermions in non-hermitian spintronics},}\ }\href@noop {} {\bibfield
  {journal} {\bibinfo  {journal} {Physical Review Letters}\ }\textbf {\bibinfo
  {volume} {123}},\ \bibinfo {pages} {097701} (\bibinfo {year}
  {2019})}\BibitemShut {NoStop}%
\bibitem [{\citenamefont {Borgnia}\ \emph {et~al.}(2020)\citenamefont
  {Borgnia}, \citenamefont {Kruchkov},\ and\ \citenamefont
  {Slager}}]{borgnia2019non}%
  \BibitemOpen
  \bibfield  {author} {\bibinfo {author} {\bibfnamefont {Dan~S}\ \bibnamefont
  {Borgnia}}, \bibinfo {author} {\bibfnamefont {Alex~Jura}\ \bibnamefont
  {Kruchkov}}, \ and\ \bibinfo {author} {\bibfnamefont {Robert-Jan}\
  \bibnamefont {Slager}},\ }\bibfield  {title} {\enquote {\bibinfo {title}
  {Non-hermitian boundary modes and topology},}\ }\href@noop {} {\bibfield
  {journal} {\bibinfo  {journal} {Physical Review Letters}\ }\textbf {\bibinfo
  {volume} {124}},\ \bibinfo {pages} {056802} (\bibinfo {year}
  {2020})}\BibitemShut {NoStop}%
\bibitem [{\citenamefont {H{\"o}ckendorf}\ \emph {et~al.}(2019)\citenamefont
  {H{\"o}ckendorf}, \citenamefont {Alvermann},\ and\ \citenamefont
  {Fehske}}]{hockendorf2019non}%
  \BibitemOpen
  \bibfield  {author} {\bibinfo {author} {\bibfnamefont {Bastian}\ \bibnamefont
  {H{\"o}ckendorf}}, \bibinfo {author} {\bibfnamefont {Andreas}\ \bibnamefont
  {Alvermann}}, \ and\ \bibinfo {author} {\bibfnamefont {Holger}\ \bibnamefont
  {Fehske}},\ }\bibfield  {title} {\enquote {\bibinfo {title} {Non-hermitian
  floquet chains as topological charge pumps},}\ }\href@noop {} {\bibfield
  {journal} {\bibinfo  {journal} {arXiv preprint arXiv:1911.11413}\ } (\bibinfo
  {year} {2019})}\BibitemShut {NoStop}%
\bibitem [{\citenamefont {Yao}\ and\ \citenamefont {Wang}(2018)}]{yao2018edge}%
  \BibitemOpen
  \bibfield  {author} {\bibinfo {author} {\bibfnamefont {Shunyu}\ \bibnamefont
  {Yao}}\ and\ \bibinfo {author} {\bibfnamefont {Zhong}\ \bibnamefont {Wang}},\
  }\bibfield  {title} {\enquote {\bibinfo {title} {Edge states and topological
  invariants of non-hermitian systems},}\ }\href {\doibase
  10.1103/PhysRevLett.121.086803} {\bibfield  {journal} {\bibinfo  {journal}
  {Phys. Rev. Lett.}\ }\textbf {\bibinfo {volume} {121}},\ \bibinfo {pages}
  {086803} (\bibinfo {year} {2018})}\BibitemShut {NoStop}%
\bibitem [{\citenamefont {Xiong}(2018)}]{xiong2018does}%
  \BibitemOpen
  \bibfield  {author} {\bibinfo {author} {\bibfnamefont {Ye}~\bibnamefont
  {Xiong}},\ }\bibfield  {title} {\enquote {\bibinfo {title} {Why does bulk
  boundary correspondence fail in some non-hermitian topological models},}\
  }\href@noop {} {\bibfield  {journal} {\bibinfo  {journal} {Journal of Physics
  Communications}\ }\textbf {\bibinfo {volume} {2}},\ \bibinfo {pages} {035043}
  (\bibinfo {year} {2018})}\BibitemShut {NoStop}%
\bibitem [{\citenamefont {Yao}\ \emph {et~al.}(2018)\citenamefont {Yao},
  \citenamefont {Song},\ and\ \citenamefont {Wang}}]{yao2018non}%
  \BibitemOpen
  \bibfield  {author} {\bibinfo {author} {\bibfnamefont {Shunyu}\ \bibnamefont
  {Yao}}, \bibinfo {author} {\bibfnamefont {Fei}\ \bibnamefont {Song}}, \ and\
  \bibinfo {author} {\bibfnamefont {Zhong}\ \bibnamefont {Wang}},\ }\bibfield
  {title} {\enquote {\bibinfo {title} {Non-hermitian chern bands},}\
  }\href@noop {} {\bibfield  {journal} {\bibinfo  {journal} {Physical review
  letters}\ }\textbf {\bibinfo {volume} {121}},\ \bibinfo {pages} {136802}
  (\bibinfo {year} {2018})}\BibitemShut {NoStop}%
\bibitem [{\citenamefont {Alvarez}\ \emph {et~al.}(2018)\citenamefont
  {Alvarez}, \citenamefont {Vargas},\ and\ \citenamefont
  {Torres}}]{alvarez2018non}%
  \BibitemOpen
  \bibfield  {author} {\bibinfo {author} {\bibfnamefont {VM~Martinez}\
  \bibnamefont {Alvarez}}, \bibinfo {author} {\bibfnamefont {JE~Barrios}\
  \bibnamefont {Vargas}}, \ and\ \bibinfo {author} {\bibfnamefont {LEF~Foa}\
  \bibnamefont {Torres}},\ }\bibfield  {title} {\enquote {\bibinfo {title}
  {Non-hermitian robust edge states in one dimension: Anomalous localization
  and eigenspace condensation at exceptional points},}\ }\href@noop {}
  {\bibfield  {journal} {\bibinfo  {journal} {Phys. Rev. B}\ }\textbf {\bibinfo
  {volume} {97}},\ \bibinfo {pages} {121401} (\bibinfo {year}
  {2018})}\BibitemShut {NoStop}%
\bibitem [{\citenamefont {Kunst}\ \emph {et~al.}(2018)\citenamefont {Kunst},
  \citenamefont {Edvardsson}, \citenamefont {Budich},\ and\ \citenamefont
  {Bergholtz}}]{kunst2018biorthogonal}%
  \BibitemOpen
  \bibfield  {author} {\bibinfo {author} {\bibfnamefont {Flore~K.}\
  \bibnamefont {Kunst}}, \bibinfo {author} {\bibfnamefont {Elisabet}\
  \bibnamefont {Edvardsson}}, \bibinfo {author} {\bibfnamefont {Jan~Carl}\
  \bibnamefont {Budich}}, \ and\ \bibinfo {author} {\bibfnamefont {Emil~J.}\
  \bibnamefont {Bergholtz}},\ }\bibfield  {title} {\enquote {\bibinfo {title}
  {Biorthogonal bulk-boundary correspondence in non-hermitian systems},}\
  }\href {\doibase 10.1103/PhysRevLett.121.026808} {\bibfield  {journal}
  {\bibinfo  {journal} {Phys. Rev. Lett.}\ }\textbf {\bibinfo {volume} {121}},\
  \bibinfo {pages} {026808} (\bibinfo {year} {2018})}\BibitemShut {NoStop}%
\bibitem [{\citenamefont {Jin}\ and\ \citenamefont {Song}(2019)}]{jin2018bulk}%
  \BibitemOpen
  \bibfield  {author} {\bibinfo {author} {\bibfnamefont {L}~\bibnamefont
  {Jin}}\ and\ \bibinfo {author} {\bibfnamefont {Z}~\bibnamefont {Song}},\
  }\bibfield  {title} {\enquote {\bibinfo {title} {Bulk-boundary correspondence
  in a non-hermitian system in one dimension with chiral inversion symmetry},}\
  }\href@noop {} {\bibfield  {journal} {\bibinfo  {journal} {Physical Review
  B}\ }\textbf {\bibinfo {volume} {99}},\ \bibinfo {pages} {081103} (\bibinfo
  {year} {2019})}\BibitemShut {NoStop}%
\bibitem [{\citenamefont {Lee}\ and\ \citenamefont
  {Thomale}(2019)}]{lee2019anatomy}%
  \BibitemOpen
  \bibfield  {author} {\bibinfo {author} {\bibfnamefont {Ching~Hua}\
  \bibnamefont {Lee}}\ and\ \bibinfo {author} {\bibfnamefont {Ronny}\
  \bibnamefont {Thomale}},\ }\bibfield  {title} {\enquote {\bibinfo {title}
  {Anatomy of skin modes and topology in non-hermitian systems},}\ }\href@noop
  {} {\bibfield  {journal} {\bibinfo  {journal} {Physical Review B}\ }\textbf
  {\bibinfo {volume} {99}},\ \bibinfo {pages} {201103} (\bibinfo {year}
  {2019})}\BibitemShut {NoStop}%
\bibitem [{\citenamefont {Yokomizo}\ and\ \citenamefont
  {Murakami}(2019)}]{yokomizo2019nonbloch}%
  \BibitemOpen
  \bibfield  {author} {\bibinfo {author} {\bibfnamefont {Kazuki}\ \bibnamefont
  {Yokomizo}}\ and\ \bibinfo {author} {\bibfnamefont {Shuichi}\ \bibnamefont
  {Murakami}},\ }\bibfield  {title} {\enquote {\bibinfo {title} {Non-bloch band
  theory of non-hermitian systems},}\ }\href {\doibase
  10.1103/PhysRevLett.123.066404} {\bibfield  {journal} {\bibinfo  {journal}
  {Phys. Rev. Lett.}\ }\textbf {\bibinfo {volume} {123}},\ \bibinfo {pages}
  {066404} (\bibinfo {year} {2019})}\BibitemShut {NoStop}%
\bibitem [{\citenamefont {Okuma}\ \emph {et~al.}()\citenamefont {Okuma},
  \citenamefont {Kawabata}, \citenamefont {Shiozaki},\ and\ \citenamefont
  {Sato}}]{nobuyuki2019gbz}%
  \BibitemOpen
  \bibfield  {author} {\bibinfo {author} {\bibfnamefont {Nobuyuki}\
  \bibnamefont {Okuma}}, \bibinfo {author} {\bibfnamefont {Kohei}\ \bibnamefont
  {Kawabata}}, \bibinfo {author} {\bibfnamefont {Ken}\ \bibnamefont
  {Shiozaki}}, \ and\ \bibinfo {author} {\bibfnamefont {Masatoshi}\
  \bibnamefont {Sato}},\ }\bibfield  {title} {\enquote {\bibinfo {title}
  {Topological origin of non-hermitian skin effects},}\ }\href@noop {} {\
  }\Eprint {http://arxiv.org/abs/1910.02878v3} {1910.02878v3} \BibitemShut
  {NoStop}%
\bibitem [{\citenamefont {Zhang}\ \emph {et~al.}()\citenamefont {Zhang},
  \citenamefont {Yang},\ and\ \citenamefont {Fang}}]{kai2019gbz}%
  \BibitemOpen
  \bibfield  {author} {\bibinfo {author} {\bibfnamefont {Kai}\ \bibnamefont
  {Zhang}}, \bibinfo {author} {\bibfnamefont {Zhesen}\ \bibnamefont {Yang}}, \
  and\ \bibinfo {author} {\bibfnamefont {Chen}\ \bibnamefont {Fang}},\
  }\bibfield  {title} {\enquote {\bibinfo {title} {Correspondence between
  winding numbers and skin modes in non-hermitian systems},}\ }\href@noop {} {\
  }\Eprint {http://arxiv.org/abs/1910.01131v1} {1910.01131v1} \BibitemShut
  {NoStop}%
\bibitem [{\citenamefont {Yang}\ \emph {et~al.}()\citenamefont {Yang},
  \citenamefont {Zhang}, \citenamefont {Fang},\ and\ \citenamefont
  {Hu}}]{zhesen2019agbz}%
  \BibitemOpen
  \bibfield  {author} {\bibinfo {author} {\bibfnamefont {Zhesen}\ \bibnamefont
  {Yang}}, \bibinfo {author} {\bibfnamefont {Kai}\ \bibnamefont {Zhang}},
  \bibinfo {author} {\bibfnamefont {Chen}\ \bibnamefont {Fang}}, \ and\
  \bibinfo {author} {\bibfnamefont {Jiangping}\ \bibnamefont {Hu}},\ }\bibfield
   {title} {\enquote {\bibinfo {title} {Auxiliary generalized brillouin zone
  method in non-hermitian band theory},}\ }\href@noop {} {\ }\Eprint
  {http://arxiv.org/abs/1912.05499v1} {1912.05499v1} \BibitemShut {NoStop}%
\bibitem [{\citenamefont {Lee}(2020)}]{lee2020many}%
  \BibitemOpen
  \bibfield  {author} {\bibinfo {author} {\bibfnamefont {Ching~Hua}\
  \bibnamefont {Lee}},\ }\bibfield  {title} {\enquote {\bibinfo {title}
  {Many-body topological and skin states without open boundaries},}\
  }\href@noop {} {\bibfield  {journal} {\bibinfo  {journal} {arXiv preprint
  arXiv:2006.01182}\ } (\bibinfo {year} {2020})}\BibitemShut {NoStop}%
\bibitem [{Note1()}]{Note1}%
  \BibitemOpen
  \bibinfo {note} {Non-Hermitian pumping also exists in fully reciprocal
  systems, where the asymmetry only appears when the effective description is
  restricted to certain momentum subspaces.}\BibitemShut {Stop}%
\bibitem [{\citenamefont {Longhi}\ \emph {et~al.}(2015)\citenamefont {Longhi},
  \citenamefont {Gatti},\ and\ \citenamefont {Della~Valle}}]{longhi2015robust}%
  \BibitemOpen
  \bibfield  {author} {\bibinfo {author} {\bibfnamefont {Stefano}\ \bibnamefont
  {Longhi}}, \bibinfo {author} {\bibfnamefont {Davide}\ \bibnamefont {Gatti}},
  \ and\ \bibinfo {author} {\bibfnamefont {Giuseppe}\ \bibnamefont
  {Della~Valle}},\ }\bibfield  {title} {\enquote {\bibinfo {title} {Robust
  light transport in non-hermitian photonic lattices},}\ }\href@noop {}
  {\bibfield  {journal} {\bibinfo  {journal} {Scientific reports}\ }\textbf
  {\bibinfo {volume} {5}},\ \bibinfo {pages} {13376} (\bibinfo {year}
  {2015})}\BibitemShut {NoStop}%
\bibitem [{\citenamefont {Midya}\ \emph {et~al.}(2018)\citenamefont {Midya},
  \citenamefont {Zhao},\ and\ \citenamefont {Feng}}]{midya2018non}%
  \BibitemOpen
  \bibfield  {author} {\bibinfo {author} {\bibfnamefont {Bikashkali}\
  \bibnamefont {Midya}}, \bibinfo {author} {\bibfnamefont {Han}\ \bibnamefont
  {Zhao}}, \ and\ \bibinfo {author} {\bibfnamefont {Liang}\ \bibnamefont
  {Feng}},\ }\bibfield  {title} {\enquote {\bibinfo {title} {Non-hermitian
  photonics promises exceptional topology of light},}\ }\href@noop {}
  {\bibfield  {journal} {\bibinfo  {journal} {Nature communications}\ }\textbf
  {\bibinfo {volume} {9}},\ \bibinfo {pages} {2674} (\bibinfo {year}
  {2018})}\BibitemShut {NoStop}%
\bibitem [{\citenamefont {Ezawa}(2019)}]{PhysRevB.100.075423}%
  \BibitemOpen
  \bibfield  {author} {\bibinfo {author} {\bibfnamefont {Motohiko}\
  \bibnamefont {Ezawa}},\ }\bibfield  {title} {\enquote {\bibinfo {title}
  {Electric circuit simulations of $n\mathrm{th}$-chern-number insulators in
  $2n$-dimensional space and their non-hermitian generalizations for arbitrary
  $n$},}\ }\href {\doibase 10.1103/PhysRevB.100.075423} {\bibfield  {journal}
  {\bibinfo  {journal} {Phys. Rev. B}\ }\textbf {\bibinfo {volume} {100}},\
  \bibinfo {pages} {075423} (\bibinfo {year} {2019})}\BibitemShut {NoStop}%
\bibitem [{\citenamefont {Hofmann}\ \emph {et~al.}(2019)\citenamefont
  {Hofmann}, \citenamefont {Helbig}, \citenamefont {Lee}, \citenamefont
  {Greiter},\ and\ \citenamefont {Thomale}}]{hofmann2019chiral}%
  \BibitemOpen
  \bibfield  {author} {\bibinfo {author} {\bibfnamefont {Tobias}\ \bibnamefont
  {Hofmann}}, \bibinfo {author} {\bibfnamefont {Tobias}\ \bibnamefont
  {Helbig}}, \bibinfo {author} {\bibfnamefont {Ching~Hua}\ \bibnamefont {Lee}},
  \bibinfo {author} {\bibfnamefont {Martin}\ \bibnamefont {Greiter}}, \ and\
  \bibinfo {author} {\bibfnamefont {Ronny}\ \bibnamefont {Thomale}},\
  }\bibfield  {title} {\enquote {\bibinfo {title} {Chiral voltage propagation
  and calibration in a topolectrical chern circuit},}\ }\href@noop {}
  {\bibfield  {journal} {\bibinfo  {journal} {Physical Review Letters}\
  }\textbf {\bibinfo {volume} {122}},\ \bibinfo {pages} {247702} (\bibinfo
  {year} {2019})}\BibitemShut {NoStop}%
\bibitem [{\citenamefont {Zhao}\ \emph {et~al.}(2019)\citenamefont {Zhao},
  \citenamefont {Qiao}, \citenamefont {Wu}, \citenamefont {Midya},
  \citenamefont {Longhi},\ and\ \citenamefont {Feng}}]{zhao2019non}%
  \BibitemOpen
  \bibfield  {author} {\bibinfo {author} {\bibfnamefont {Han}\ \bibnamefont
  {Zhao}}, \bibinfo {author} {\bibfnamefont {Xingdu}\ \bibnamefont {Qiao}},
  \bibinfo {author} {\bibfnamefont {Tianwei}\ \bibnamefont {Wu}}, \bibinfo
  {author} {\bibfnamefont {Bikashkali}\ \bibnamefont {Midya}}, \bibinfo
  {author} {\bibfnamefont {Stefano}\ \bibnamefont {Longhi}}, \ and\ \bibinfo
  {author} {\bibfnamefont {Liang}\ \bibnamefont {Feng}},\ }\bibfield  {title}
  {\enquote {\bibinfo {title} {Non-hermitian topological light steering},}\
  }\href@noop {} {\bibfield  {journal} {\bibinfo  {journal} {Science}\ }\textbf
  {\bibinfo {volume} {365}},\ \bibinfo {pages} {1163--1166} (\bibinfo {year}
  {2019})}\BibitemShut {NoStop}%
\bibitem [{\citenamefont {Longhi}(2019{\natexlab{a}})}]{longhi2019nonlaser}%
  \BibitemOpen
  \bibfield  {author} {\bibinfo {author} {\bibfnamefont {Stefano}\ \bibnamefont
  {Longhi}},\ }\bibfield  {title} {\enquote {\bibinfo {title} {Non-hermitian
  topological phase transition in pt-symmetric mode-locked lasers},}\
  }\href@noop {} {\bibfield  {journal} {\bibinfo  {journal} {Optics letters}\
  }\textbf {\bibinfo {volume} {44}},\ \bibinfo {pages} {1190--1193} (\bibinfo
  {year} {2019}{\natexlab{a}})}\BibitemShut {NoStop}%
\bibitem [{\citenamefont {Schomerus}(2020)}]{schomerus2019nonreciprocal}%
  \BibitemOpen
  \bibfield  {author} {\bibinfo {author} {\bibfnamefont {Henning}\ \bibnamefont
  {Schomerus}},\ }\bibfield  {title} {\enquote {\bibinfo {title} {Nonreciprocal
  response theory of non-hermitian mechanical metamaterials: Response phase
  transition from the skin effect of zero modes},}\ }\href@noop {} {\bibfield
  {journal} {\bibinfo  {journal} {Physical Review Research}\ }\textbf {\bibinfo
  {volume} {2}},\ \bibinfo {pages} {013058} (\bibinfo {year}
  {2020})}\BibitemShut {NoStop}%
\bibitem [{\citenamefont {Brandenbourger}\ \emph {et~al.}(2019)\citenamefont
  {Brandenbourger}, \citenamefont {Locsin}, \citenamefont {Lerner},\ and\
  \citenamefont {Coulais}}]{brandenbourger2019non}%
  \BibitemOpen
  \bibfield  {author} {\bibinfo {author} {\bibfnamefont {Martin}\ \bibnamefont
  {Brandenbourger}}, \bibinfo {author} {\bibfnamefont {Xander}\ \bibnamefont
  {Locsin}}, \bibinfo {author} {\bibfnamefont {Edan}\ \bibnamefont {Lerner}}, \
  and\ \bibinfo {author} {\bibfnamefont {Corentin}\ \bibnamefont {Coulais}},\
  }\bibfield  {title} {\enquote {\bibinfo {title} {Non-reciprocal robotic
  metamaterials},}\ }\href@noop {} {\bibfield  {journal} {\bibinfo  {journal}
  {Nature communications}\ }\textbf {\bibinfo {volume} {10}},\ \bibinfo {pages}
  {1--8} (\bibinfo {year} {2019})}\BibitemShut {NoStop}%
\bibitem [{\citenamefont {Fruchart}\ \emph {et~al.}(2020)\citenamefont
  {Fruchart}, \citenamefont {Zhou},\ and\ \citenamefont
  {Vitelli}}]{fruchart2019dualities}%
  \BibitemOpen
  \bibfield  {author} {\bibinfo {author} {\bibfnamefont {Michel}\ \bibnamefont
  {Fruchart}}, \bibinfo {author} {\bibfnamefont {Yujie}\ \bibnamefont {Zhou}},
  \ and\ \bibinfo {author} {\bibfnamefont {Vincenzo}\ \bibnamefont {Vitelli}},\
  }\bibfield  {title} {\enquote {\bibinfo {title} {Dualities and non-abelian
  mechanics},}\ }\href@noop {} {\bibfield  {journal} {\bibinfo  {journal}
  {Nature}\ }\textbf {\bibinfo {volume} {577}},\ \bibinfo {pages} {636--640}
  (\bibinfo {year} {2020})}\BibitemShut {NoStop}%
\bibitem [{\citenamefont {Helbig}\ \emph {et~al.}(2020)\citenamefont {Helbig},
  \citenamefont {Hofmann}, \citenamefont {Imhof}, \citenamefont {Abdelghany},
  \citenamefont {Kiessling}, \citenamefont {Molenkamp}, \citenamefont {Lee},
  \citenamefont {Szameit}, \citenamefont {Greiter},\ and\ \citenamefont
  {Thomale}}]{tobias2019observation}%
  \BibitemOpen
  \bibfield  {author} {\bibinfo {author} {\bibfnamefont {T}~\bibnamefont
  {Helbig}}, \bibinfo {author} {\bibfnamefont {T}~\bibnamefont {Hofmann}},
  \bibinfo {author} {\bibfnamefont {S}~\bibnamefont {Imhof}}, \bibinfo {author}
  {\bibfnamefont {M}~\bibnamefont {Abdelghany}}, \bibinfo {author}
  {\bibfnamefont {T}~\bibnamefont {Kiessling}}, \bibinfo {author}
  {\bibfnamefont {LW}~\bibnamefont {Molenkamp}}, \bibinfo {author}
  {\bibfnamefont {CH}~\bibnamefont {Lee}}, \bibinfo {author} {\bibfnamefont
  {A}~\bibnamefont {Szameit}}, \bibinfo {author} {\bibfnamefont
  {M}~\bibnamefont {Greiter}}, \ and\ \bibinfo {author} {\bibfnamefont
  {R}~\bibnamefont {Thomale}},\ }\bibfield  {title} {\enquote {\bibinfo {title}
  {Generalized bulk--boundary correspondence in non-hermitian topolectrical
  circuits},}\ }\href@noop {} {\bibfield  {journal} {\bibinfo  {journal}
  {Nature Physics}\ ,\ \bibinfo {pages} {1--4}} (\bibinfo {year}
  {2020})}\BibitemShut {NoStop}%
\bibitem [{\citenamefont {Hofmann}\ \emph {et~al.}(2020)\citenamefont
  {Hofmann}, \citenamefont {Helbig}, \citenamefont {Schindler}, \citenamefont
  {Salgo}, \citenamefont {Brzezi{\'n}ska}, \citenamefont {Greiter},
  \citenamefont {Kiessling}, \citenamefont {Wolf}, \citenamefont {Vollhardt},
  \citenamefont {Kaba{\v{s}}i} \emph {et~al.}}]{tobias2019reciprocal}%
  \BibitemOpen
  \bibfield  {author} {\bibinfo {author} {\bibfnamefont {Tobias}\ \bibnamefont
  {Hofmann}}, \bibinfo {author} {\bibfnamefont {Tobias}\ \bibnamefont
  {Helbig}}, \bibinfo {author} {\bibfnamefont {Frank}\ \bibnamefont
  {Schindler}}, \bibinfo {author} {\bibfnamefont {Nora}\ \bibnamefont {Salgo}},
  \bibinfo {author} {\bibfnamefont {Marta}\ \bibnamefont {Brzezi{\'n}ska}},
  \bibinfo {author} {\bibfnamefont {Martin}\ \bibnamefont {Greiter}}, \bibinfo
  {author} {\bibfnamefont {Tobias}\ \bibnamefont {Kiessling}}, \bibinfo
  {author} {\bibfnamefont {David}\ \bibnamefont {Wolf}}, \bibinfo {author}
  {\bibfnamefont {Achim}\ \bibnamefont {Vollhardt}}, \bibinfo {author}
  {\bibfnamefont {Anton}\ \bibnamefont {Kaba{\v{s}}i}},  \emph {et~al.},\
  }\bibfield  {title} {\enquote {\bibinfo {title} {Reciprocal skin effect and
  its realization in a topolectrical circuit},}\ }\href@noop {} {\bibfield
  {journal} {\bibinfo  {journal} {Physical Review Research}\ }\textbf {\bibinfo
  {volume} {2}},\ \bibinfo {pages} {023265} (\bibinfo {year}
  {2020})}\BibitemShut {NoStop}%
\bibitem [{\citenamefont {Ghatak}\ \emph {et~al.}()\citenamefont {Ghatak},
  \citenamefont {Brandenbourger}, \citenamefont {van Wezel},\ and\
  \citenamefont {Coulais}}]{ananya2019observation}%
  \BibitemOpen
  \bibfield  {author} {\bibinfo {author} {\bibfnamefont {Ananya}\ \bibnamefont
  {Ghatak}}, \bibinfo {author} {\bibfnamefont {Martin}\ \bibnamefont
  {Brandenbourger}}, \bibinfo {author} {\bibfnamefont {Jasper}\ \bibnamefont
  {van Wezel}}, \ and\ \bibinfo {author} {\bibfnamefont {Corentin}\
  \bibnamefont {Coulais}},\ }\bibfield  {title} {\enquote {\bibinfo {title}
  {Observation of non-hermitian topology and its bulk-edge correspondence},}\
  }\href@noop {} {\ }\Eprint {http://arxiv.org/abs/1907.11619v1} {1907.11619v1}
  \BibitemShut {NoStop}%
\bibitem [{\citenamefont {Xiao}\ \emph {et~al.}()\citenamefont {Xiao},
  \citenamefont {Deng}, \citenamefont {Wang}, \citenamefont {Zhu},
  \citenamefont {Wang}, \citenamefont {Yi},\ and\ \citenamefont
  {Xue}}]{lei2019observation}%
  \BibitemOpen
  \bibfield  {author} {\bibinfo {author} {\bibfnamefont {Lei}\ \bibnamefont
  {Xiao}}, \bibinfo {author} {\bibfnamefont {Tianshu}\ \bibnamefont {Deng}},
  \bibinfo {author} {\bibfnamefont {Kunkun}\ \bibnamefont {Wang}}, \bibinfo
  {author} {\bibfnamefont {Gaoyan}\ \bibnamefont {Zhu}}, \bibinfo {author}
  {\bibfnamefont {Zhong}\ \bibnamefont {Wang}}, \bibinfo {author}
  {\bibfnamefont {Wei}\ \bibnamefont {Yi}}, \ and\ \bibinfo {author}
  {\bibfnamefont {Peng}\ \bibnamefont {Xue}},\ }\bibfield  {title} {\enquote
  {\bibinfo {title} {Observation of non-hermitian bulk-boundary correspondence
  in quantum dynamics},}\ }\href@noop {} {\ }\Eprint
  {http://arxiv.org/abs/1907.12566v1} {1907.12566v1} \BibitemShut {NoStop}%
\bibitem [{Note2()}]{Note2}%
  \BibitemOpen
  \bibinfo {note} {It is distinct from the non-Hermitian SSH model which also
  possesses $\protect \mathbb {Z}_2$ topology.}\BibitemShut {Stop}%
\bibitem [{Note3()}]{Note3}%
  \BibitemOpen
  \bibinfo {note} {These conditions are generically sufficient unless there is
  additional symmetry obstruction, see Ref.~\cite
  {nobuyuki2019gbz}.}\BibitemShut {Stop}%
\bibitem [{\citenamefont {Li}\ \emph {et~al.}(2020{\natexlab{a}})\citenamefont
  {Li}, \citenamefont {Lee},\ and\ \citenamefont {Gong}}]{li2019topology}%
  \BibitemOpen
  \bibfield  {author} {\bibinfo {author} {\bibfnamefont {Linhu}\ \bibnamefont
  {Li}}, \bibinfo {author} {\bibfnamefont {Ching~Hua}\ \bibnamefont {Lee}}, \
  and\ \bibinfo {author} {\bibfnamefont {Jiangbin}\ \bibnamefont {Gong}},\
  }\bibfield  {title} {\enquote {\bibinfo {title} {Topological switch for
  non-hermitian skin effect in cold-atom systems with loss},}\ }\href@noop {}
  {\bibfield  {journal} {\bibinfo  {journal} {Physical Review Letters}\
  }\textbf {\bibinfo {volume} {124}},\ \bibinfo {pages} {250402} (\bibinfo
  {year} {2020}{\natexlab{a}})}\BibitemShut {NoStop}%
\bibitem [{\citenamefont {Longhi}(2019{\natexlab{b}})}]{longhi2019probing}%
  \BibitemOpen
  \bibfield  {author} {\bibinfo {author} {\bibfnamefont {Stefano}\ \bibnamefont
  {Longhi}},\ }\bibfield  {title} {\enquote {\bibinfo {title} {Probing
  non-hermitian skin effect and non-bloch phase transitions},}\ }\href@noop {}
  {\bibfield  {journal} {\bibinfo  {journal} {Physical Review Research}\
  }\textbf {\bibinfo {volume} {1}},\ \bibinfo {pages} {023013} (\bibinfo {year}
  {2019}{\natexlab{b}})}\BibitemShut {NoStop}%
\bibitem [{Note4()}]{Note4}%
  \BibitemOpen
  \bibinfo {note} {A quasi-reciprocal system is not necessarily reciprocal,
  since it can still be non-reciprocal while exhibiting unbroken bulk-boundary
  correspondence i.e. with Hermitian flux and no gain/loss.}\BibitemShut
  {Stop}%
\bibitem [{Note5()}]{Note5}%
  \BibitemOpen
  \bibinfo {note} {Except for subextensive topological modes}\BibitemShut
  {NoStop}%
\bibitem [{Note6()}]{Note6}%
  \BibitemOpen
  \bibinfo {note} {Up to exponentially small corrections in system size, see
  Refs.~\cite {alvarez2018non,lee2019anatomy}.}\BibitemShut {Stop}%
\bibitem [{sup()}]{suppmat}%
  \BibitemOpen
  \bibfield  {title} {\enquote {\bibinfo {title} {Supplemental materials},}\
  }\href@noop {} {\bibinfo  {journal} {Supplemental Materials}\ }\BibitemShut
  {NoStop}%
\bibitem [{Note7()}]{Note7}%
  \BibitemOpen
\bibfield  {journal} {  }\bibinfo {note} {Non-Hermitian pumping occurs whenever
  $t^*_n\not =t_n\not =t_{-n}$ for at least one $n$.}\BibitemShut {Stop}%
\bibitem [{Note8()}]{Note8}%
  \BibitemOpen
  \bibinfo {note} {Some exceptions arise in Floquet systems, with rapid
  quenching behavior giving rise to high temporal and spatial harmonics or
  dynamical instabilities~\cite
  {mikami2016brillouin,zhou2018recipe,li2018realistic,lee2020ultrafast}.}\BibitemShut
  {Stop}%
\bibitem [{Note9()}]{Note9}%
  \BibitemOpen
  \bibinfo {note} {See for instance Ref.~\cite {he2001exponential,lee2015free}
  for a treatment of how Fourier coefficient decay rates depend on complex
  analytic structure.}\BibitemShut {Stop}%
\bibitem [{Note10()}]{Note10}%
  \BibitemOpen
  \bibinfo {note} {Negative powers of $z$ can be made positive through an
  appropriate multiplicative factor.}\BibitemShut {Stop}%
\bibitem [{\citenamefont {Kitaev}(2001)}]{kitaev}%
  \BibitemOpen
  \bibfield  {author} {\bibinfo {author} {\bibfnamefont {A~Yu}\ \bibnamefont
  {Kitaev}},\ }\bibfield  {title} {\enquote {\bibinfo {title} {Unpaired
  majorana fermions in quantum wires},}\ }\href@noop {} {\bibfield  {journal}
  {\bibinfo  {journal} {Physics-Uspekhi}\ }\textbf {\bibinfo {volume} {44}},\
  \bibinfo {pages} {131} (\bibinfo {year} {2001})}\BibitemShut {NoStop}%
\bibitem [{\citenamefont {Li}\ \emph {et~al.}(2016)\citenamefont {Li},
  \citenamefont {Yang},\ and\ \citenamefont {Chen}}]{li2016Z2}%
  \BibitemOpen
  \bibfield  {author} {\bibinfo {author} {\bibfnamefont {Linhu}\ \bibnamefont
  {Li}}, \bibinfo {author} {\bibfnamefont {Chao}\ \bibnamefont {Yang}}, \ and\
  \bibinfo {author} {\bibfnamefont {Shu}\ \bibnamefont {Chen}},\ }\bibfield
  {title} {\enquote {\bibinfo {title} {Topological invariants for phase
  transition points of one-dimensional z 2 topological systems},}\ }\href@noop
  {} {\bibfield  {journal} {\bibinfo  {journal} {The European Physical Journal
  B}\ }\textbf {\bibinfo {volume} {89}},\ \bibinfo {pages} {195} (\bibinfo
  {year} {2016})}\BibitemShut {NoStop}%
\bibitem [{Note11()}]{Note11}%
  \BibitemOpen
  \bibinfo {note} {More possibilities exist for $k$-dependent non-Hermitian
  terms, i.e. $\protect \qopname \relax o{sin}k\protect \tmspace +\thinmuskip
  {.1667em} \sigma _z$.}\BibitemShut {Stop}%
\bibitem [{\citenamefont {Jiang}\ \emph {et~al.}(2018)\citenamefont {Jiang},
  \citenamefont {Yang},\ and\ \citenamefont {Chen}}]{Jiang2018nochiral}%
  \BibitemOpen
  \bibfield  {author} {\bibinfo {author} {\bibfnamefont {Hui}\ \bibnamefont
  {Jiang}}, \bibinfo {author} {\bibfnamefont {Chao}\ \bibnamefont {Yang}}, \
  and\ \bibinfo {author} {\bibfnamefont {Shu}\ \bibnamefont {Chen}},\
  }\bibfield  {title} {\enquote {\bibinfo {title} {Topological invariants and
  phase diagrams for one-dimensional two-band non-hermitian systems without
  chiral symmetry},}\ }\href {\doibase 10.1103/PhysRevA.98.052116} {\bibfield
  {journal} {\bibinfo  {journal} {Phys. Rev. A}\ }\textbf {\bibinfo {volume}
  {98}},\ \bibinfo {pages} {052116} (\bibinfo {year} {2018})}\BibitemShut
  {NoStop}%
\bibitem [{\citenamefont {Brody}(2013)}]{brody2013biorthogonal}%
  \BibitemOpen
  \bibfield  {author} {\bibinfo {author} {\bibfnamefont {Dorje~C}\ \bibnamefont
  {Brody}},\ }\bibfield  {title} {\enquote {\bibinfo {title} {Biorthogonal
  quantum mechanics},}\ }\href@noop {} {\bibfield  {journal} {\bibinfo
  {journal} {Journal of Physics A: Mathematical and Theoretical}\ }\textbf
  {\bibinfo {volume} {47}},\ \bibinfo {pages} {035305} (\bibinfo {year}
  {2013})}\BibitemShut {NoStop}%
\bibitem [{\citenamefont {Herviou}\ \emph {et~al.}(2019)\citenamefont
  {Herviou}, \citenamefont {Regnault},\ and\ \citenamefont
  {Bardarson}}]{herviou2019entanglement}%
  \BibitemOpen
  \bibfield  {author} {\bibinfo {author} {\bibfnamefont {Loic}\ \bibnamefont
  {Herviou}}, \bibinfo {author} {\bibfnamefont {Nicolas}\ \bibnamefont
  {Regnault}}, \ and\ \bibinfo {author} {\bibfnamefont {Jens~H}\ \bibnamefont
  {Bardarson}},\ }\bibfield  {title} {\enquote {\bibinfo {title} {Entanglement
  spectrum and symmetries in non-hermitian fermionic non-interacting models},}\
  }\href@noop {} {\bibfield  {journal} {\bibinfo  {journal} {SciPost Physics}\
  }\textbf {\bibinfo {volume} {7}} (\bibinfo {year} {2019})}\BibitemShut
  {NoStop}%
\bibitem [{\citenamefont {Lee}\ and\ \citenamefont
  {Qi}(2014)}]{lee2014lattice}%
  \BibitemOpen
  \bibfield  {author} {\bibinfo {author} {\bibfnamefont {Ching~Hua}\
  \bibnamefont {Lee}}\ and\ \bibinfo {author} {\bibfnamefont {Xiao-Liang}\
  \bibnamefont {Qi}},\ }\bibfield  {title} {\enquote {\bibinfo {title} {Lattice
  construction of pseudopotential hamiltonians for fractional chern
  insulators},}\ }\href@noop {} {\bibfield  {journal} {\bibinfo  {journal}
  {Physical Review B}\ }\textbf {\bibinfo {volume} {90}},\ \bibinfo {pages}
  {085103} (\bibinfo {year} {2014})}\BibitemShut {NoStop}%
\bibitem [{\citenamefont {Lee}\ \emph {et~al.}(2017)\citenamefont {Lee},
  \citenamefont {Claassen},\ and\ \citenamefont {Thomale}}]{lee2017band}%
  \BibitemOpen
  \bibfield  {author} {\bibinfo {author} {\bibfnamefont {Ching~Hua}\
  \bibnamefont {Lee}}, \bibinfo {author} {\bibfnamefont {Martin}\ \bibnamefont
  {Claassen}}, \ and\ \bibinfo {author} {\bibfnamefont {Ronny}\ \bibnamefont
  {Thomale}},\ }\bibfield  {title} {\enquote {\bibinfo {title} {Band structure
  engineering of ideal fractional chern insulators},}\ }\href@noop {}
  {\bibfield  {journal} {\bibinfo  {journal} {Phys. Rev. B}\ }\textbf {\bibinfo
  {volume} {96}},\ \bibinfo {pages} {165150} (\bibinfo {year}
  {2017})}\BibitemShut {NoStop}%
\bibitem [{\citenamefont {Li}\ \emph {et~al.}(2020{\natexlab{b}})\citenamefont
  {Li}, \citenamefont {Tohyama}, \citenamefont {Iitaka}, \citenamefont {Su},\
  and\ \citenamefont {Zeng}}]{li2020detection}%
  \BibitemOpen
  \bibfield  {author} {\bibinfo {author} {\bibfnamefont {Zhi}\ \bibnamefont
  {Li}}, \bibinfo {author} {\bibfnamefont {Takami}\ \bibnamefont {Tohyama}},
  \bibinfo {author} {\bibfnamefont {Toshiaki}\ \bibnamefont {Iitaka}}, \bibinfo
  {author} {\bibfnamefont {Haibin}\ \bibnamefont {Su}}, \ and\ \bibinfo
  {author} {\bibfnamefont {Haibo}\ \bibnamefont {Zeng}},\ }\bibfield  {title}
  {\enquote {\bibinfo {title} {Detection of quantum geometric tensor by
  nonlinear optical response},}\ }\href@noop {} {\bibfield  {journal} {\bibinfo
   {journal} {arXiv preprint arXiv:2007.02481}\ } (\bibinfo {year}
  {2020}{\natexlab{b}})}\BibitemShut {NoStop}%
\bibitem [{\citenamefont {Zhang}\ \emph {et~al.}(2019)\citenamefont {Zhang},
  \citenamefont {Wang},\ and\ \citenamefont {Gong}}]{zhang2019quantum}%
  \BibitemOpen
  \bibfield  {author} {\bibinfo {author} {\bibfnamefont {Da-Jian}\ \bibnamefont
  {Zhang}}, \bibinfo {author} {\bibfnamefont {Qing-hai}\ \bibnamefont {Wang}},
  \ and\ \bibinfo {author} {\bibfnamefont {Jiangbin}\ \bibnamefont {Gong}},\
  }\bibfield  {title} {\enquote {\bibinfo {title} {Quantum geometric tensor in
  pt-symmetric quantum mechanics},}\ }\href@noop {} {\bibfield  {journal}
  {\bibinfo  {journal} {Physical Review A}\ }\textbf {\bibinfo {volume} {99}},\
  \bibinfo {pages} {042104} (\bibinfo {year} {2019})}\BibitemShut {NoStop}%
\bibitem [{\citenamefont {Lee}\ \emph {et~al.}(2019)\citenamefont {Lee},
  \citenamefont {Li},\ and\ \citenamefont {Gong}}]{lee2019hybrid}%
  \BibitemOpen
  \bibfield  {author} {\bibinfo {author} {\bibfnamefont {Ching~Hua}\
  \bibnamefont {Lee}}, \bibinfo {author} {\bibfnamefont {Linhu}\ \bibnamefont
  {Li}}, \ and\ \bibinfo {author} {\bibfnamefont {Jiangbin}\ \bibnamefont
  {Gong}},\ }\bibfield  {title} {\enquote {\bibinfo {title} {Hybrid
  higher-order skin-topological modes in nonreciprocal systems},}\ }\href@noop
  {} {\bibfield  {journal} {\bibinfo  {journal} {Physical review letters}\
  }\textbf {\bibinfo {volume} {123}},\ \bibinfo {pages} {016805} (\bibinfo
  {year} {2019})}\BibitemShut {NoStop}%
\bibitem [{\citenamefont {Mu}\ \emph {et~al.}(2019)\citenamefont {Mu},
  \citenamefont {Lee}, \citenamefont {Li},\ and\ \citenamefont
  {Gong}}]{mu2019emergent}%
  \BibitemOpen
  \bibfield  {author} {\bibinfo {author} {\bibfnamefont {Sen}\ \bibnamefont
  {Mu}}, \bibinfo {author} {\bibfnamefont {Ching~Hua}\ \bibnamefont {Lee}},
  \bibinfo {author} {\bibfnamefont {Linhu}\ \bibnamefont {Li}}, \ and\ \bibinfo
  {author} {\bibfnamefont {Jiangbin}\ \bibnamefont {Gong}},\ }\bibfield
  {title} {\enquote {\bibinfo {title} {Emergent fermi surface in a many-body
  non-hermitian fermionic chain},}\ }\href@noop {} {\bibfield  {journal}
  {\bibinfo  {journal} {arXiv preprint arXiv:1911.00023}\ } (\bibinfo {year}
  {2019})}\BibitemShut {NoStop}%
\bibitem [{\citenamefont {Lee}\ \emph {et~al.}(2020)\citenamefont {Lee},
  \citenamefont {Lee},\ and\ \citenamefont {Yang}}]{lee2019many}%
  \BibitemOpen
  \bibfield  {author} {\bibinfo {author} {\bibfnamefont {Eunwoo}\ \bibnamefont
  {Lee}}, \bibinfo {author} {\bibfnamefont {Hyunjik}\ \bibnamefont {Lee}}, \
  and\ \bibinfo {author} {\bibfnamefont {Bohm-Jung}\ \bibnamefont {Yang}},\
  }\bibfield  {title} {\enquote {\bibinfo {title} {Many-body approach to
  non-hermitian physics in fermionic systems},}\ }\href@noop {} {\bibfield
  {journal} {\bibinfo  {journal} {Physical Review B}\ }\textbf {\bibinfo
  {volume} {101}},\ \bibinfo {pages} {121109} (\bibinfo {year}
  {2020})}\BibitemShut {NoStop}%
\bibitem [{\citenamefont {Mikami}\ \emph {et~al.}(2016)\citenamefont {Mikami},
  \citenamefont {Kitamura}, \citenamefont {Yasuda}, \citenamefont {Tsuji},
  \citenamefont {Oka},\ and\ \citenamefont {Aoki}}]{mikami2016brillouin}%
  \BibitemOpen
  \bibfield  {author} {\bibinfo {author} {\bibfnamefont {Takahiro}\
  \bibnamefont {Mikami}}, \bibinfo {author} {\bibfnamefont {Sota}\ \bibnamefont
  {Kitamura}}, \bibinfo {author} {\bibfnamefont {Kenji}\ \bibnamefont
  {Yasuda}}, \bibinfo {author} {\bibfnamefont {Naoto}\ \bibnamefont {Tsuji}},
  \bibinfo {author} {\bibfnamefont {Takashi}\ \bibnamefont {Oka}}, \ and\
  \bibinfo {author} {\bibfnamefont {Hideo}\ \bibnamefont {Aoki}},\ }\bibfield
  {title} {\enquote {\bibinfo {title} {Brillouin-wigner theory for
  high-frequency expansion in periodically driven systems: Application to
  floquet topological insulators},}\ }\href@noop {} {\bibfield  {journal}
  {\bibinfo  {journal} {Physical Review B}\ }\textbf {\bibinfo {volume} {93}},\
  \bibinfo {pages} {144307} (\bibinfo {year} {2016})}\BibitemShut {NoStop}%
\bibitem [{\citenamefont {Zhou}\ and\ \citenamefont
  {Gong}(2018)}]{zhou2018recipe}%
  \BibitemOpen
  \bibfield  {author} {\bibinfo {author} {\bibfnamefont {Longwen}\ \bibnamefont
  {Zhou}}\ and\ \bibinfo {author} {\bibfnamefont {Jiangbin}\ \bibnamefont
  {Gong}},\ }\bibfield  {title} {\enquote {\bibinfo {title} {Recipe for
  creating an arbitrary number of floquet chiral edge states},}\ }\href@noop {}
  {\bibfield  {journal} {\bibinfo  {journal} {Physical Review B}\ }\textbf
  {\bibinfo {volume} {97}},\ \bibinfo {pages} {245430} (\bibinfo {year}
  {2018})}\BibitemShut {NoStop}%
\bibitem [{\citenamefont {Li}\ \emph {et~al.}(2018)\citenamefont {Li},
  \citenamefont {Lee},\ and\ \citenamefont {Gong}}]{li2018realistic}%
  \BibitemOpen
  \bibfield  {author} {\bibinfo {author} {\bibfnamefont {Linhu}\ \bibnamefont
  {Li}}, \bibinfo {author} {\bibfnamefont {Ching~Hua}\ \bibnamefont {Lee}}, \
  and\ \bibinfo {author} {\bibfnamefont {Jiangbin}\ \bibnamefont {Gong}},\
  }\bibfield  {title} {\enquote {\bibinfo {title} {Realistic floquet semimetal
  with exotic topological linkages between arbitrarily many nodal loops},}\
  }\href@noop {} {\bibfield  {journal} {\bibinfo  {journal} {Physical review
  letters}\ }\textbf {\bibinfo {volume} {121}},\ \bibinfo {pages} {036401}
  (\bibinfo {year} {2018})}\BibitemShut {NoStop}%
\bibitem [{\citenamefont {Lee}\ and\ \citenamefont
  {Longhi}(2020)}]{lee2020ultrafast}%
  \BibitemOpen
  \bibfield  {author} {\bibinfo {author} {\bibfnamefont {Ching~Hua}\
  \bibnamefont {Lee}}\ and\ \bibinfo {author} {\bibfnamefont {Stefano}\
  \bibnamefont {Longhi}},\ }\bibfield  {title} {\enquote {\bibinfo {title}
  {Ultrafast and anharmonic rabi oscillations between non-bloch-bands},}\
  }\href@noop {} {\bibfield  {journal} {\bibinfo  {journal} {arXiv preprint
  arXiv:2003.10763}\ } (\bibinfo {year} {2020})}\BibitemShut {NoStop}%
\bibitem [{\citenamefont {He}\ and\ \citenamefont
  {Vanderbilt}(2001)}]{he2001exponential}%
  \BibitemOpen
  \bibfield  {author} {\bibinfo {author} {\bibfnamefont {Lixin}\ \bibnamefont
  {He}}\ and\ \bibinfo {author} {\bibfnamefont {David}\ \bibnamefont
  {Vanderbilt}},\ }\bibfield  {title} {\enquote {\bibinfo {title} {Exponential
  decay properties of wannier functions and related quantities},}\ }\href@noop
  {} {\bibfield  {journal} {\bibinfo  {journal} {Phys. Rev. Lett.}\ }\textbf
  {\bibinfo {volume} {86}},\ \bibinfo {pages} {5341} (\bibinfo {year}
  {2001})}\BibitemShut {NoStop}%
\bibitem [{\citenamefont {Lee}\ and\ \citenamefont {Ye}(2015)}]{lee2015free}%
  \BibitemOpen
  \bibfield  {author} {\bibinfo {author} {\bibfnamefont {Ching~Hua}\
  \bibnamefont {Lee}}\ and\ \bibinfo {author} {\bibfnamefont {Peng}\
  \bibnamefont {Ye}},\ }\bibfield  {title} {\enquote {\bibinfo {title}
  {Free-fermion entanglement spectrum through wannier interpolation},}\
  }\href@noop {} {\bibfield  {journal} {\bibinfo  {journal} {Physical Review
  B}\ }\textbf {\bibinfo {volume} {91}},\ \bibinfo {pages} {085119} (\bibinfo
  {year} {2015})}\BibitemShut {NoStop}%
\end{thebibliography}%
\clearpage
\pagebreak

\onecolumngrid

\begin{center}
\textbf{\large Appendix for ``Unraveling non-Hermitian pumping: emergent spectral singularities and anomalous responses" }
\end{center}

\appendix
\setcounter{figure}{0}
\renewcommand{\thefigure}{A\arabic{figure}}


\section{Various OBC singularity types \GJ{vs} non-reciprocal length scales}
\label{sec:various}
Here, we supplement the main text discussions of the various OBC singularity classes with more details.

\subsection{Reciprocal ($t_n= t_{-n}$) case}
Consider a characteristic polynomial (eigenenergy equation) of the form
\begin{equation}
F(E)=\sum_n t_nz^n+t_{-n}z^{-n},
\label{poly}
\end{equation}
$z=e^{ik}$. It is instructive to first prove the absence of non-Hermitian pumping i.e. the skin effect when $t_n= t_{-n}$ for all $n$. Recall that non-Hermitian pumping occurs when $|z_1|=|z_2|\neq 1$, where $z_1,z_2$ are the pair of roots of Eq.~\ref{poly} closest to the unit circle. When $t_n= t_{-n}$ for all $n$, one can always turn Eq.~\ref{poly} into a simpler polynomial through the substitution
\begin{equation}
u=z+\frac1{z}.
\label{u}
\end{equation}
Assuming non-vanishing terms from only a single $n$, doing so gives $z_1,z_2=\left(-u\pm\sqrt{u^2-4}\right)/2$, which are of equal magnitude iff $u$ and $\sqrt{u^2-4}$ differ by a phase of $\pi/2$, i.e.
\begin{equation}
\sqrt{u^2-4}=i\,ru
\end{equation}
where $r$ is a real multiplier. This implies that $u=\frac{2}{\sqrt{1+r^2}}$ or $0<u<2$. Evidently then, $z+\frac1{z}=2\cos k =u$ will always have a real $k$ solution, thereby obviating any skin effect. In the case of multiple $n$, a similar analysis applies for the more complicated resultant polynomial in $u$.

\subsection{One length scale}

For completeness, we review and elaborate on the simplest case where non-Hermitian pumping affects only NN couplings. Due to the presence of only one length scale, non-Hermitian pumping can be completely ``gauged away'' with a $k$-independent $\kappa$.

With only one length scale, the RHS of Eq.~\ref{poly} contains only two dissimilar terms $t_\pm$, in addition to a constant term containing $t_0$. To be concrete, consider the non-reciprocal SSH model with $H_\text{SSH}(z)=(t_-+z)\sigma_++(t_++z^{-1})\sigma_-$ where $t_\pm=t\pm\gamma$. Eq.~\ref{poly} takes the form
\begin{equation}
E^2=t_+z+\frac{t_-}{z}+t_+t_-+1
\label{poly2}
\end{equation}
which can be expressed as
\begin{equation}
\frac{E^2}{\sqrt{t_+t_-}}-2\cosh\log\sqrt{t_+t_-}=u'=z'+\frac1{z'}
\end{equation}
where $z'=\sqrt{\frac{t_+}{t_-}}z$, i.e. $k'=k+i\log\sqrt{\frac{t_-}{t_+}}$. This is manifestly of the form Eq.~\ref{u} with $u'$ defined as $\frac{E^2}{\sqrt{t_+t_-}}-2\cosh\log\sqrt{t_+t_-}$, except that $k$ is now deformed into $k'$ by a constant imaginary displacement $i\log\sqrt{\frac{t_-}{t_+}}$.

Note that this result applies to any system obeying Eq.~\ref{poly2}, and not just the non-reciprocal SSH model. Physically, the complex deformation of $k$ (or rescaling of $z$) corresponds to a spatial exponential rescaling that counteracts the mode accumulation from the pumping. Evidently, it will no longer work when more than one non-reciprocal length scale is at play, as studied below.

\subsection{Two length scales}
A minimal characteristic polynomial with more than one length scale is given by
\begin{equation}
F(E)=z^2+\frac{b}{z}.
\label{poly3}
\end{equation}
Here $F(E)$ is a function of $E$ that absorbs the constant term, if any, and rescales the coefficient of $z^2$ to unity. GJ{Its} exact form is immaterial for the branching topology of the skin spectrum - it is the algebraic form of the polynomial in $z$ that matters.

\begin{figure}
 \centering
\subfloat[]{\includegraphics[width=.27\linewidth]{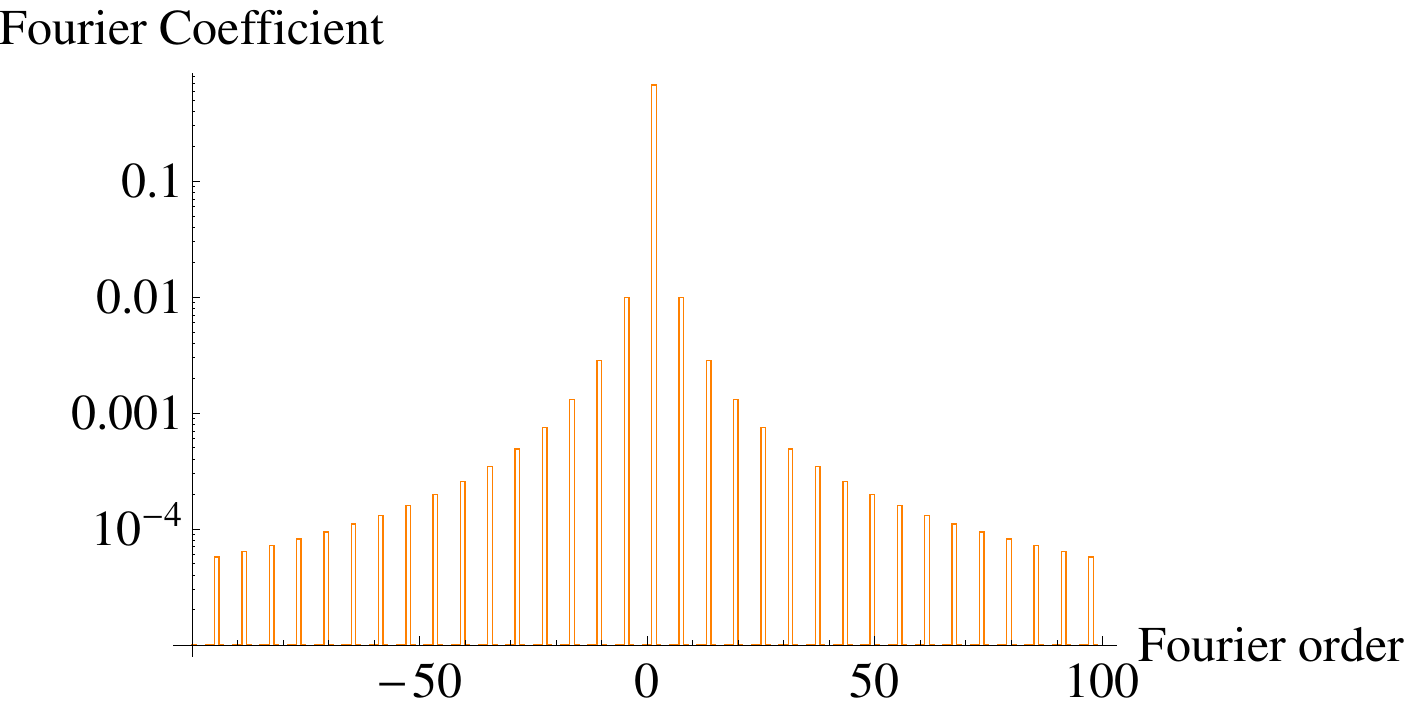}}
\subfloat[]{\includegraphics[width=.18\linewidth]{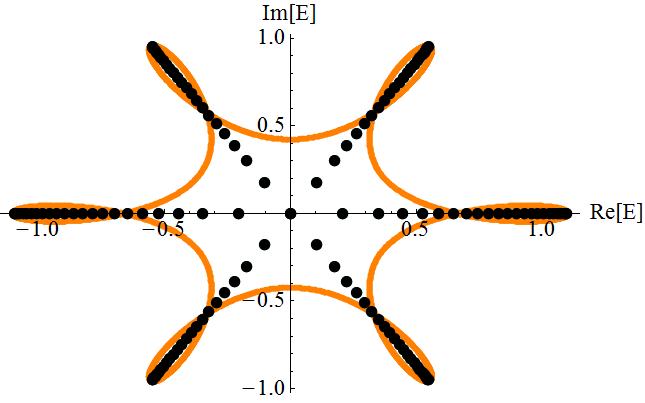}}
\subfloat[]{\includegraphics[width=.18\linewidth]{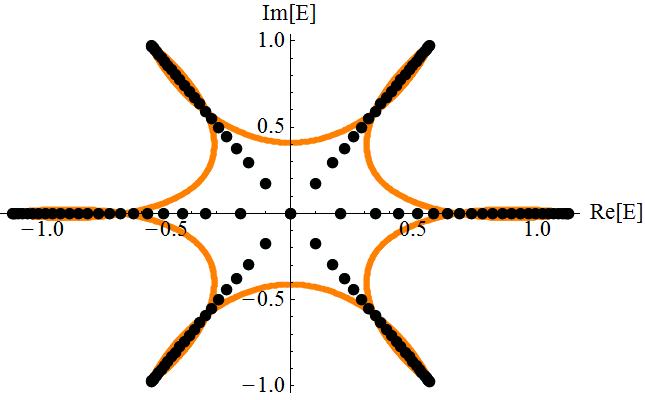}}
\subfloat[]{\includegraphics[width=.18\linewidth]{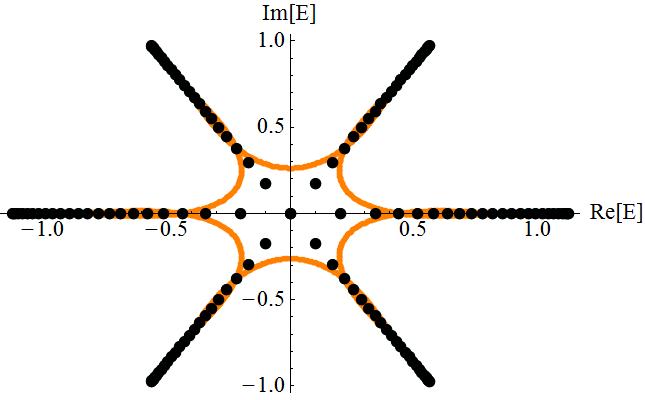}}
\subfloat[]{\includegraphics[width=.18\linewidth]{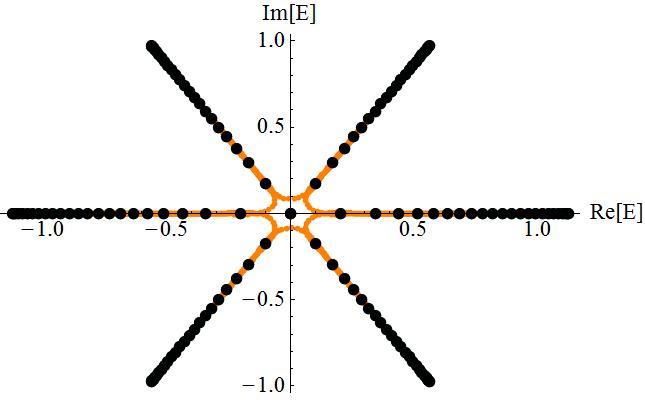}}
\caption{(a) Distribution of Fourier coefficients in the coupling approximation to the deformed (surrogate) OBC spectrum (Eqs.~\ref{newk} and \ref{fe}) with $b=0.5$, showcasing the real-space distribution of the non-local change of basis. (b-e) Corresponding approximations of the OBC spectrum  ith $3,5,10$ and $100$ harmonics respectively. }
 \label{cubic_approx}
\end{figure}

Using Cardano's formula, the three roots of Eq.~\ref{poly3} are
\begin{align}
z_{1,2}&=\frac{2\sqrt{3}F(E)+\sqrt[3]{\sqrt{12}D^2}\pm i\left(\sqrt[3]{18D^2}-6F(E)\right)}{-2\sqrt[3]{36\sqrt{3}D}},\notag\\
z_3&=\frac{\sqrt[3]{12}F(E)+\sqrt[3]{D^2}}{\sqrt[3]{18D}},
\end{align}
where $D=\sqrt{81b^2-12F(E)^3}-9b$. Again, to obtain the OBC modes $\bar \epsilon(k)$, we need $|z_1|=|z_2|$ (and permutations), which occur when
\begin{equation}
\sqrt[3]{18D^2}-6F(E)=r\left(2\sqrt{3}F(E)+\sqrt[3]{\sqrt{12}D^2}\right),
\end{equation}
with $r$ being a real multiplier. This can be inverted to yield
\begin{equation}
F(E)=\frac{r^2-3}{\sqrt[3]{(1+r^2)^2}}\left(\frac{b}{2}\right)^{2/3}\omega_j,
\end{equation}
where $\omega^3=1$. The three branches with $j=0,1,2$ are selected such that they are indeed the solution to Eq.~\ref{poly3} closest to the unit circle. Note that the precise functional dependence on $r$ is not important, $r$ being an auxiliary multiplier. Rather, what is important is the maximal and minimal range of $F(E)$; in this case, the main observation is that $F(E)$ fans out as three straight lines from the origin, as illustrated in Fig.~\ref{cubic_approx} for $F(E)=E^2$ (the power of $2$ gives $2\times 3=6$ straight skin mode branches.) Given the genericity of Eq.~\ref{poly3}, we have established that the cubic (Y-shaped) junction of OBC skin modes as a hallmark of non-Hermitian pumping with 2 length scales.

\subsubsection{Deformation of the momentum}
Now that we have established the skin spectrum which describes a bona-fide quasi-reciprocal lattice by virtue of its imperviously to non-Hermitian pumping, we shall obtain the complex momentum deformation that transforms the PBC spectrum in the the OBC skin spectrum. In general, this can be done by substituting $z\rightarrow |z|e^{i\theta}$ for fixed real momentum $\theta\in \mathcal{R}$ in the characteristic polynomial. This will yield $\kappa =-\log|z|$. From Eq.~\ref{poly3}, we solve for $\text{Im}\left[\left(|z|^2e^{2i\theta}+\frac{b}{|z|}e^{-i\theta}\right)\omega^{-j}\right]=0$ to obtain
\begin{equation}
z\rightarrow z\sqrt{\frac{b}{2}\sec\left(k-2\pi j /3\right)}
\end{equation}
or
\begin{equation}
k\rightarrow k-\frac{i}{3}\log\left|{\frac{b}{2\cos\left(k-2\pi j /3\right)}}\right|.
\label{newk}
\end{equation}
This is illustrated in Fig.~1e of the main text. 
Note that the form of $F(E)$ does not have to explicitly appear in this deformation.

\subsubsection{Coupling approximation to deformed basis}
To relate the above $k$-dependent deformation to an effective physical coupling lattice, one can perform a Fourier decomposition of the characteristic polynomial in terms of of the deformed $z\rightarrow ze^{-\kappa(k)}=ze^{-\kappa(-i\log z)}$. For instance, with $b=0.5$, an approximation of Eq.~\ref{poly3} up to 5 Fourier harmonics via Eq.~\ref{newk} gives
\begin{equation}
F(E)=-0.051z^5+0.492z^2+\frac{0.700}{z}-\frac{0.047}{z^4}.
\label{fe}
\end{equation}
Further physical understanding can be obtained by studying the modification to a single NN coupling $z$: In the above example,
\begin{equation}
z\rightarrow z\left[0.677 -0.030\left(z^3+\frac1{z^3}\right)+0.010\left(z^6+\frac1{z^6}\right)\right]
\end{equation}
up to six additional harmonics. This is a real-space rescaling consisting of multiple scales: for instance, a coupling over $X$ sites is not just suppressed by a factor of $0.677^X$, but is also approximately equivalent to the superposition of many other terms with different rates of exponential suppression wit $X$. Shown in Fig.~\ref{cubic_approx} is an illustration of the Fourier approximation, where one sees that the spectrum converges to the quasi-reciprocal OBC spectrum as more harmonics are added. The convergence is power-law due to the non-analyticity of $\kappa(k)$.

\subsection{Generic cases}
\begin{figure}
 \centering
\includegraphics[width=.4\linewidth]{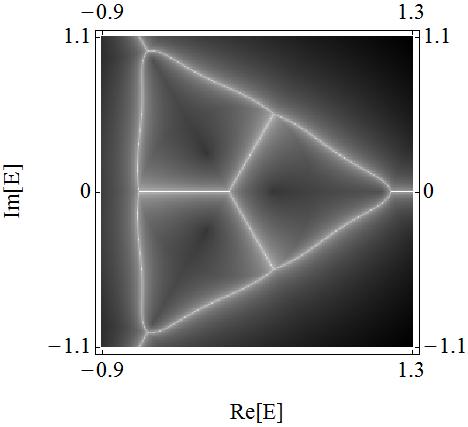}
\includegraphics[width=.59\linewidth]{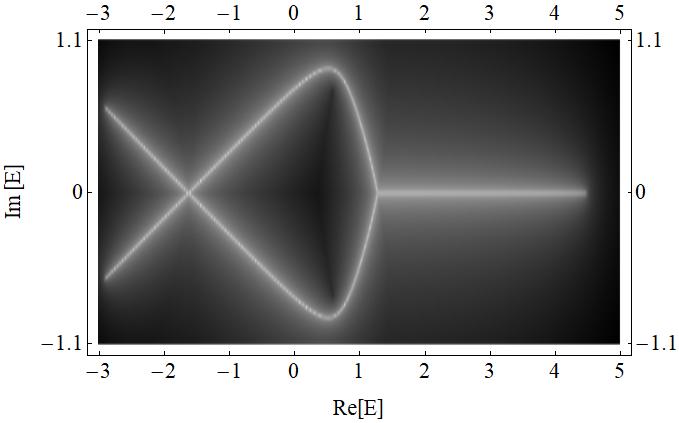}
\caption{Density plots of the gap between the 2nd and 3rd smallest $\kappa$ solutions as a function of $E$, for (a) $E^2-\frac{0.7E}{z^2}-z^2-\frac1{2z}=0$ (case (f) in Fig.~2 of the main text), and (b) $E=(z^3+2z^2+z+z^{-1}+4z^{-2})/2$ as in Ref.~\cite{kai2019gbz}. Light/dark regions denote small/large gaps. The OBC spectra coalesce along the lines of vanishing $\kappa$ (imaginary) gap.
}
 \label{gen6gap}
\end{figure}

The precise prediction of their graph structure of the OBC spectrum from $P(E,z)$ is an open problem. Generically, the OBC spectrum lie along intersections of the various solutions of $\kappa(k)$ in energy space, as shown in Fig.~\ref{gen6gap} for two illustrative cases mentioned in the maub text. 
As we can see, the OBC spectrum accumulate along branch-like trenches, but their global topology depends also on the PBC spectral loops. The reader is encouraged to refer to Refs.~\cite{lee2019anatomy,lee2018tidal,kai2019gbz} for further discussions. 

\section{Analytic treatment of the extended non-Hermitian Kitaev model}
\label{sec:kitaev}

The non-Hermitian extension of the extended Kitaev model~\cite{kitaev,li2016Z2} realizes a prime example of a non-Hermitian topological phase (D-class) that minimally require NN and NNN couplings and hence more than one reciprocal length scales.  The model Hamiltonian is given by $H=\sigma\cdot \bold h$, where
\begin{eqnarray}
h_x&=&d_x+ig_x\notag\\
h_y&=&d_y+ig_y\notag\\
h_z&=&d_z,
\end{eqnarray}
with $d_x=\Delta_2\sin\phi\sin 2k$, $d_y=\Delta_2\cos\phi\sin 2k+\Delta_1 \sin k$, $d_z=m-t_1\cos k -t_2 \cos 2k$.
The two terms $ig_x\sigma_x$ and $ig_y\sigma_y$ are the only constant non-Hermitian terms that do not violate particle-hole symmetry (PHS)
\begin{eqnarray}
\sigma_xH^*(k)\sigma_x=-H(-k),
\end{eqnarray}
which protects the $Z_2$ topology inherited from the Hermitian version of this model. To ensure no residual $Z$-topology~\cite{li2016Z2}, we also break chiral symmetry via the parameter $\phi$, which represents a phase difference between the NN and NNN parings $\Delta_1$ and $\Delta_2$.

To analytically solve for its OBC skin modes, one considers the complex continuation of $z=e^{ik}$, and finds values of $E$ for which the two roots $z_\mu,z_\nu$ of the eigenenergy equation $E^2=h_x^2+h_y^2+h_z^2$ nearest to the real circle satisfies $|z_\mu|=|z_\nu|$. The spectrum will converge to these energy loci during the PBC-OBC interpolation. Here, the eigenenergy equation explicitly takes the following form
\begin{eqnarray}
E&=&\frac{\Delta_1^2+\Delta_2^2+t_1^2+t_2^2}{2}- g_x^2- g_y^2+ m^2 \notag\\
&&\ +\frac{t_2^2-\Delta_2^2}{4}\left(z^4+\frac1{z^4}\right)+\frac{t_1 t_2-\Delta_1 \Delta_2 \cos \phi }{2}\left(z^3+\frac1{z^3}\right)\notag\\
&&\ +\frac{t_1 t_2+\Delta_1 \Delta_2 \cos \phi-2 m t_1 }{2}\left(z+\frac1{z}\right)\notag\\
&&\ +\Delta_1g_y\left(z-\frac1{z}\right)+\frac{t_1^2-\Delta_1^2-4 t_2m }{4}\left(z^2+\frac1{z^2}\right)\notag\\
&&\ +\Delta_2(g_y \cos \phi +g_x \sin\phi )\left(z^2-\frac1{z^2}\right).
\label{EE}
\end{eqnarray}
\subsection{Reduction to analytically tractable form for special cases}
In order to explore analytic solutions, we shall consider special cases in which the cubic and quartic terms vanish. That occurs when the parameters satisfy the constraints $t_2=\Delta_2$ and $t_1=\Delta_1\cos\phi$. Also, we shall set $m=1$ without loss of generality, which can always be satisfied via a global rescaling. Doing so, the above expression simplifies to
\begin{eqnarray}
E^2&=&\frac{\Delta_1^2(1+ \cos \phi ^2)}{2}- \left(g_x^2+g_y^2-\Delta_2^2-1\right)\notag\\
&&+\Delta_1 (g_y- \cos \phi+\Delta_2\cos\phi )z\notag\\
&&-\frac{\Delta_1 (g_y+\cos \phi -\Delta_2 \cos \phi )}{z} \notag\\
&&+z^2 \left(-\frac{\Delta_1^2}{4}\sin^2\phi-\Delta_2 +\Delta_2(g_x  \sin\phi+g_y\cos\phi) \right)\notag \\
&&+\frac1{z^2}\left(-\frac1{4}\Delta_1^2\sin^2\phi- \Delta_2- \Delta_2 (g_x \sin\phi+ g_y \cos \phi)\right).
\end{eqnarray}
Although we have reduced the above from an 8-th order to a 4-th polynomial in $z$, an additional constraint is still needed for a simple analytic solution. We will like to substitute the terms linear in $z$ and $z^{-1}$ by a single variable $u$, i.e.
\begin{equation}
u=u_+z+\frac{u_-}{z},\qquad u_\pm=\Delta_1[(\Delta_2-1)\cos\phi\pm g_y]
\label{uz}
\end{equation}
such that $u^2=u_+^2z^2+\frac{u_-^2}{z^2}+2u_+u_-$ reproduces the rest of $E^2$ up to a linear transformation i.e.
\begin{equation}
E^2=\Sigma u^2+u+u_0,
\label{Eu}
\end{equation}
where $\Sigma$ and $u_0$ will be computed shortly. This is only possible if $\Delta_1$ is chosen to satisfy
\begin{equation}
\left(\frac{u_-}{u_+}\right)^2=\frac{-\frac1{4}\Delta_1^2\sin^2\phi- \Delta_2- \Delta_2 (g_x \sin\phi+ g_y \cos \phi)}{-\frac{1}{4}\Delta_1^2\sin^2\phi-\Delta_2 +\Delta_2(g_x  \sin\phi+g_y\cos\phi)}
\end{equation}
or, more explicitly, (defining $A=(\Delta_2-1)\cos\phi$)
\begin{equation}
\Delta_1^2=-\frac{2\Delta_2\left[2Ag_y+(g_y^2+A^2)(g_y\cos \phi+g_x\sin\phi)\right]}{g_yA\sin^2\phi}
\label{D12}
\end{equation}
In order to avoid complications, we require that $\Delta_1,\Delta_2$ are real. This possible along some interval within $0<\Delta_2<1$ giving a positive value to the RHS of Eq.~\ref{D12}. With some labor, we can also show that
\begin{equation}
\Sigma=\frac{\sin^2 \phi (g_y \cos\phi+g_x \sin \phi)}{4 (\Delta_2-1)^2 \cos^2\phi (g_x+g_y \cos\phi)+4 g_y \left(g_x g_y+\left(g_y^2+2( \Delta_2-1)\right) \cos\phi\right)},
\label{Sig}
\end{equation}
\begin{equation}
u_0=\frac{-2 g_x (g_y^2 + ( \Delta_2-1)^2) \Delta_2 \cos\phi\sin\phi +
 g_y\sin^2\phi (-1 + g_y^2 +g_x^2  +  \Delta_2 (1+g_y^2-g_x^2 + 3 (-1 + \Delta_2) \Delta_2)) -
    2 \Delta_2 (-1 + g_y^2 + \Delta_2^2) }{g_y (\Delta_2-1)}.
\label{u0}
\end{equation}
To summarize, if we enforce Eq.~\ref{D12} as well as $t_2=\Delta_2$ and $t_1=\Delta_1\cos\phi$, the eigenvalue equation Eq.~\ref{EE} will reduce to Eq.~\ref{Eu} with $u$ given in terms of $z$ via Eq.~\ref{uz}, and $\Sigma$ and $u_0$ given by Eqs.~\ref{Sig} and \ref{u0} respectively. After adhering to these constraints we still have 4 free parameters left: $\Delta_2,\phi,g_x,g_y$.

\subsection{OBC skin mode solutions of surrogate system}
\begin{figure}
 \centering
\subfloat[]{\includegraphics[width=.27\linewidth]{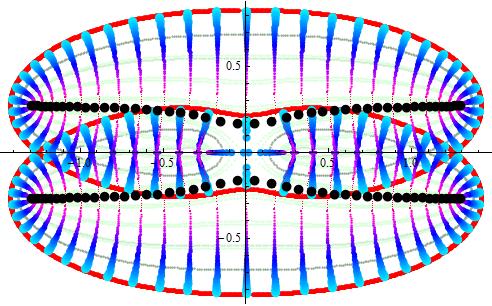}}
\subfloat[]{\includegraphics[width=.27\linewidth]{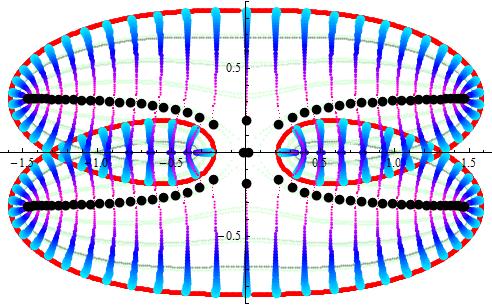}}
\subfloat[]{\includegraphics[width=.32\linewidth]{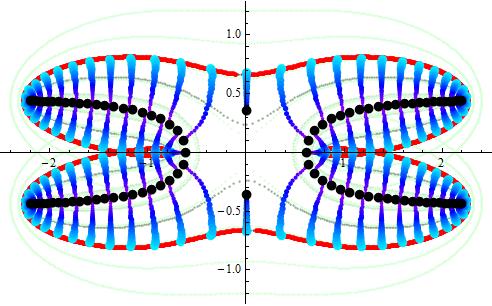}}
 \caption{OBC skin mode eigenenergies (black) enclosed by PBC eigenenergies (red) for three illustrative cases: (a) imaginary line gap without topological modes, (c) real line gap with isolated in-gap topological modes and (b) their intermediate case extremely close to the phase transition, where the OBC eigenenergies almost touch the $E=0$ origin. Parameters used are: (a) $\Delta_2=t_2=0.5,\phi=\pi/3,\Delta_1=0.64144,t_1=0.3207,g_x=g_y=0.6$, (b) $\Delta_2=t_2=0.55\approx 0.548,\phi=\pi/3,\Delta_1=0.8503,t_1=0.4252,g_x=g_y=0.6$, (c) $\Delta_2=t_2=0.7,\phi=\pi/3,\Delta_1=1.664,t_1=0.832,g_x=g_y=0.6$. Faint light green curves represent contours of constant $\kappa(k)$.}
 \label{spec}
  \end{figure}

The roots of Eq.~\ref{EE} are given via Eq.~\ref{uz} as
\begin{equation}
z_\pm=\frac{u\pm\sqrt{u^2-4u_+u_-}}{2u_+}
\end{equation}
for each $u$ satisfying Eq.~\ref{Eu}. To find the skin mode loci satisfying $|z_+|=|z_-|$, the key observation is that $u$ has to be purely imaginary if $u_+,u_-$ are real. This is very similar to the case of Eq.~\ref{poly2} considered in the previous section. More concretely, note that, in the large regime where $u_\pm$ are real and of opposite signs (indication of large non-reciprocity), $|z_+|=|z_-|$ requires $0>u^2\geq 4u_+u_-$, which is in the regime of imaginary $u$. Then, enforcing $\text{Re}\,u=\left(u_+|z|+\frac{u_-}{|z|}\right)\cos(\text{Re}\,k)=0$ yields the condition
\begin{equation}
k\rightarrow k-i\log|z|=k-\frac{i}{2}\log\left|-\frac{u_-}{u_+}\right|\label{kappa_SM}
\end{equation}
where the PBC modes shall be continued into skin modes. After solving for $u$, $E$ can be obtained via the additional step Eq.~\ref{Eu} ($E=\sqrt{u_0+iv-\Sigma v^2}$ where $v\in \mathcal{R}$), as computed in Fig.~\ref{spec}.

\subsection{Topological phase transition}
Topological phase transitions occur when the OBC skin modes intersect. Setting $E=0$ in Eq.~\ref{Eu} and noting that $u$ has to be purely imaginary, it is not hard to see that $u_0=0$ is the condition for skin gap closure i.e.
\begin{equation}
-2 g_x (g_y^2 + ( \Delta_2-1)^2) \Delta_2 \cos\phi\sin\phi +
 g_y\sin^2\phi (-1 + g_y^2 +g_x^2  +  \Delta_2 (1+g_y^2-g_x^2 + 3 (-1 + \Delta_2) \Delta_2)) -
    2 \Delta_2 (-1 + g_y^2 + \Delta_2^2) =0.
\label{u00}
\end{equation}
For the parameters of Fig.~\ref{spec}, it occurs around $\Delta_2=t_2\approx 0.548$. For generic parameter cases that are not analytical tractable, we will need to numerically solve for the skin modes to find when they intersect.

\subsection{Topological properties from pseudospin vectors}

We next characterize the above-mentioned topological transition in terms of the topological properties of the pseudospin expectation vector. In a non-Hermitian system, one can construct four possible pseudospin expectations out of the left (L) and right (R) eigenvectors
\begin{eqnarray}
\bar{S}_{i,\pm}^{\alpha\beta}=\langle\bar{\psi}^\alpha_\pm|\sigma_i|\bar{\psi}^\beta_\pm\rangle,
\end{eqnarray}
where $i=x,y,z$, $\pm$ label the two eigenvectors and $\alpha,\beta$ indicate whether the $L$ or $R$ eigenvector was used. Of the four expectation vectors, only two types are qualitatively distinct: the expectation $\bar{\bold S}^{RR}_\pm$ or $\bar{\bold S}^{LL}_\pm$ or a single R or L eigenvector, and the biorthogonal expectation $\bar{\bold S}^{LR}_\pm$ (or its conjugate). Since we are interested in the OBC topological boundary modes, we shall only work within the eigenspace of the surrogate Hamiltonian $\bar{H}(k)=H(k+i\kappa)$ obtained via complex momentum deformation.

For a geometric interpretation of the topology~\cite{Jiang2018nochiral}, we shall first consider the pseudospin expectation of the left or right eigenvectors, which IS always real. Writing $\bar{E}=\sqrt{\bar{h}_x^2+\bar{h}_y^2+\bar{h}_z^2}$, the pseudospin components are given by
\begin{eqnarray}
&&\bar{S}_{x,\pm}^{RR}=\langle\bar{\psi}^R_\pm|\sigma_x|\bar{\psi}^R_\pm\rangle,\nonumber\\
&&\propto\pm\left(\frac{\bar{h}^*_x+i\bar{h}^*_y}{\bar{E}^*}+\frac{\bar{h}_x-i\bar{h}_y}{\bar{E}}\right)-\left(\frac{\bar{h}^*_x+i\bar{h}^*_y}{\bar{E}^*}\frac{\bar{h}_z}{\bar{E}}+\frac{\bar{h}_x-i\bar{h}_y}{\bar{E}}\frac{\bar{h}^*_z}{\bar{E}^*}\right),\nonumber\\
&&\bar{S}_{y,\pm}^{RR}=\langle\bar{\psi}^R_\pm|\sigma_y|\bar{\psi}^R_\pm\rangle\nonumber\\
&&\propto\pm\left(\frac{-i\bar{h}^*_x+\bar{h}^*_y}{\bar{E}^*}+\frac{i\bar{h}_x+\bar{h}_y}{\bar{E}}\right)-\left(\frac{-i\bar{h}^*_x+\bar{h}^*_y}{\bar{E}^*}\frac{\bar{h}_z}{\bar{E}}+\frac{i\bar{h}_x+\bar{h}_y}{\bar{E}}\frac{\bar{h}^*_z}{\bar{E}^*}\right),\nonumber\\
&&\bar{S}_{z,\pm}^{RR}=\langle\bar{\psi}^R_\pm|\sigma_z|\bar{\psi}^R_\pm\rangle\nonumber\\
&&\propto\frac{\bar{h}_x\bar{h}^*_x+\bar{h}_y\bar{h}^*_y-\bar{h}_z\bar{h}^*_z}{\bar{E}\bar{E}^*}-1+i\frac{\bar{h}_y^*\bar{h}_x-\bar{h}^*_x\bar{h}_y}{\bar{E}\bar{E}^*}\pm\left(\frac{\bar{h}_z}{\bar{E}}+\frac{\bar{h}^*_z}{\bar{E}^*}\right),
\label{SRR}
\end{eqnarray}
and
\begin{eqnarray}
&&\bar{S}_{x,\pm}^{LL}=\langle\bar{\psi}^L_\pm|\sigma_x|\bar{\psi}^L_\pm\rangle,\nonumber\\
&&\propto\pm\left(\frac{\bar{h}_x+i\bar{h}_y}{\bar{E}}+\frac{\bar{h}^*_x-i\bar{h}^*_y}{\bar{E}^*}\right)-\left(\frac{\bar{h}_x+i\bar{h}_y}{\bar{E}}\frac{\bar{h}^*_z}{\bar{E}^*}+\frac{\bar{h}^*_x-i\bar{h}^*_y}{\bar{E}^*}\frac{\bar{h}_z}{\bar{E}}\right),\nonumber\\
&&\bar{S}_{y,\pm}^{LL}=\langle\bar{\psi}^L_\pm|\sigma_y|\bar{\psi}^L_\pm\rangle\nonumber\\
&&\propto\pm\left(\frac{-i\bar{h}_x+\bar{h}_y}{\bar{E}}+\frac{i\bar{h}^*_x+\bar{h}^*_y}{\bar{E}^*}\right)-\left(\frac{-i\bar{h}_x+\bar{h}_y}{\bar{E}}\frac{\bar{h}^*_z}{\bar{E}^*}+\frac{i\bar{h}^*_x+\bar{h}^*_y}{\bar{E}^*}\frac{\bar{h}_z}{\bar{E}}\right),\nonumber\\
&&\bar{S}_{z,\pm}^{LL}=\langle\bar{\psi}^R_\pm|\sigma_z|\bar{\psi}^R_\pm\rangle\nonumber\\
&&\propto\frac{\bar{h}_x\bar{h}^*_x+\bar{h}_y\bar{h}^*_y-\bar{h}_z\bar{h}^*_z}{\bar{E}\bar{E}^*}-1+i\frac{\bar{h}_y\bar{h}^*_x-\bar{h}_x\bar{h}^*_y}{\bar{E}\bar{E}^*}\pm\left(\frac{\bar{h}_z}{\bar{E}}+\frac{\bar{h}^*_z}{\bar{E}^*}\right),
\end{eqnarray}
which corresponds to $\bar{\bold S}^{RR}_\pm$ for $H^\dagger$, which exhibits the same topology. 

We next specialize to $\bar{\bold S}^{RR}_\pm$ without loss of generality, and consider its behavior at high symmetric points. Since PHS also holds for the surrogate Hamiltonian, $\sigma_x \bar{H}_D(k)\sigma_x=-\bar{H}^*_D(-k)$,
the right eigenvectors satisfy 
\begin{eqnarray}
|\bar\psi^R_{\alpha}(k)\rangle=\sigma_x|\bar\psi^R_{\alpha'}(-k)\rangle^*,\label{eq_D_Rvector}
\end{eqnarray}
with eigenenergies 
\begin{eqnarray}
\bar{E}_\alpha(k)=-\bar{E}_{\alpha'}^*(-k),\label{eq_D_energy}
\end{eqnarray}
and $\alpha,\alpha'\in\{+,-\}$ labelling possibly different band indices. In terms of the pseudospin vector components,
\begin{eqnarray}
\bar{S}_{x,\alpha}^{RR}(k)=\bar{S}_{x,\alpha'}^{RR}(-k),~\bar{S}_{y,\alpha}^{RR}(k)=\bar{S}_{y,\alpha'}^{RR}(-k),~\bar{S}_{z,\alpha}^{RR}(k)=-\bar{S}_{z,\alpha'}^{RR}(-k),
\end{eqnarray}
i.e. the $\alpha$-band at $k$ is symmetric to the $\alpha'$-band at $-k$ about the equator of their Bloch sphere. Qualitatively different possibilities arise depending on whether $\alpha$ is the same as $\alpha'$ at the high symmetric points  $k=0$ and $\pi$. 
At these two points $k=k_0$, PHS allows only purely imaginary $\bar{h}_{x,y}(k_0)$ and real $\bar{h}_z(k_0)$, such that the eigenenergy $\bar{E}(k_0)=\sqrt{\bar{h}_x^2+\bar{h}_y^2+\bar{h}_z^2}$ is also either real or purely imaginary. As PHS always ensures that $\bar{E}_+(k)=-\bar{E}_-(k)$, Eq. \ref{eq_D_energy} yield two distinct scenarios:
\begin{itemize}
\item $\alpha=\alpha'$ when $\bar{E}_\alpha(k_0)$ is imaginary, suggesting that each $\bar{\bm S}_{\pm}^{RR}$ trajectory is symmetric to itself. In particular, from Eq.~\ref{SRR}, we have
\begin{eqnarray}
\bar{S}_{x,\pm}^{RR}(k_0)&\propto&\pm\frac{2\bar{h}_x}{\bar{E(k_0)}}-\frac{2i\bar{h}_y\bar{h}_z}{\bar{E}(k_0)^2},\nonumber\\
\bar{S}_{y,\pm}^{RR}(k_0)&\propto&\pm\frac{2\bar{h}_y}{\bar{E(k_0)}}+\frac{2i\bar{h}_x\bar{h}_z}{\bar{E}(k_0)^2},\nonumber\\
\bar{S}_{z,\pm}^{RR}(k_0)&=&0,
\end{eqnarray}
indicating that at the high symmetric points $k_0=0$ or $\pi$, the pseudospin vectors for both bands lie on the equator, but are not symmetric to each other.
\item$\alpha\neq \alpha'$ when $\bar{E}_\alpha(k_0)$ is real, suggesting that the two trajectories of $\bar{\bm S}_{\pm}^{RR}$ are symmetric to each other. Eq.~\ref{SRR} yields normalized pseudospin vectors
\begin{eqnarray}
\bar{\bm S}_{\pm}^{RR}(k_0)=\frac{1}{\sqrt{1-4\bar{h}_x^2-4\bar{h}_y^2}}(\frac{-2i\bar{h}_y}{\bar{E}(k_0)},\frac{2i\bar{h}_x}{\bar{E}(k_0)},\mp1),
\end{eqnarray}
with the two bands corresponding to two pseudospin vectors symmetric about the equator.
\end{itemize}

These two types of behaviors, distinguished simply by ${\rm Sign}[\bar{E}^2(k_0)]$ at $k_0=0$ and $\pi$, are respectively illustrated in Fig. 4b and 4a,c of the main text. 
These two situations are also characterized by imaginary and real line gaps respectively. Note that $\bar{E}(0)$ and $\bar{E}(\pi)$ must simultaneously be both imaginary or real, since otherwise the four eigenenergies at these two points shall lie on each of the positive and negative branches of the real and imaginary axis respectively, giving rise to a full OBC spectrum assuming a loop enclosing the origin, in contradiction to the quasi-reciprocity of the surrogate Hamiltonian.

To further determine the existence of topological boundary modes,
we need to turn to the biorthogonal pseudospin expectation vector $\bar{\bm S}_{\pm}^{LR}$ defined by
\begin{eqnarray}
\bar{S}_{x,\pm}^{LR}=\langle\bar{\psi}^L_\pm|\sigma_x|\bar{\psi}^R_\pm\rangle=\pm \bar{h}_x/\bar{E},\nonumber\\
\bar{S}_{y,\pm}^{LR}=\langle\bar{\psi}^L_\pm|\sigma_y\bar{\psi}^R_\pm\rangle=\pm \bar{h}_y/\bar{E},\nonumber\\
\bar{S}_{z,\pm}^{LR}=\langle\bar{\psi}^L_\pm|\sigma_z|\bar{\psi}^R_\pm\rangle=\pm \bar{h}_z/\bar{E},\nonumber\\
\end{eqnarray}
which is however generally complex for non-Hermitian systems and not \GJ{even} directly visualizable on a Bloch sphere. As discussed in the main text and Ref.~\cite{Li2019geometric}, it \GJ{nevertheless} allows the computation of the topological invariant $\nu=\text{Sign}\{[\text{Re}[\bar S_z^{LR}(0)]\text{Re}[\bar S_z^{LR}(\pi)]\}$, which takes values of $0,-1$ and $1$. As evident from the above argument, the $\nu=0$ case is the scenario with imaginary line gap corresponding to Fig.~\ref{spec}(a) of the main text. The $\nu=\mp 1$ cases are scenarios with real line gaps with/without topological boundary modes \GJ{(see Fig.~\ref{spec}c) as an example).}  


\section{Further details of the extended non-Hermitian Chern model}
\label{sec:chern}
The Chern model considered in our work is
\begin{equation}
H_\text{Ch}(\bold k)=(v+z^{-1})\sigma_+(u+z-vz^2)\sigma_- + \sin k_y \sigma_z,
\end{equation}
where $u=M+\cos k_y -\mu$ and $v=v_0(M+\cos k_y+\mu)$. Containing both linear and quadratic terms in $z,z^{-1}$, it has both NN and NNN couplings. It is specially designed such that $E^2=(v+z^{-1})(u+z-vz^2)+\sin^2k_y$ only has $z^2$, $z^{-1}$ and constant terms, thereby reducing to Eq.~\ref{poly3} with $k_y$-dependent coefficients and amenable to analytic treatment. Being fundamentally unlike the NN SSH model (in fact $u\sigma_++v\sigma_-$ is equivalent to the non-Hermitian SSH model in $k_y$), its phase diagram has never been studied. In real space, its various hihgly non-reciprocal couplings across $-2$ sites to $2$ sites in both $x$ and $y$ directions are
\begin{equation}
\left(
\begin{array}{ccccc}
 0 & 0 & 0 & 0 & 0 \\
 0 & 0 & v_0/2 & 0 & 0 \\
 0 & 1 & M v_0+\mu  v_0 & 0 & 0 \\
 0 & 0 & v_0/2 & 0 & 0 \\
 0 & 0 & 0 & 0 & 0 \\
\end{array}
\right)
\end{equation}
from pseudospin A to B and
\begin{equation}
\left(
\begin{array}{ccccc}
 0 & 0 & 0 & 0 & 0 \\
 0 & 0 & 1/2 & 0 & -v_0/2 \\
 0 & 0 & M-\mu  & 1 & -M v_0-\mu  v_0 \\
 0 & 0 & 1/2 & 0 & -v_0/2 \\
 0 & 0 & 0 & 0 & 0 \\
\end{array}
\right)
\end{equation}
from pseudospin B to A. Between the same pseudospin sublattice, we simply have Hermitian NN $\pm i/2$ couplings from $\sin k_y \sigma_z$. 

\begin{figure}
 \centering
\subfloat[]{\includegraphics[width=.32\linewidth]{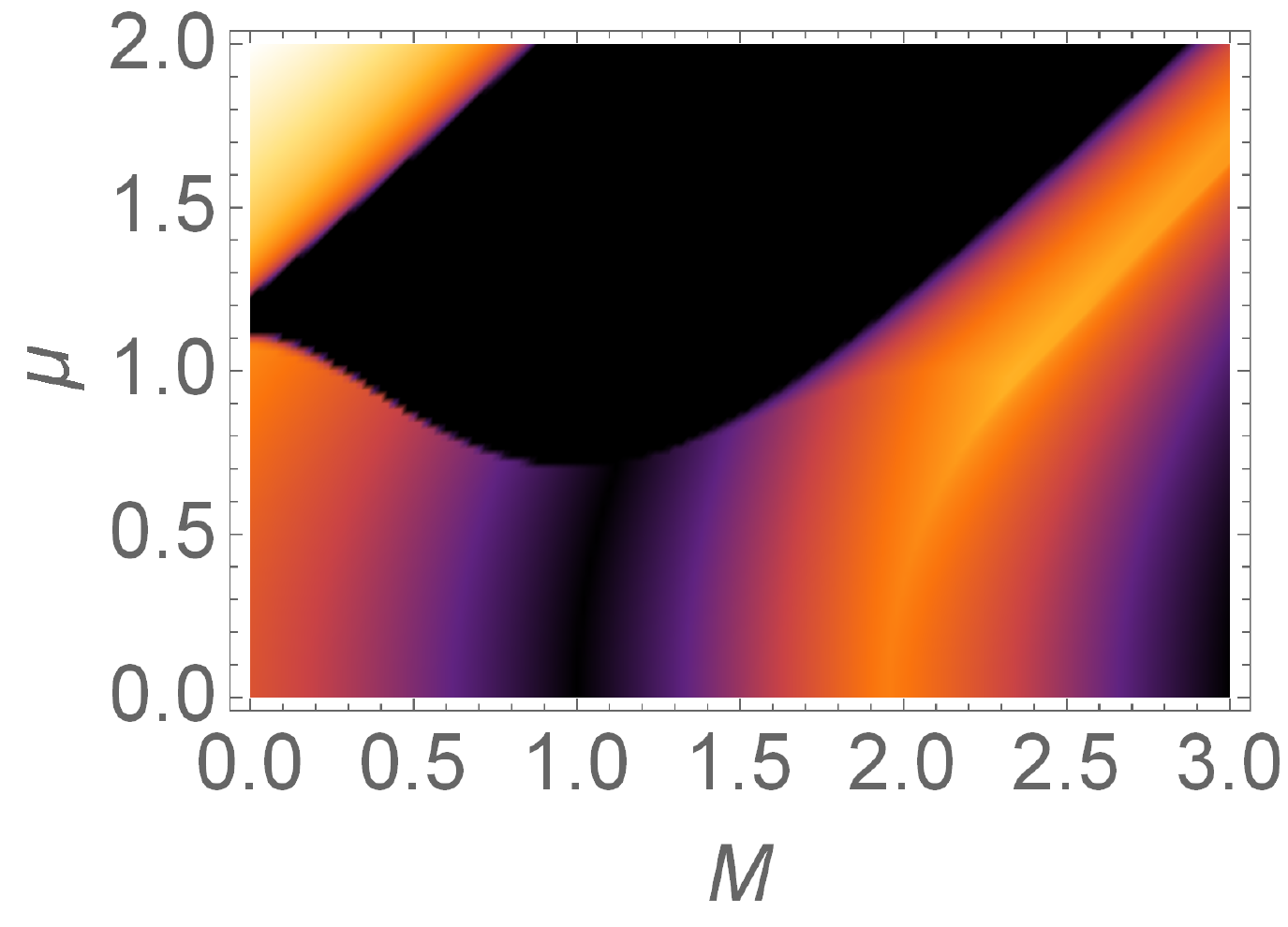}}
\subfloat[]{\includegraphics[width=.32\linewidth]{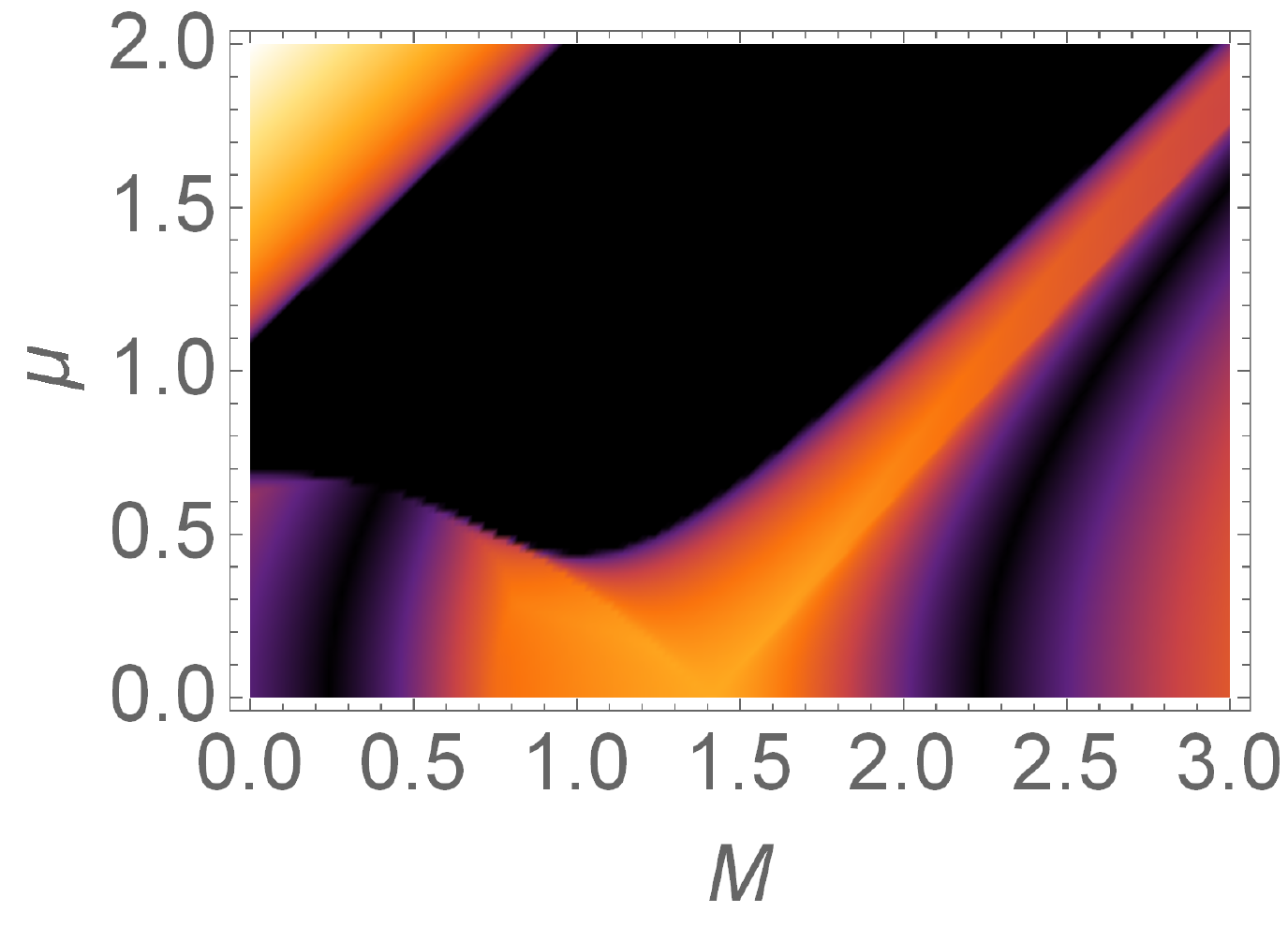}}
\subfloat[]{\includegraphics[width=.32\linewidth]{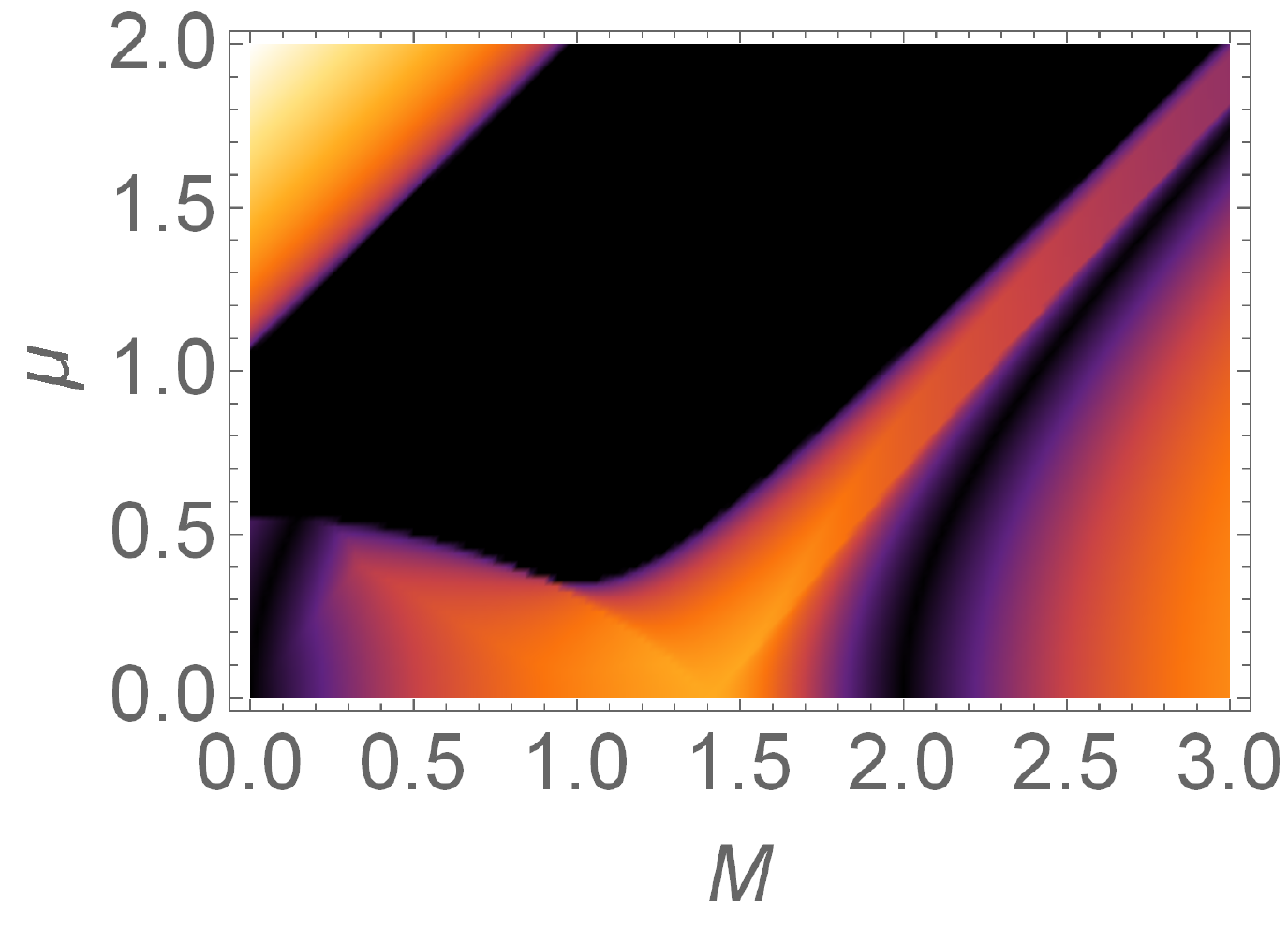}}
\subfloat[]{\includegraphics[width=.04\linewidth]{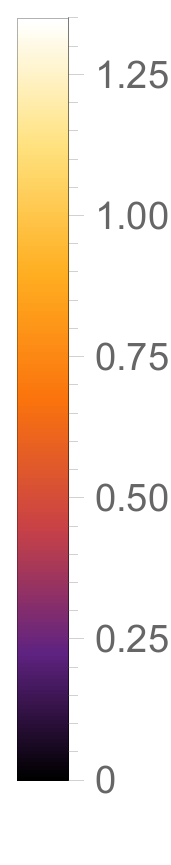}}
\caption{$|\kappa_y|$ diagrams for a) $v_0=0.5$, b) $v_0=1.3$ and c) $v_0=2$. Gapless regions (black), which demarcate the different topological phases, can form extended 2D regions, unlike in Hermitian systems where they can only be 1D boundaries separating different phase regions.}
 \label{fig:phase_kappa}
\end{figure}

The OBC spectrum takes the form of 
\begin{equation}
F_\text{Ch}(E)=\frac{1+\sin^2k_y+uv-E^2}{v^2}=z^2+\frac{b}{z}
\end{equation}
where $b=-u/v^2$, with $z$ deformed according to $z\rightarrow z\sqrt[3]{\frac{b}{2\cos(k-2\pi j/3)}}$ where $j=0,1,2$ depends on the branch. Simplifying, we can show that for the OBC gap to close,
\begin{equation}
4(2-\cos^2k_y+v_0((M+\cos k_y)^2-\mu^2))^3-27v_0^2((M+\cos k_y)^2-\mu^2)^2=0.
\end{equation}
Gapless regions correspond to parameter sets where a real $k_y$ exists. If not, the system is gapped, characterized by a nonzero $\kappa_y=\text{Im}\, k_y$, as plotted in Fig.~\ref{fig:phase_kappa}.

\end{document}